\documentclass{article}
\usepackage{amsmath,amsfonts,amssymb,amsthm}
\usepackage{bm,color,graphicx,subfig,enumerate,geometry}
\usepackage{placeins}
\usepackage{natbib,hyperref}
\usepackage[capitalize,nameinlink]{cleveref}
\renewcommand{\Re}{\operatorname{Re}}
\renewcommand{\Im}{\operatorname{Im}}
\renewcommand{\i}{\mathrm{i}}
\DeclareMathOperator{\mat}{mat}
\DeclareMathOperator{\sn}{sn}
\DeclareMathOperator{\cn}{cn}
\DeclareMathOperator{\dn}{dn}
\renewcommand{\i}{\mathrm{i}}
\newtheorem{theorem}{Theorem}[section]
\newtheorem{corollary}[theorem]{Corollary}
\newtheorem{lemma}[theorem]{Lemma}

\newtheorem{remark}[theorem]{Remark}

\title{Elliptic analogues of KP and BKP hierarchies and their tau functions}
\author{%
  Changyan Shi and Bao-Feng Feng\thanks{Corresponding author,
    \href{mailto:baofeng.feng@utrgv.edu}{baofeng.feng@utrgv.edu}.}\\[0.5em]
  {\small School of Mathematical and Statistical Sciences}\\
  {\small The University of Texas Rio Grande Valley}\\
  {\small Edinburg, TX 78539, USA}%
}
\date{}
\begin{document}
\maketitle
\begin{abstract}
We develop a tau-function framework for soliton solutions on elliptic
backgrounds associated with the Kadomtsev--Petviashvili (KP) hierarchy
and its B-type counterpart (BKP). Starting from discrete KP (dKP) and discrete BKP (dBKP) equations of elliptic type, we construct Casorati and Gram determinant
tau functions for the elliptic dKP through Darboux and binary Darboux transformations, and elliptic Pfaffian tau functions for dBKP through the Moutard transformation. Moreover, the higher-order discrete KP and BKP hierarchies together with their tau functions are constructed via Pfaffian identities. 
Then, through Miwa transformations, the continuous elliptic KP, modified KP, and BKP hierarchies and their tau functions are provided. 
Suitable constraints on spectral parameters or wave numbers yield reductions to the Korteweg--de Vries (KdV), focusing and defocusing modified Korteweg--de Vries (mKdV), and Sawada--Kotera (SK) equations, together with their solutions on elliptic backgrounds. 
We also construct KP line-soliton solutions and obtain lump solutions for the KP-I equation.
For selected determinant families, we establish sufficient
conditions for real-valuedness and global nonsingularity.
The resulting framework provides a common approach to elliptic
tau functions, their discrete and continuous bilinear identities,
and the construction of solutions through reductions.
\end{abstract}

\section{Introduction}
In the 1980s, inspired by Hirota's bilinear method \cite{Hirota2004Direct}, a framework for the integrable systems, which is called the Kadomtsev-Petviashvili (KP) theory, was developed by the Kyoto School led by Sato \cite{SatoSato1983,JimboMiwa1983,MiwaJimboDate2000}.
The KP hierarchy usually contains a sequence of positive flows $(x_1, \cdots, x_n, \cdots)$ which can be reduced to the Korteweg–de Vries (KdV) and Boussinesq equations by 2- and 3-reductions, respectively. Soon after, it was shown that the KP hierarchy can be extended to a hierarchy that also contains a sequence of negative flows $(x_{-1}, \cdots, x_{-n}, \cdots)$, which is called the \emph{KP-Toda hierarchy} \cite{UenoTakasaki1984Toda}. The KP-Toda hierarchy reduces to the sine-Gordon, the nonlinear Schr\"odinger (NLS) equations and many other soliton equations by reductions.

The KP hierarchy was built upon the  Lie algebra of $A_\infty$,  which is the infinite
dimensional analogue of the classical Lie algebras $A_n$, so it is also called the KP hierarchy of A type or simply the AKP hierarchy. Based on the Lie algebra of $B_\infty$, $C_\infty$ and $D_\infty$, the KP hierarchies of 
B, C and D type, or simply the BKP, CKP and DKP hierarchies were also constructed by the Kyoto School \cite{JimboMiwa1983}.  
The tau function, which is a single scalar function that encodes a solution of the entire hierarchy
and related with the highest weight representations of the infinite dimensional Lie algebras, plays an important role in many applications such as the construction of soliton solution.
Each hierarchy is characterized by the tau function and the associated infinitely many  bilinear equations.  The tau functions in the BKP and DKP hierarchies  have the Pfaffian structure  whereas in  the AKP and CKP hierarchies they take the form of determinant.  
Since their introduction, the range of applications of the tau functions associated to various integrable systems  has expanded well beyond the original scope. Applications have been found in random matrix theory and the associated theory of multi-orthogonal polynomials \cite{AdlerVanMoerbekeVanhaecke2009}, enumerative geometry \cite{Kontsevich1992}, combinatorics \cite{GouldenJackson2008}, symplectic geometry \cite{Givental2001}, theory of isomonodromic deformations \cite{JimboMiwaUeno1981} and integrable probability \cite{Okounkov2001}.

 Theory of discrete integrable systems forms a very important branch for the whole theory of integrable systems \cite{HietarintaJoshiNijhoff2016}.
The Toda lattice provided an early example of a nonlinear
lattice treated by the inverse scattering method
\cite{Flaschka1974InverseScattering,Manakov1975}.
For broader accounts of nonlinear waves and lattice solitons,
we refer to Toda's books
\cite{Toda1995Waves,Toda2000NonlinearWaves}.
The discrete KP hierarchies of A, B, C and D type play a fundamental role in integrable systems:
they are the generating equations for the corresponding hierarchies of continuous integrable equations
and encode the basic algebraic structures associated with these hierarchies. At the level of hierarchy prior to reductions, the discrete variables and the continuous variables are connected by the Miwa transformation \cite{Miwa1982Hirota}. The discrete and continuous hierarchies sit at the same footing and are characterized by tau functions satisfying infinitely many Hirota's bilinear equations.

The discrete KP  equation 
\begin{equation}\label{discrete_kp}
  a_1(a_2 - a_3)\tau_1 \tau_{23} + a_2(a_3 - a_1)\tau_2 \tau_{13} + a_3(a_1 - a_2)\tau_3 \tau_{12} = 0
\end{equation}
was introduced by Hirota in 1981 \cite{Hirota1981Discrete}, and Miwa subsequently connected it systematically with the KP hierarchy through the Miwa transformation \cite{Miwa1982Hirota}. Thus 
the discrete KP equation is often called the Hirota-Miwa equation and is described as a generating equation for the KP/Toda hierarchies through the Miwa transformation.

The discrete KP equation acts as a master equation from which many lower-dimensional discrete integrable equations can be obtained by reductions. 
Reductions of the Hirota--Miwa equation include discrete Toda- and
KdV-type systems \cite{Hirota1981Discrete,DateJimboMiwa1983III}.
Suitable further reductions yield finite-dimensional integrable maps
and three-term Gale--Robinson recurrences, including Somos-4 and
Somos-5
\cite{FominZelevinsky2002,FordyHone2014,HoneKouloukasWard2017}.
In \cite{MiyakeOhtaSatsuma1990,OhtaHirotaTsujimotoImai1993}, the authors constructed the soliton solutions to the discrete and continuous KP hierarchies. Many familiar \(1+1\)-dimensional soliton equations such as the KdV, modified KdV, sine-Gordon and the NLS equations arise from the KP-Toda hierarchy by a reduction procedure called the KP-Toda reduction \cite{Hirota1981Discrete,DateKashiwaraMiwa1981Vertex,DateJimboMiwa1983III,DateJimboMiwa1983IV,JimboMiwa1983}.

The continuous BKP hierarchy was introduced systematically by Date, Jimbo, Kashiwara, and Miwa using \(\tau\)-functions, vertex operators, Clifford algebras, and an infinite-dimensional orthogonal-group structure \cite{DateJimboKashiwaraMiwa1982BKP}. They also showed that the Sawada–Kotera equation arises as a reduction of BKP.
The discrete BKP (Miwa) equation originated in the discrete-soliton framework developed by Date, Jimbo and Miwa in the early 1980s \cite{DateJimboMiwa1982} \cite{Miwa1982Hirota}.
Hirota developed the Pfaffian technique for continuous BKP \cite{Hirota1989BKP}. Tsujimoto and Hirota (1996) established a systematic Pfaffian representation for the discrete BKP hierarchy \cite{TsujimotoHirota1996Pfaffian}.

The lattice CKP equation belongs to the \(C_\infty\)-type KP family, alongside lattice AKP and BKP. Unlike the Hirota–Miwa AKP equation, its \(\tau\)-function equation is not simply bilinear; it has a quartic structure which was firstly derived by \cite{Schief2003Lattice}.
The coupled KP hierarchy was introduced by 
Hirota and Ohta \cite{HirotaOhta1991} through Hirota bilinear equations. A discrete version of the coupled KP hierarchy, also called discrete DKP hierarchy, was derived in \cite{GilsonNimmoTsujimoto2001} by pfaffianizing the ordinary discrete KP equation.
The Wronski- and Gram-Pfaffian solutions were developed in \cite{Ohta2004Special} \cite{LiOhta2005}.

Finite-gap solutions provide a fundamental link between integrable
equations, algebraic geometry, and Riemann theta functions. For the KdV
equation, a real, nonsingular finite-gap solution is associated with a
Schr\"odinger operator whose spectrum consists of finitely many bands
separated by open gaps; the corresponding solution is expressed through
the Riemann theta function of a hyperelliptic spectral
curve \cite{ItsMatveev1975,DubrovinMatveevNovikov1976}.
Soliton solutions arise through suitable degenerations in which spectral
bands shrink to points, geometrically corresponding to the pinching of
handles of the spectral curve. With an appropriate scaling of the phase
parameters, complete degeneration reduces the theta function to a finite
sum of exponentials. More generally, pinching \(N\) handles of a
genus-\((N+1)\) spectral curve produces a nodal curve whose normalization
has genus one. Under suitable reality and regularity conditions, this
partial degeneration yields an \(N\)-soliton solution on an elliptic,
or cnoidal, background, with the limiting tau function expressed as a
finite sum of exponential terms multiplied by shifted genus-one theta
functions \cite{BertolaJenkinsTovbis2023}.


As a matter of fact, elliptic periodic solutions play an important role in integrable systems
because they describe nonlinear wave trains within a framework that also
includes small-amplitude trigonometric waves and isolated solitons as
limiting cases. They therefore provide a natural setting for understanding
the transition between periodic and localized wave motion. Their
significance is also historical: Toda's search for exact solutions
expressed through Jacobi elliptic functions, together with the use of
their addition formula, led him to identify the exponential interaction
underlying the Toda lattice. This discovery illustrates how the analytic
structure of elliptic functions can guide the construction of integrable
nonlinear models \cite{Toda1967,Toda1995Waves,Toda2000NonlinearWaves}.
Soliton solutions on elliptic backgrounds extend this framework to the interaction of localized excitations with nonlinear periodic waves. Such
solutions allow the propagation speeds, internal oscillations, and
induced phase shifts of localized disturbances to be studied explicitly.
For example, the KdV equation admits bright and dark breathers on a cnoidal
background, which propagate respectively faster and slower than the
underlying wave and produce a phase shift across the localized excitation.
These solutions also help explain soliton interactions with dispersive
shock waves, whose oscillatory structure can be described by modulated
cnoidal waves \cite{Hoefer_2023}.
A related motivation
comes from the focusing nonlinear Schr\"odinger equation, where
modulational instability generates expanding regions of slowly modulated
elliptic waves. Interactions with these regions can transmit or trap an
incident soliton and, for suitable spectral parameters, generate an
additional localized wake. Studying solitons on elliptic backgrounds thus
provides explicit models for understanding how nonlinear wave trains
influence the transport and localization of coherent excitations \cite{BiondiniMantzavinos2017}.

A natural question arises: how to construct the general soliton solution on the elliptic background? Actually, in the literature,   
soliton solutions on elliptic backgrounds can be constructed mainly through three complementary approaches: finite-gap degeneration, Darboux transformations, and elliptic direct linearisation. In the finite-gap approach, one starts from a higher-genus Riemann-theta solution and partially degenerates the spectral curve, with the surviving genus-one component providing the elliptic background and the pinched handles producing solitons; this viewpoint is exemplified by Bertola, Jenkins, and Tovbis \cite{BertolaJenkinsTovbis2023}. 

Early spectral and algebro-geometric treatments of periodic solutions
of the KdV equation and the Toda lattice were given by Date and Tanaka
\cite{DateTanaka1976Inverse,DateTanaka1976Periodic}. In the Darboux approach developed extensively by Ling and
collaborators, one starts directly from a genus-one elliptic seed solution and introduces discrete spectral data through successive Darboux–B\"acklund transformation, producing solitons, breathers, and rogue waves on the elliptic background 
\cite{FengLingTakahashi2020,LingSun2023SineGordon,LingSun2023mKdV,LingSun2024NLS,LingSun2024EllipticRogue,LingTang2026DNLS,FengTangYu2026}.
A spectral and Darboux-transformation approach to rogue waves on periodic backgrounds was developed by  Pelinovsky and his collaborators.   Starting from Jacobi-elliptic (dn)- and (cn)-type standing waves, they explicitly characterize the associated Zakharov–Shabat spectral problem and use its special eigenfunctions to construct rogue waves on elliptic backgrounds, thereby generalizing the Peregrine solution beyond the constant-background case  for the focusing nonlinear Schr\"odinger equation \cite{ChenPelinovsky2018NLS}. Their work further establishes a close connection between the Lax spectrum, modulational instability of the periodic background, and the existence and amplification of localized rogue waves, and extends the construction to more general double-periodic backgrounds \cite{ChenPelinovskyWhite2019,ChenPelinovskyWhite2020}.  Chen and Pelinovsky extended the spectral and Darboux-transformation theory of rogue waves on periodic backgrounds to focusing mKdV \cite{ChenPelinovsky2018mKdV},
the derivative NLS  \cite{ChenPelinovsky2021DNLS} and 
the Ablowitz–Ladik lattice  equations
\cite{ChenPelinovsky2018mKdV,ChenPelinovsky2021DNLS}.
Li, Geng, and collaborators have developed a systematic approach for constructing localized nonlinear waves on periodic and theta-function backgrounds for a broad class of continuous and semi-discrete integrable systems, including the sine–Gordon, Yajima–Oikawa, KPI, Fokas–Lenells, nonlinear Schrödinger, and semi-discrete KdV equations. A characteristic feature of their approach is to combine algebro-geometric methods with Darboux transformations \cite{LiGeng2023KPI,LiGengGeng2023FokasLenells,LiGeng2023DerivativeYO,LingSun2023SineGordon}.
 In the direct-linearization approach, used for elliptic KP and lattice integrable systems by Nijhoff and collaborators, the elliptic background is incorporated into a Lamé/elliptic Cauchy kernel, while a finite discrete spectral measure generates the soliton dressing. Although their starting points differ, all three approaches encode essentially the same structure: a genus-one background supplemented by discrete spectral degrees of freedom representing localized nonlinear excitations.

Although Hirota's bilinear method and the KP reduction method are powerful tools for constructing multisoliton solutions, further work is needed to develop the KP reduction method into a systematic approach to soliton solutions on elliptic backgrounds. In this setting, the elementary exponential factors of the usual soliton construction are replaced by Lam\'e-type factors involving Weierstrass functions. The bilinear 
equations of the integrable hierarchy
must be adapted using elliptic-function identities~\cite{Li2022Elliptic}. Beyond obtaining explicit solutions for individual equations, a systematic approach requires elliptic tau functions satisfying the relevant hierarchy identities, together with compatible reduction conditions that can be implemented at the level of their spectral data.

Li and Zhang~\cite{Li2022Elliptic} developed a bilinear framework for elliptic soliton solutions, treating the KdV equation in detail and extending the construction to the KP hierarchy. They derived tau functions in Hirota's form, constructed vertex operators that generate these functions, and established the associated bilinear identities. Their work connects the addition identities of Weierstrass functions and the Lam\'e spectral problem with the algebraic structure of soliton tau functions. 
These results provide an elliptic counterpart of the classical Hirota and vertex-operator description of solitons.

Kakei~\cite{Kakei2023Solutions} subsequently described a class of solutions of the KP hierarchy on an elliptic background within the pseudo-differential operator framework. Starting from a stationary elliptic background, he applied successive Darboux transformations and recovered the determinantal solutions of Li and Zhang. This construction gives an operator interpretation of the elliptic tau functions and explains how the soliton degrees of freedom arise through dressing the background solution. 
These developments connect explicit elliptic soliton formulae with both the bilinear and the operator formulations of the KP hierarchy.



An earlier development in the discrete setting was due to Yoo-Kong and Nijhoff~\cite{yoo2013elliptic}, who constructed elliptic $(N,N')$-soliton solutions for a family of lattice KP-type equations using elliptic Cauchy matrices. Their framework incorporates an elliptic seed into the matrix kernel and provides determinant tau functions and associated solutions of the bilinear, potential, modified, and Schwarzian lattice KP equations.
In particular, they proposed a  discrete KP equation of elliptic type
\begin{equation}\label{elliptic_dkp_intro}
  \sigma(a_1)\sigma(a_2 - a_3)\tau_1 \tau_{23} + \sigma(a_2)\sigma(a_3 - a_1)\tau_2 \tau_{13} + \sigma(a_3)\sigma(a_1 - a_2)\tau_3 \tau_{12} = 0,
\end{equation}
where $\sigma$ is the Weierstrass sigma function.
The accompanying determinant solutions supply an explicit starting point for studying discrete AKP equations and soliton interactions on elliptic backgrounds.

The connection with the BKP equation was explored by Li, Nijhoff, Sun, and Zhang~\cite{li2021symmetric}. They showed that, when the discrete AKP tau function possesses appropriate reflection symmetries in the lattice variables and their parameters, the discrete BKP (Miwa) equation can be expressed as a linear combination of the discrete AKP equation and its reflected forms. Consequently, these two equations share a class of tau functions satisfying the required symmetries. The authors extended this connection to certain higher-order discrete equations and, by considering symmetric elliptic Cauchy-matrix tau functions, obtained a discrete BKP equation with elliptic coefficients
\begin{align}
    \begin{split}\label{elliptic_dbkp_intro}
        &\sigma(a_1 - a_2)\sigma(a_2 - a_3)\sigma(a_1 - a_3)\tau_{123}\tau - \sigma(a_1 - a_2)\sigma(a_2 + a_3)\sigma(a_1 + a_3)\tau_{12}\tau_3 \\
    &+ \sigma(a_1 - a_3)\sigma(a_2 + a_1)\sigma(a_2 + a_3)\tau_{13}\tau_2 - \sigma(a_2 - a_3)\sigma(a_1 + a_2)\sigma(a_1 + a_3)\tau_{23}\tau_1 = 0.
    \end{split}
\end{align}
However, this construction concerns a common class of AKP and BKP solutions; an elliptic Pfaffian construction for BKP was not provided in that work. Developing such a construction, together with the corresponding higher-order bilinear identities, is a central motivation for the present study.

All above results just mentioned  provide important ingredients for extending the KP reduction method to elliptic backgrounds. Turning these ingredients into a broadly applicable construction procedure nevertheless requires further development. The higher-order bilinear identities, determinant and Pfaffian representations, and spectral reduction conditions must be brought into a common framework that connects the discrete and continuous hierarchies. In addition, reductions must be compatible with the elliptic background, and the resulting solutions must be examined for reality and regularity. These requirements motivate the development of a method in which both the reduced equations and their solutions can be derived systematically from the underlying tau functions.

In the present paper, building on these developments by Nijhoff, Zhang, and their collaborators, we address several of these requirements by constructing elliptic analogues of the discrete AKP and BKP hierarchies and deriving their tau functions in determinant and Pfaffian forms, respectively. For the AKP hierarchy, Darboux and binary Darboux transformations yield Casorati and Gram determinant solutions. For the BKP hierarchy, the Moutard transformation produces elliptic Pfaffian solutions. We establish the associated higher-order discrete bilinear equations through determinant and Pfaffian identities, thereby placing the explicit solutions within their respective hierarchies.

We then apply the Miwa transformation to derive the continuous elliptic KP, modified KP, and BKP hierarchies together with their tau functions. 
The resulting tau functions provide a common basis for constructing solutions through suitable reductions and choices of spectral data. We illustrate this approach with line solitons of the KP equation, KP-I lump solutions, 
and soliton solutions of the KdV, focusing and defocusing mKdV, and Sawada--Kotera equations on elliptic backgrounds. These constructions constitute a step toward a systematic KP reduction method for elliptic backgrounds. Extending its applicability to a wider range of equations will require further development of the underlying hierarchies, their admissible reductions, and the associated reality and regularity conditions.

The remainder of the paper is organized as follows. Section~2 presents the discrete AKP hierarchy of elliptic type, its Lax formulation, and its Casorati and Gram determinant tau functions obtained through Darboux and binary Darboux transformations. Section~3 develops the discrete BKP hierarchy of elliptic type and constructs its tau functions using the Moutard transformation and Pfaffian identities. In Section~4, the Miwa transformation is employed to derive the continuous elliptic KP, modified KP, and BKP hierarchies together with their tau functions. Section~5 illustrates the reduction approach through KP line solitons, KP-I lump solutions, and soliton solutions of the KdV, focusing and defocusing mKdV, and Sawada--Kotera equations on elliptic backgrounds. Section~6 summarizes the main results and discusses the further developments needed to broaden the scope of the method. Appendix~A collects the definitions, conventions, and identities for the Weierstrass, Jacobi elliptic, and theta functions used in the paper.
\section{Discrete KP hierarchy of the elliptic type}

\subsection{Lax formulation and Darboux transformation to the discrete KP equation of the elliptic type}
In \cite{yoo2013elliptic,li2021symmetric}, they studied the discrete KP (also known as the Hirota-Miwa) equation in elliptic form \eqref{elliptic_dkp_intro}
\begin{align}
    &\Phi_{a_2}(-a_3)\tau_1 \tau_{23} + \Phi_{a_3}(-a_1)\tau_2 \tau_{13} + \Phi_{a_1}(-a_2)\tau_3 \tau_{12} = 0,\label{elliptic_dkp}
\end{align}where  \(\Phi_a (b) := \frac{\sigma(a+b)}{\sigma(a)\sigma(b)}\) is called Lam\'e-type function \cite{NijhoffSunZhang2023} or the elliptic Cauchy kernel \cite{zabrodin2024elliptic},
\(\tau = \tau(n_1, n_2, n_3)\) and \(f_i := f|_{n_i \to n_i+1}, i = 1,2,3\).

In this subsection, we would like to present the Lax formulation and introduce the Darboux and binary Darboux transformations for deriving the solution to Eq.~\eqref{elliptic_dkp}. One can check the following linear problem
\begin{align}\label{edkp_lin_prob_tau}
    \frac{\tau \tau_{ij}}{\tau_i \tau_j}\frac{\sigma(a_j-a_i)}{\sigma(a_i)\sigma(a_j)}\phi_{ij} &= \phi_j - \phi_i, \quad 1\leq i<j \leq 3,
\end{align}
is the Lax pair for \eqref{elliptic_dkp} under the compatibility condition \((\phi_{12})_3 = (\phi_{13})_2 = (\phi_{23})_1\). Let us also introduce the associated adjoint linear problem
\begin{align}\label{edkp_adj_lin_prob_tau}
    \frac{\tau_{i'j'} \tau}{\tau_{i'} \tau_{j'}}\frac{\sigma(a_j-a_i)}{\sigma(a_i)\sigma(a_j)}\psi_{i'j'} &= \psi_{j'} - \psi_{i'}, \quad \text{or}\quad \frac{\tau \tau_{ij}}{\tau_i \tau_j}\frac{\sigma(a_j-a_i)}{\sigma(a_i)\sigma(a_j)}\psi = \psi_i - \psi_j, \quad 1\leq i<j \leq 3,
\end{align}
where \(f_{i'}:=f|_{n_i \to n_i-1}, i = 1,2,3\).

The Hirota–Miwa equation then follows essentially from the Fay trisecant identity for the Riemann theta function.
The following addition identity of the Weierstrass sigma function will be utilized throughout the paper:
\begin{align}
    \begin{split}\label{three_term_sigma}
    &\sigma(x + a)\sigma(x - a)\sigma(y + b)\sigma(y - b) - \sigma(x + b)\sigma(x - b)\sigma(y + a)\sigma(y - a) \\
    &= \sigma(x + y)\sigma(x - y)\sigma(a + b)\sigma(a - b).
  \end{split}
\end{align}
Define \(\xi=\sum_{i=1}^3 n_i a_i\), and put \(x = \frac{\xi}{2}+a_1, \ y = \frac{\xi}{2}+a_2, \ a = \frac{\xi}{2}, \ b = \frac{\xi}{2}+a_3\) in \eqref{three_term_sigma}; we have the following seed solution to Eq.~\eqref{elliptic_dkp}.
For the genus one case, the seed solution is the Weierstrass sigma function. This is exactly why the sigma-function addition identity occurs so naturally in elliptic lattice KP equations. 
\begin{lemma}\label{lma:vac_edkp}
  \(\sigma(\xi)\) is a seed solution to Eq.~\eqref{elliptic_dkp}, where \(\xi=\sum_{i=1}^3 n_i a_i\).
\end{lemma}
Subsequently, with the above seed solution, the linear and adjoint linear problems \eqref{edkp_lin_prob_tau}-\eqref{edkp_adj_lin_prob_tau} become
\begin{align}
    \frac{\sigma(\xi) \sigma(\xi_{ij})}{\sigma(\xi_i) \sigma(\xi_j)}\frac{\sigma(a_j-a_i)}{\sigma(a_i)\sigma(a_j)}\phi_{ij} &= \phi_j - \phi_i, \label{edkp_lax_pair}\\
    \frac{\sigma(\xi) \sigma(\xi_{ij})}{\sigma(\xi_i) \sigma(\xi_j)}\frac{\sigma(a_j-a_i)}{\sigma(a_i)\sigma(a_j)}\psi &= \psi_i - \psi_j. \label{edkp_adj_lax_pair}
\end{align}
\begin{lemma}\label[lemma]{prop:eigen_func_edkp}
  The following eigenfunctions
  \begin{align*}
    \phi(p) := \Phi_{\xi}(p) \prod_{l=1}^3\Phi_{a_l} (p)^{-n_l}, \quad \psi(q) := \Phi_{\xi}(q) \prod_{l=1}^3\Phi_{a_l} (-q)^{n_l}
  \end{align*}
  satisfy the above linear problems.
\end{lemma}
\begin{proof}
    For the linear problem, let us consider
    \begin{align*}
      \phi_j - \phi_i &= \left(\prod_{l=1}^3\Phi_{a_l} (p)^{-n_l}\right) \left[\frac{\Phi_{\xi_j}(p)}{\Phi_{a_j}(p)} - \frac{\Phi_{\xi_i}(p)}{\Phi_{a_i}(p)}\right]\\
                      &= \left(\prod_{l=1}^3\Phi_{a_l} (p)^{-n_l}\right) \frac{1}{\Phi_{a_i}(p)\Phi_{a_j}(p)} \left[\Phi_{\xi_j}(p)\Phi_{a_i}(p) -  \Phi_{\xi_i}(p)\Phi_{a_j}(p)\right].
    \end{align*}
    With \(\Phi_{\xi_j}(p)\Phi_{a_i}(p) = \Phi_{\xi_{ij}}(p) \left[\zeta(p) +
        \zeta(\xi_j) + \zeta(a_i) - \zeta(p+\xi_{ij})\right]\),
    this gives
    \begin{align*}
      \phi_j - \phi_i &= \left(\prod_{l=1}^3\Phi_{a_l} (p)^{-n_l}\right) \frac{\Phi_{\xi_{ij}}(p)}{\Phi_{a_i}(p)\Phi_{a_j}(p)} \left[\zeta(\xi_j) + \zeta(a_i) + \zeta(-a_j) - \zeta(\xi_i)\right] \\
                      &= \phi_{ij} \frac{\Phi_{a_i}(\xi_j)\Phi_{a_i}(-a_j)}{\Phi_{a_i}(\xi_j-a_j)} = \phi_{ij} \frac{\sigma(a_j-a_i)}{\sigma(a_i)\sigma(a_j)}\frac{\sigma(\xi)\sigma(\xi_{ij})}{\sigma(\xi_i)\sigma(\xi_j)}.
    \end{align*}
    Similarly, for the adjoint linear problem, we have
    \begin{align*}
      \psi_i - \psi_j &= \left(\prod_{l=1}^3 \Phi_{a_l} (-q)^{n_l}\right)
                        \left[\Phi_{\xi_i}(q) \Phi_{a_i} (-q)- \Phi_{\xi_j}(q) \Phi_{a_j} (-q)\right]  \\
                      &= -\left(\prod_{l=1}^3 \Phi_{a_l} (-q)^{n_l}\right)
                        \left[\Phi_{\xi_i}(q) \Phi_{-a_i} (q)-\Phi_{\xi_j}(q) \Phi_{-a_j} (q)\right] \\
                      &= -\left(\prod_{l=1}^3 \Phi_{a_l} (-q)^{n_l}\right) \Phi_{\xi}(q) \left[\zeta(\xi_i)+\zeta(-a_i)+\zeta(a_j)-\zeta(\xi_j)\right] \\
                      &= -\psi \frac{\Phi_{a_j}(\xi_i)\Phi_{a_j}(-a_i)}{\Phi_{a_j}(\xi_i-a_i)} =  \psi \frac{\sigma(a_j-a_i)}{\sigma(a_i)\sigma(a_j)}\frac{\sigma(\xi)\sigma(\xi_{ij})}{\sigma(\xi_i)\sigma(\xi_j)}.
    \end{align*}
    This completes the proof of \cref{prop:eigen_func_edkp}.
\end{proof}

\begin{theorem}
    Based on the \(N\)-fold Darboux transformation to the discrete KP equation \cite{nimmo1997darboux}, we can derive the Casorati determinant solution of the elliptic type to \eqref{elliptic_dkp}. Given the seed solutions \(\varphi_0\) and \(\tau_0\) that solve the equations \eqref{edkp_lin_prob_tau} and \eqref{elliptic_dkp}, respectively, and a series of eigenfunctions \(\varphi^1(n),\varphi^2(n), \ldots, \varphi^N(n)\) of \eqref{edkp_lin_prob_tau}, the transformation
    \begin{align*}
        \varphi^0 \to \frac{\begin{vmatrix}
            \Phi^+(0) & \Phi^+(-1) & \cdots & \Phi^+(-N)
        \end{vmatrix}}{\begin{vmatrix}
            \Phi(0) & \Phi(-1) & \cdots & \Phi(1-N)
        \end{vmatrix}}, \quad \tau_0 \to \tau_0 \begin{vmatrix}
            \Phi(0) & \Phi(-1) & \cdots & \Phi(1-N)
        \end{vmatrix} 
    \end{align*}
    leaves the Lax pair \eqref{edkp_lin_prob_tau} invariant, where 
    \begin{align*}
        \Phi^+(n) = (\varphi^1(n), \varphi^2(n), \ldots, \varphi^N(n), \varphi^0(n))^t,\quad 
        \Phi(n) = (\varphi^1(n), \varphi^2(n), \ldots, \varphi^N(n))^t.
    \end{align*}
    Here, \(s>3\) is a special direction such that for any distinct \(i,j,k \in \{1,2,3,s\}\), \(\tau(n_i, n_j, n_k)\) satisfies \eqref{elliptic_dkp}, where \(\varphi(n) = \varphi|_{n_s \to n_s + n}\). To get the explicit expression of the soliton solution, we pick
    \begin{align*}
        &\varphi^j(n) := \alpha_j\phi(p_j;n) + \beta_j\phi(q_j;n), \quad \phi(p;n):= \left[\Phi_{b} (p)\right]^{-n} \Phi_{\xi(n)}(p) \prod_{l=1}^3 \left[\Phi_{a_l} (p)\right]^{-n_l},\\
        &\tau_0 = \sigma(\xi), \quad \xi = bn + \sum_{i=1}^3 a_i n_i.
    \end{align*}
    As a result, the general \(N\)-soliton solution to \eqref{elliptic_dkp} in Casorati form is written as
    \begin{align*}
        \tau &= \sigma(\xi) \begin{vmatrix}
                 \varphi^1(0) & \varphi^1(-1) & \cdots & \varphi^1(1-N) \\
                 \varphi^2(0) & \varphi^2(-1) & \cdots & \varphi^2(1-N) \\
                 \vdots & \vdots &  & \vdots \\
                 \varphi^N(0) & \varphi^N(-1) & \cdots & \varphi^N(1-N)
             \end{vmatrix}.
    \end{align*}
\end{theorem}

Moreover, as discussed in \cite{nimmo1997darboux}, with given eigenfunctions satisfying \eqref{edkp_lax_pair}-\eqref{edkp_adj_lax_pair}, there must exist a discrete potential \(m = m(\phi,\psi)\) satisfying
\begin{align}\label{pot_m}
    \Delta_i m = \psi \phi_i, \quad i = 1,2,3,
\end{align}
where \(\Delta_i f := f_i - f\).
\begin{lemma}\label[lemma]{prop:pot_m}
    The discrete potential
    \begin{align*}
        m = m(\phi, \psi) = c - \Phi_{\xi}(p+q)\prod_{l=1}^3 \left[\frac{\Phi_{a_l} (-q)}{\Phi_{a_l} (p)}\right]^{n_l}
    \end{align*}
    satisfies \eqref{pot_m}.
\end{lemma}
\begin{proof}
    Since \(\Phi_{\xi}(p+q) \Phi_{a_1}(q) = \Phi_{\xi}(p) \Phi_{\xi_1}(q) +
      \Phi_{\xi_1}(p+q)\Phi_{a_1}(-p)\), we have
    \begin{align*}
      m_i - m &= -\left[\Phi_{\xi_i} (p+q) \frac{\Phi_{a_i} (-q)}{\Phi_{a_i} (p)} - \Phi_{\xi} (p+q)\right] \prod_{l=1}^3 \left[\frac{\Phi_{a_l} (-q)}{\Phi_{a_l} (p)}\right]^{n_l}\\
                                &= -\frac{\left(\Phi_{\xi_i} (p+q)\Phi_{a_i} (-q) - \Phi_{\xi} (p+q) \Phi_{a_i} (p)\right) }{\Phi_{a_i} (p)}\prod_{l=1}^3 \frac{\Phi_{a_l} (-q)^{n_l}}{\Phi_{a_l} (p)^{n_l}}\\
                                &= \Phi_{\xi}(q) \Phi_{\xi_i}(p)\prod_{l=1}^3 \frac{\Phi_{a_l} (-q)^{n_l}}{\Phi_{a_l} (p)^{n_l}}\frac{1}{\Phi_{a_i} (p)} = \phi_i \psi.
    \end{align*}
    This completes the proof of \cref{prop:pot_m}.
\end{proof}

\begin{theorem}
    By use of the \(N\)-fold binary Darboux transformation to the discrete KP equation \cite{nimmo1997darboux}, the Gram determinant solution of the elliptic type to \eqref{elliptic_dkp} can be derived. Given the eigenfunctions \(\phi(p_1), \ldots, \phi(p_N)\) satisfying the linear problem \eqref{edkp_lin_prob_tau} and \(\psi(q_1), \ldots, \psi(q_N)\) satisfying the adjoint linear problem \eqref{edkp_adj_lin_prob_tau}, the transformation 
    \begin{align*}
    &\phi_0 \to \frac{\begin{vmatrix} \phi_0 &m_{0,j} \\ \phi_i & m_{i,j} \end{vmatrix}_{1\leq i,j\leq N}
      }{|m_{i,j}|_{1\leq i,j\leq N}}, \quad \psi_0 \to \frac{\begin{vmatrix} \psi_0 & \psi_j \\ m_{i,0} & m_{i,j} \end{vmatrix}_{1\leq i,j\leq N}
      }{|m_{i,j}|_{1\leq i,j\leq N}}, \quad \tau_0 \to \tau_0|m_{i,j}|_{1\leq i,j\leq N} ,
    \end{align*}
    leaves the Lax pair and adjoint Lax pair \eqref{edkp_lin_prob_tau}-\eqref{edkp_adj_lin_prob_tau} invariant, where this time the subscripts mean
    \begin{align*}
      \phi_j := \phi(p_j), \quad \psi_j := \psi(q_j), \quad m_{i,j}:=m(\phi(p_i), \psi(q_j)), \quad m_{0,j}:=m(\phi_0,\psi_j),\quad m_{i,0}:=m(\phi_i,\psi_0).
    \end{align*}
    Based on this, we pick
    \begin{align*}
        &\phi_j := \Phi_{\xi}(p_j) \prod_{l=1}^3\Phi_{a_l} (p_j)^{-n_l}, \quad \psi_j := \Phi_{\xi}(q_j) \prod_{l=1}^3\Phi_{a_l} (-q_j)^{n_l}, \quad \tau_0 = \sigma(\xi),\\
        &m_{k,l} := c_{kl} - \Phi_{\xi}(p_k+q_l)\prod_{i=1}^3 \left[\frac{\Phi_{a_i} (-q_l)}{\Phi_{a_i} (p_k)}\right]^{n_i}, \quad \xi = \sum_{i=1}^3 a_i n_i,
    \end{align*}
    to derive the general \(N\)-soliton solution to \eqref{elliptic_dkp} in Gram determinant form
    \begin{align*}
        \tau &= \sigma(\xi) \begin{vmatrix}
                 m_{1,1} & m_{1,2} & \cdots & m_{1,N} \\
                 m_{2,1} & m_{2,2} & \cdots & m_{2,N} \\
                 \vdots & \vdots &  & \vdots \\
                 m_{N,1} & m_{N,2} & \cdots & m_{N,N}
             \end{vmatrix},
    \end{align*}
    which recovers the results in Refs.~\cite{yoo2013elliptic,li2021symmetric}. 
\end{theorem}

\begin{remark}
    The normalized discrete KP equation
    \begin{align*}
      \tau_1 \tau_{23} - \tau_2 \tau_{13} + \tau_3 \tau_{12} = 0
    \end{align*}
    is obtained by putting the transformation
    \[
      \tau \to \left(\frac{\sigma(a_3 -
          a_2)}{\sigma(a_2)\sigma(a_3)}\right)^{-n_2n_3}\left(\frac{\sigma(a_3
          -
          a_1)}{\sigma(a_1)\sigma(a_3)}\right)^{-n_1n_3}\left(\frac{\sigma(a_2
          - a_1)}{\sigma(a_1)\sigma(a_2)}\right)^{-n_1n_2} \tau
    \]
    in Eq.~\eqref{elliptic_dkp}.
\end{remark}

\subsection{General soliton solutions to the discrete KP hierarchy of the elliptic type}
The general discrete KP hierarchy of the elliptic type is given in this form
\begin{align}\label{edkp_hierarchy}
  & \sum_{i\in A}(-1)^{i-1} \prod_{\substack{j<k\\ j,k\in A\setminus \{i\}}} \Phi_{a_j}(-a_{k}) \prod_{\nu\in B} \Phi_{a_{i}}(-a_{\nu}) \tau_{A \setminus \{i\}} \tau_{B \cup \{i\}}  =0 
\end{align}
together with the generalized \(\tau = \tau(n_1, \ldots, n_{K})\) in the discrete variables \(n_j\) with associated parameters \(a_j\), where \(K=I+J\), \(0\leq J\leq I-2\), and the index sets are \(A=\{1,\ldots,I\}\) and \(B=\{I+1,\ldots,K\}\), and for  \(S\subseteq\{1,\ldots,K\}\) we denote
\begin{align*}
    \tau_S &:= \tau|_{n_j\to n_j+1} \text{ for all } j\in S, \quad \tau_\varnothing = \tau.
\end{align*}
As an example, the discrete KP in elliptic form \eqref{elliptic_dkp} is a special case of \eqref{edkp_hierarchy} at \(I=3, J=0\). More examples include
\begin{align}
    I=3, J=1:\nonumber\\
    \begin{split}
        & \Phi_{a_2}(-a_3) \Phi_{a_1}(-a_4) \tau_{23} \tau_{14} - \Phi_{a_1}(-a_3) \Phi_{a_2}(-a_4) \tau_{13} \tau_{24} + \Phi_{a_1}(-a_2) \Phi_{a_3}(-a_4) \tau_{12} \tau_{34} = 0,
    \end{split} \\
    I=4, J=0:\nonumber\\
    \begin{split}
        & \Phi_{a_2}(-a_3) \Phi_{a_2}(-a_4) \Phi_{a_3}(-a_4) \tau_{1} \tau_{234} - \Phi_{a_1}(-a_3) \Phi_{a_1}(-a_4) \Phi_{a_3}(-a_4) \tau_{2} \tau_{134} \\
        & \quad + \Phi_{a_1}(-a_2) \Phi_{a_1}(-a_4) \Phi_{a_2}(-a_4) \tau_{3} \tau_{124} - \Phi_{a_1}(-a_2) \Phi_{a_1}(-a_3) \Phi_{a_2}(-a_3) \tau_{4} \tau_{123} = 0,
    \end{split} \\
    &\cdots\nonumber
\end{align}

\begin{theorem}\label[theorem]{lma:casorati_edkp_hierarchy}
    The Casorati determinant solution to \eqref{edkp_hierarchy} is given by
    \begin{align}\label{edkp_tau_casorati}
        \tau = \sigma(\xi) \begin{vmatrix}
            \Phi(0) & \Phi(-1) & \Phi(-2) & \cdots & \Phi(1-N)
        \end{vmatrix},
    \end{align}
    where \(\xi = b n+\sum_{l=1}^K a_l n_l\), \(\Phi(n)=(\varphi^1(n), \varphi^2(n), \ldots, \varphi^N(n))^t\), and
    \begin{align*}
        \varphi^j(n) &:= \alpha_j\phi(p_j;n) + \beta_j\phi(q_j;n), \quad \phi(p;n):= \left[\Phi_{b} (p)\right]^{-n} \Phi_{\xi(n)}(p) \prod_{l=1}^K \left[\Phi_{a_l} (p)\right]^{-n_l}.
    \end{align*}
\end{theorem}

To show the above theorem, let us firstly show this.
\begin{lemma}
    The difference formula for the \(\tau\) function in Casorati determinant form \eqref{edkp_tau_casorati} with one increment in \(K\) different directions is
    \begin{align}
        &\tau(n_1 +1, n_2+1,\ldots, n_K+1)\frac{\sigma^{K-1}(\xi)}{\prod^K_{j=1}  \sigma(\xi_j)} 
        \prod^K_{i<j}\Phi_{a_j}(-a_i)\nonumber \\
        ={}&\begin{vmatrix}
        \Phi_{1}(0) & \Phi_2(0) & \cdots & \Phi_K(0) & \Phi(0) & \cdots & \Phi(1+K-N)
    \end{vmatrix}.\label{edkp_caso_general}
    \end{align}
\end{lemma}
\begin{proof}
     Start with \(\xi = b n_s + \sum_{l=1}^K a_l n_l\), which includes a special direction \(n_s\)(\(s>K+1\)) such that for any distinct \(i,j,k \in \{1,\ldots, K,s\}\), \(\tau(n_i, n_j, n_k)\) satisfies \eqref{elliptic_dkp}. From \eqref{edkp_lax_pair},
    \begin{align*}
        U_{ij}\Phi_{ij} = \Phi_j - \Phi_i, \quad U_{ij} := -\frac{\sigma(\xi)\sigma(\xi_{ij})}{\sigma(\xi_i)\sigma(\xi_j)}\Phi_{a_j}(-a_i), \quad \text{for } i<j,
    \end{align*}
    and the shifted Lax pair between \(\phi_s\) and \(\phi_i, i \in \{1,\ldots, K\}\) is
    \begin{align}\label{rel_s_and_i}
     &\Phi_i(-n) = \Phi(1-n) - V(-n)\Phi_i(1-n), \quad V(n) := U_{is}|_{n_s\to n_s+n}.
    \end{align}
    We show \eqref{edkp_caso_general} by induction. Consider the base case
    \begin{align*}
    \tau(n_1+1)/\sigma(\xi_1) = \begin{vmatrix}
    \Phi_1(0) & \Phi_1(-1) & \Phi_1(-2) & \cdots & \Phi_1(2-N) & \Phi_1(1-N)
    \end{vmatrix},
    \end{align*}
    which can be written as the following by the use of \eqref{rel_s_and_i}
    \begin{align*}
    \tau(n_1+1)/\sigma(\xi_1) = \begin{vmatrix}
    \Phi_1(0) & \Phi(0) & \Phi(-1) & \cdots & \Phi(3-N) & \Phi(2-N)
    \end{vmatrix}.
    \end{align*}
    And for the \(K+1\) case, with \(s>K+1\), assuming \eqref{edkp_caso_general} holds, we have
    \begin{align*}
        &\tau(n_1 +1, n_2+1,\ldots, n_K+1, n_{K+1}+1)\frac{\sigma^{K-1}(\xi_{K+1})}{\prod^{K}_{j=1}  \sigma(\xi_{j,K+1})} 
        \prod^{K}_{i<j}\Phi_{a_j}(-a_i) \\
        ={}&\begin{vmatrix}
        \Phi_{1,K+1}(0) & \Phi_{2,K+1}(0) & \cdots & \Phi_{K,K+1}(0) & \Phi_{K+1}(0) & \cdots & \Phi_{K+1}(1+K-N)
        \end{vmatrix} \\
        ={}&\begin{vmatrix}
        \frac{-\Phi_{1}(0)}{U_{1,K+1}} & \frac{-\Phi_{2}(0)}{U_{2,K+1}} & \cdots & \frac{-\Phi_{K}(0)}{U_{K,K+1}} & \Phi_{K+1}(0) & \Phi(0)-V(-1)\Phi_{K+1}(0) & \cdots & \Phi_{K+1}(1+K-N)
        \end{vmatrix}\\
        ={}&\frac{(-)^K}{\prod_{j=1}^K U_{j,K+1}}\begin{vmatrix}
        \Phi_{1}(0) & \Phi_{2}(0) & \cdots & \Phi_{K+1}(0) & \Phi(0) & \cdots & \Phi(2+K-N)
        \end{vmatrix}.
    \end{align*}
    Hence
    \begin{align*}
        &\tau(n_1 +1, n_2+1,\ldots, n_K+1, n_{K+1}+1) \frac{\sigma^{K}(\xi)}{\prod^{K+1}_{j=1}  \sigma(\xi_{j})} 
        \prod^{K+1}_{i<j}\Phi_{a_j}(-a_i) \\
        ={}& \begin{vmatrix}
        \Phi_{1}(0) & \Phi_{2}(0) & \cdots & \Phi_{K+1}(0) & \Phi(0) & \cdots & \Phi(2+K-N)
        \end{vmatrix}.
    \end{align*}
\end{proof}

Let us show \cref{lma:casorati_edkp_hierarchy} for now.
\begin{proof}
    Note that the Pl\"ucker relations of determinants \cite{ohta1988elementary} are obtained by the Laplace expansion of 
    \begin{align*}
        \det \begin{pmatrix}
            \Xi_A & 0 \\
            \Xi_A & \Xi_B
        \end{pmatrix} = \det \begin{pmatrix}
            \Xi_A & 0 \\
            0 & \Xi_B
        \end{pmatrix} = 0,
    \end{align*} 
    where \(\Xi_A\) and \(\Xi_B\) are \(N\times(N+1)\) and \(N\times(N-1)\) matrices defined as 
    \begin{align*}
        \Xi_A&=\begin{pmatrix}
            \Phi_{1}(0) & \Phi_{2}(0) & \cdots & \Phi_{I}(0) & \Phi(0) & \cdots & \Phi(I-N)
        \end{pmatrix}, \\
        \Xi_B&=\begin{pmatrix}
            \Phi_{I+1}(0) & \Phi_{I+2}(0) & \cdots & \Phi_{K}(0) & \Phi(0) & \cdots & \Phi(2+J-N)
        \end{pmatrix}.
    \end{align*}
    Applying the expansion gives
    \begin{align*}
        &\sum_{i=1}^I (-1)^{i-1} \begin{vmatrix}
            \Phi_1(0) & \cdots & \Phi_{i-1}(0) & \Phi_{i+1}(0) & \cdots & \Phi_I(0) & \Phi(0) & \cdots & \Phi(I-N)
        \end{vmatrix} \\
        &\quad \times \begin{vmatrix}
            \Phi_i(0) & \Phi_{I+1}(0) & \cdots & \Phi_{K}(0) & \Phi(0) & \cdots & \Phi(2+J-N)
        \end{vmatrix} = 0.
    \end{align*}
    This is equivalent to \eqref{edkp_hierarchy} with the result of \eqref{edkp_caso_general}.
\end{proof}

\begin{theorem}\label[theorem]{lma:gram_edkp_hierarchy}
    The Gram determinant solution to \eqref{edkp_hierarchy} is given by
    \begin{align}\label{edkp_tau_gram}
        \tau &= \sigma(\xi) \det\left(M\right),
    \end{align}
    where \(\xi = \sum_{l=1}^K a_l n_l\) and \(M\) is an \(N\times N\) matrix whose elements are defined as
    \begin{align*}
        m_{k,l} &=  c_{kl} - \Phi_{\xi}(p_k+q_l)\prod_{i=1}^K \left[\frac{\Phi_{a_i} (-q_l)}{\Phi_{a_i} (p_k)}\right]^{n_i}.
    \end{align*}
    This can be expressed using a pfaffian
    \begin{align*}
        &\tau = \sigma(\xi) (1,2,\ldots, N, N^*, \ldots, 2^*, 1^*),
    \end{align*}
    where the entries satisfy \((k,l) = 0, \ (k^*, l^*) = 0, \ (k, l^*) = m_{k,l}\) for \(1\leq k,l\leq N\).
\end{theorem}

Let us show the following result first.
\begin{lemma}
    The difference formula for the \(\tau\) function in Gram determinant form \eqref{edkp_tau_gram} with one increment in \(K\) different directions is
    \begin{align}\label{edkp_gram_general}
      &\tau(n_1 +1, n_2+1,\ldots, n_K+1) \frac{\sigma^{K-1}(\xi)}{\prod^K_{j=1}  \sigma(\xi_j)} \prod^K_{i<j}\Phi_{a_j}(-a_i)\nonumber\\
      &=\begin{vmatrix}
        M & \Phi^1 & \cdots & \Phi^K \\
        -\Psi^{(K-1)} & (-)^{K-1} \prod_{i=0}^{K-2}  U_{1 s}^{(i)} & \cdots & (-)^{K-1} \prod_{i=0}^{K-2}  U_{K s}^{(i)} \\ 
        \vdots & \vdots & \ddots & \vdots \\
        -\Psi & 1 & \cdots & 1
      \end{vmatrix},
    \end{align}
    where \(\Phi=(\phi(p_1), \phi(p_2), \ldots, \phi(p_N))^t\), \(\Psi=(\psi(q_1), \psi(q_2), \ldots, \psi(q_N))\), and
    \begin{align*}
      &U_{ij} := -\frac{\sigma(\xi)\sigma(\xi_{ij})}{\sigma(\xi_i)\sigma(\xi_j)}\Phi_{a_j}(-a_i), \quad U_{ij}^{(n),h}=U_{ij}|_{n_s\to n_s+n,n_h\to n_h+1}, \\
      &\Phi^{ij}=\Phi|_{n_i\to n_i+1, n_j\to n_j+1}, \quad \Psi^{(n),i}=\Psi|_{n_s\to n_s+n, n_i\to n_i+1}.
    \end{align*}
\end{lemma}
\begin{proof}
    Start with \(\xi = bn_s+\sum_{l=1}^K a_l n_l\). For the direction \(n_s\)(\(s>K+1\)) and any distinct \(i,j,k \in \{1,\ldots, K,s\}\), \(\tau(n_i, n_j, n_k)\) satisfies \eqref{elliptic_dkp}.
    From \eqref{edkp_lax_pair}-\eqref{edkp_adj_lax_pair}, we have
    \begin{align*}
      U_{ij}\Phi^{ij} = \Phi^j - \Phi^i, \quad U_{is}^{(n-1)}\Psi^{(n-1)} = \Psi^{(n-1),i} - \Psi^{(n)},
    \end{align*}
    where \(1\leq i<j\leq K\).
    We show \eqref{edkp_gram_general}
    by induction: the base case 
    \begin{align*}
      \tau(n_1+1)/\sigma(\xi_1) = \det_{1\leq k,l\leq N}\Big(m_1(\phi(p_k),\psi(q_l))\Big) = \det_{1\leq k,l\leq N}\Big(m(\phi(p_k),\psi(q_l)) + \phi_1(p_k) \psi(q_l)\Big) = \begin{vmatrix}
       M & \Phi^1 \\ -\Psi & 1
    \end{vmatrix},
    \end{align*}
    and the \(K+1\) case, with \(s>K+1\)
    \begin{align*}
      &\tau(n_1 +1, n_2+1,\ldots, n_K+1, n_{K+1}+1) \frac{\sigma^{K-1}(\xi_{K+1})}{\prod^{K}_{j=1}  \sigma(\xi_{j,K+1})} 
        \prod^{K}_{i<j}\Phi_{a_j}(-a_i) \times (-)^K \prod_{j=1}^K U_{j,K+1}\\
      &=\begin{vmatrix}
        M & -U_{1,K+1}\Phi^{1,K+1} & \cdots & -U_{K,K+1}\Phi^{K,K+1} & \Phi^{K+1} \\
          -\Psi^{(K-1),K+1} & (-)^{K} U_{1,K+1}\prod_{i=0}^{K-2}  U_{1 s}^{(i),K+1} & \cdots & (-)^{K} U_{K,K+1}\prod_{i=0}^{K-2}  U_{K s}^{(i),K+1} & 0 \\
          \vdots & \vdots & \ddots & \vdots & \vdots \\
          -\Psi^{K+1} & -U_{1,K+1} & \cdots & -U_{K,K+1} & 0 \\
          -\Psi & 0 & \cdots & 0 & 1
      \end{vmatrix}\\
      &=\begin{vmatrix}
        M & \Phi^{1} & \cdots & \Phi^{K} & \Phi^{K+1} \\
          -\Psi^{(K-1),K+1} & (-)^{K} U_{1,K+1}\prod_{i=0}^{K-2}  U_{1 s}^{(i),K+1} & \cdots & (-)^{K} U_{K,K+1}\prod_{i=0}^{K-2}  U_{K s}^{(i),K+1} & 0 \\ 
          \vdots & \vdots & \ddots & \vdots & \vdots \\
          -\Psi^{K+1}+U_{K+1,s}\Psi & -U_{1,K+1}-U_{K+1,s} & \cdots & -U_{K,K+1}-U_{K+1,s} & -U_{K+1,s} \\
          -\Psi & 1 & \cdots & 1 & 1
        \end{vmatrix}.
    \end{align*}
    The \(\Psi^{(L),K+1}\) terms can be treated iteratively. Moreover, through
    \begin{align}
      \begin{split}\label{rel_akp_uij}
        &(-)^{L+1} U_{1,K+1}\prod_{i=0}^{L-1}  U_{1 s}^{(i),K+1} + (-)^{L} (-U_{K+1,s}^{(L)}) \prod_{i=0}^{L-1} U_{1s}^{(i)} \\
        ={}& (-)^{L+1}\left(\frac{\sigma(a_{s}-a_{1})}{\sigma(a_{s})\sigma(a_1)}\right)^L\frac{\sigma(\xi)\sigma(\xi_{1,Ls})}{\sigma(\xi_1)}\\
        &\times\left[\frac{\sigma(a_{K+1}-a_{1})}{\sigma(a_{K+1})\sigma(a_1)}\frac{\sigma(\xi_{1,Ls,K+1})}{\sigma(\xi_{Ls,K+1})\sigma(\xi_{1,Ls})}+\frac{\sigma(a_s-a_{K+1})}{\sigma(a_{K+1})\sigma(a_s)}\frac{\sigma(\xi_{1,(L+1)s,K+1})}{\sigma(\xi_{K+1,Ls})\sigma(\xi_{(L+1)s})}\right] \\
        ={}& (-)^{L+1}\prod_{i=0}^{L} U_{1s}^{(i)},
      \end{split}
    \end{align}
    in the end, one can get
    \begin{align*}
      &\tau(n_1 +1, n_2+1,\ldots, n_K+1, n_{K+1}+1) \frac{\sigma^{K}(\xi)}{\prod^{K+1}_{j=1}  \sigma(\xi_{j})} 
        \prod^{K+1}_{i<j}\Phi_{a_j}(-a_i) \\
      ={}&\begin{vmatrix}
        M & \Phi^{1} & \cdots & \Phi^{K} & \Phi^{K+1} \\
        -\Psi^{(K)} & (-)^{K} \prod_{i=0}^{K-1}  U_{1 s}^{(i)} & \cdots & (-)^{K} \prod_{i=0}^{K-1}  U_{K s}^{(i)} & (-)^{K} \prod_{i=0}^{K-1} U_{K+1,s}^{(i)} \\ 
        \vdots & \vdots & \ddots & \vdots & \vdots \\
        -\Psi^{(1)} & -U_{1,s} & \cdots & -U_{K,s} & -U_{K+1,s} \\
        -\Psi & 1 & \cdots & 1 & 1
    \end{vmatrix}.
    \end{align*}
\end{proof}

And now we are able to show \cref{lma:gram_edkp_hierarchy}.
\begin{proof}
    We define
    \begin{align*}
      &(d_\nu, k) = -\phi_\nu(p_k), \quad (b_\nu, l^*) = -\psi_{(\nu)}(q_{l}):=\psi(q_l)|_{n_s\to n_s+\nu}, \quad (b_0, d_\nu) = 1, \quad (b_n, d_\nu) = (-)^n \prod_{i=0}^{n-1}  U_{\nu s}^{(i)}, \\
      &(d_\nu, d_\mu) = (b_\nu, b_\mu) = 0, \quad (d_\nu, k^*) = 0, \quad (b_\nu, k) = 0,
    \end{align*}
    and the pfaffian expression of the difference formula \eqref{edkp_gram_general} is obtained \cite{Ohta2004Special}
    \begin{align}\label{edkp_pfa_general}
        &\tau(n_1 +1, n_2+1,\ldots, n_K+1) \frac{\sigma^{K-1}(\xi)}{\prod^K_{j=1}  \sigma(\xi_j)} \prod^K_{i<j}\Phi_{a_j}(-a_i) \nonumber\\
        ={}& (b_{K-1}, b_{K-2}, \ldots, b_0, d_K, d_{K-1}, \ldots, d_1, 1, \ldots, N, N^*, \ldots, 1^*).
    \end{align}
    We are now able to derive the hierarchy \eqref{edkp_hierarchy} using pfaffians. Note that
    \begin{align*}
      &(b_n, b_{I-2}, \ldots, b_0, d_I, \ldots, d_1) = \sum_{i=1}^{I} (-1)^{i-1} (b_n, d_i)  (b_{I-2}, \ldots, b_0, d_I, \ldots, \hat{d}_i, \ldots, d_1) = 0,\\
      &(b_J, \ldots, b_0, d_K, \ldots, d_{I+1}, d_i) = \sum_{n=0}^{J} (-1)^{J-n} (b_J, \ldots, \hat{b}_n, \ldots, b_0, d_K, \ldots, d_{I+1}) (b_n, d_i), 
    \end{align*}
    for \(0 \leq n \leq I-2, 1\leq i \leq I\). Hence
    \begin{align*}
      &\sum_{n=0}^{J} (-1)^{J-n} (b_J, \ldots, \hat{b}_n, \ldots, b_0, d_K, \ldots, d_{I+1})(b_n, b_{I-2}, \ldots, b_0, d_I, \ldots, d_1) \\
      ={}& \sum_{i=1}^{I} (-1)^{i-1} (b_{I-2}, \ldots, b_0, d_I, \ldots, \hat{d}_i, \ldots, d_1)(b_J, \ldots, b_0, d_K, \ldots, d_{I+1}, d_i) = 0.
    \end{align*}
    Using the golden theorem of the pfaffians \cite{Ohta2004Special}, we have 
    \begin{align*}
      &\sum_{i=1}^{I} (-1)^{i-1} (b_{I-2}, \ldots, b_0, d_I, \ldots, \hat{d}_i, \ldots, d_1,1, \ldots, N, N^*, \ldots, 1^*)\\
      &\quad\times(b_J, \ldots, b_0, d_K, \ldots, d_{I+1}, d_i,1, \ldots, N, N^*, \ldots, 1^*) = 0.
    \end{align*}
    By \eqref{edkp_pfa_general}, the above equality is equivalent to \eqref{edkp_hierarchy}.
\end{proof}

\section{Discrete BKP hierarchy of the elliptic type}
\subsection{Lax formulation and Moutard transformation to the discrete BKP equation of the elliptic type}
In \cite{li2021symmetric}, the authors proposed a discrete BKP equation of the elliptic type \eqref{elliptic_dbkp_intro}, which can be rewritten as 
\begin{align}\label{elliptic_dbkp}
\begin{split}
    &\Phi_{a_1}(-a_2) \Phi_{a_2}(-a_3) \Phi_{a_3}(-a_1) \tau_{123} \tau + \Phi_{a_1}(-a_2) \Phi_{a_2}(a_3) \Phi_{a_3}(a_1) \tau_{12}\tau_3 \\
&\quad + \Phi_{a_3}(-a_1) \Phi_{a_1}(a_2) \Phi_{a_2}(a_3) \tau_{13}\tau_2 + \Phi_{a_2}(-a_3) \Phi_{a_1}(a_2) \Phi_{a_3}(a_1) \tau_{23}\tau_1  = 0.
\end{split}
\end{align}
In this section, we will derive the general soliton solution to \eqref{elliptic_dbkp} by considering its Lax pair. 
\begin{align}
    \label{edbkp_lax_pair_tau}
    \frac{\sigma(a_j - a_i)}{\sigma(a_i + a_j)} \frac{\tau\tau_{ij}}{\tau_i \tau_j} (\chi_{ij} - \chi) = \chi_j - \chi_i, \quad 1\leq i<j \leq3.
\end{align}
This linear problem is compatible (\((\chi_{12})_3 = (\chi_{13})_2 = (\chi_{23})_1\)) if \(\tau\) satisfies Eq.~\eqref{elliptic_dbkp}.

\begin{lemma}
    \(\sigma^2(\xi)\) is a seed solution to Eq.~\eqref{elliptic_dbkp}, where \(\xi=\sum_{i=1}^3 n_i a_i\).
\end{lemma}
\begin{proof}
    The following equalities are a direct result of the three-term identity:    \begin{align*}
        \sigma(\xi_{12})\sigma(\xi_{3}) &= \frac{1}{\sigma(a_1 - a_2) \sigma(a_3)}\left[\sigma(a_2)\sigma(a_1 - a_3)\sigma(\xi_2)\sigma(\xi_{13}) - \sigma(a_1)\sigma(a_2 - a_3)\sigma(\xi_1)\sigma(\xi_{23})\right], \\
        \sigma(\xi)\sigma(\xi_{123}) &= \frac{1}{\sigma(a_1 - a_2) \sigma(a_3)} \left[\sigma(a_1)\sigma(a_2 + a_3)\sigma(\xi_2)\sigma(\xi_{13}) - \sigma(a_2)\sigma(a_1 + a_3)\sigma(\xi_1)\sigma(\xi_{23})\right],
    \end{align*}
    obtained by taking \((x,y,a,b)\) in \eqref{three_term_sigma} to be
    \[
     \textstyle{(\xi + \frac{a_1 + a_2 + a_3}{2}, \frac{a_1 + a_2 - a_3}{2}, \frac{a_1 - a_2 + a_3}{2}, \frac{a_1 - a_2 - a_3}{2}), \quad \text{or}\quad (\xi + \frac{a_1 + a_2 + a_3}{2}, \frac{a_1 + a_2 + a_3}{2},\frac{a_1 - a_2 - a_3}{2}, \frac{-a_1 + a_2 - a_3}{2})}.
    \]
    With the help of \(\sigma(x+y)\sigma(x-y)=\sigma^2(x)\sigma^2(y)(\wp(y) - \wp(x))\), we have
    {\allowdisplaybreaks\begin{align*}
        & \sigma(a_1 - a_2)\sigma(a_2 - a_3)\sigma(a_1 - a_3) \sigma^2(\xi)\sigma^2(\xi_{123}) - \sigma(a_1 - a_2)\sigma(a_2 + a_3)\sigma(a_1 + a_3)\sigma^2(\xi_{12})\sigma^2(\xi_{3}) \\
        ={}& \frac{1}{\sigma(a_1 - a_2) \sigma^2(a_3)}\\
        &\times \left\{\sigma(a_1 - a_3)\sigma(a_2 + a_3) \sigma^2(\xi_2) \sigma^2(\xi_{13}) \left[\sigma(a_2 + a_3) \sigma(a_2 - a_3)\sigma^2(a_1) - \sigma(a_1 + a_3)\sigma(a_1 - a_3) \sigma^2(a_2)\right]\right. \\
        &\left. \quad + \sigma(a_2 - a_3)\sigma(a_1 + a_3)\sigma^2(\xi_1) \sigma^2(\xi_{23})\left[\sigma(a_1 + a_3) \sigma(a_1 - a_3)\sigma^2(a_2) - \sigma(a_2 + a_3) \sigma(a_2 - a_3) \sigma^2(a_1)\right]\right\} \\
        ={}& \frac{1}{\sigma(a_1 - a_2) \sigma^2(a_3)}\\
        &\times \left\{\sigma(a_1 - a_3)\sigma(a_2 + a_3) \sigma^2(\xi_2) \sigma^2(\xi_{13})\sigma^2(a_1)\sigma^2(a_2)\sigma^2(a_3) \left[(\wp(a_3) - \wp(a_2)) - (\wp(a_3) - \wp(a_1))\right]\right. \\
        &\quad \left.+\sigma(a_2 - a_3)\sigma(a_1 + a_3)\sigma^2(\xi_1) \sigma^2(\xi_{23})\sigma^2(a_1)\sigma^2(a_2)\sigma^2(a_3)\left[(\wp(a_3) - \wp(a_1)) - (\wp(a_3) - \wp(a_2))\right]\right\}  \\
        ={}& \frac{1}{\sigma(a_1 - a_2) } \left\{ \sigma(a_1 - a_3)\sigma(a_2 + a_3) \sigma^2(\xi_2) \sigma^2(\xi_{13})\sigma^2(a_1)\sigma^2(a_2)(\wp(a_1) - \wp(a_2))\right. \\
        &\quad \left.+\sigma(a_2 - a_3)\sigma(a_1 + a_3)\sigma^2(\xi_1) \sigma^2(\xi_{23})\sigma^2(a_1)\sigma^2(a_2)(\wp(a_2) - \wp(a_1))\right\} \\
        ={}& -\sigma(a_1 - a_3)\sigma(a_2 + a_1)\sigma(a_2 + a_3) \sigma^2(\xi_2) \sigma^2(\xi_{13}) \\
        &+\sigma(a_2 - a_3)\sigma(a_1 + a_2)\sigma(a_1 + a_3)\sigma^2(\xi_1) \sigma^2(\xi_{23}).
    \end{align*}}
    This is nothing but the elliptic dBKP.
\end{proof}

With the seed solution, the Lax pair \eqref{edbkp_lax_pair_tau} becomes
\begin{align}\label{edbkp_lax_pair}
     \frac{\sigma(a_j - a_i)}{\sigma(a_j + a_i)} \frac{\sigma^2(\xi)\sigma^2(\xi_{ij})}{\sigma^2(\xi_{i})\sigma^2(\xi_{j})} (\chi_{ij} - \chi) = \chi_j - \chi_i.
\end{align}

\begin{lemma}\label[lemma]{prop:eigen_func_edbkp}
    The following eigenfunction
    \begin{align}
      \chi(p) = \Phi_\xi^2 (p) \prod_{l=1}^3 \left[\frac{\Phi_{a_l}(-p)}{\Phi_{a_l}(p)}\right]^{n_l}
    \end{align}
    solves Eq.~\eqref{edbkp_lax_pair}.    
\end{lemma}
\begin{proof}
    Let us calculate $\chi_j - \chi_i$ and $\chi_{ij} - \chi$ for \(i<j\), respectively. Note that via \(\Phi_\kappa(x)\Phi_\kappa(y) = \Phi_\kappa(x + y)\bigl[\zeta(\kappa) + \zeta(x) + \zeta(y) - \zeta(\kappa + x + y)\bigr]\), we have
    \begin{align*}
        (\Phi_{\xi_i}(p))^2\Phi_{a_i}(-p)\Phi_{a_j}(p) ={}& -\Phi_{\xi}(p)\left[\zeta(\xi_i) + \zeta(-a_i) + \zeta(p) - \zeta(p+\xi)\right] \\
        &\times \Phi_{\xi_{ij}}(p)\left[\zeta(\xi_i) + \zeta(a_j) + \zeta(p) - \zeta(p+\xi_{ij})\right],
    \end{align*}
    and subsequently
    {\allowdisplaybreaks\begin{align*}
        \chi_j - \chi_i ={}& \left(\prod_{l=1}^3 \left[\frac{\Phi_{a_l}(-p)}{\Phi_{a_l}(p)}\right]^{n_l}\right) \frac{1}{\Phi_{a_j}(p) \Phi_{a_i}(p)}\left[(\Phi_{\xi_j}(p))^2\Phi_{a_j}(-p)\Phi_{a_i}(p) - (\Phi_{\xi_i}(p))^2\Phi_{a_i}(-p)\Phi_{a_j}(p)\right]\\
        ={}& \left(\prod_{l=1}^3 \left[\frac{\Phi_{a_l}(-p)}{\Phi_{a_l}(p)}\right]^{n_l}\right) \frac{\Phi_{\xi_{ij}}(p)\Phi_{\xi}(p)}{\Phi_{a_j}(p) \Phi_{a_i}(p)}  \left(\zeta(\xi_j) - \zeta(\xi_i) + \zeta(a_i) - \zeta(a_j)\right) \\
        &\times \left(\zeta(p+\xi_{ij}) + \zeta(p+\xi) - 2\zeta(p) - \zeta(\xi_i) - \zeta(\xi_j)\right) \\
        ={}& \left(\prod_{l=1}^3 \left[\frac{\Phi_{a_l}(-p)}{\Phi_{a_l}(p)}\right]^{n_l}\right) \frac{\Phi_{\xi_{ij}}(p)\Phi_{\xi}(p)}{\Phi_{a_j}(p) \Phi_{a_i}(p)} \frac{\Phi_{a_i}(-a_j) \Phi_{a_i}(\xi_j)}{\Phi_{a_i}(\xi)}\\
        &\times \left( \zeta(p+\xi) - \zeta(p) - \zeta(\xi_i) + \zeta(a_i) + \zeta(p+\xi_{ij}) - \zeta(p) - \zeta(\xi_j) - \zeta(a_i)\right) \\
        ={}& \left(\prod_{l=1}^3 \left[\frac{\Phi_{a_l}(-p)}{\Phi_{a_l}(p)}\right]^{n_l}\right) \frac{\Phi_{\xi_{ij}}(p)\Phi_{\xi}(p)}{\Phi_{a_j}(p) \Phi_{a_i}(p)} \frac{\Phi_{a_i}(-a_j) \Phi_{a_i}(\xi_j)}{\Phi_{a_i}(\xi)} \left( \frac{\Phi_{a_i}(-p)\Phi_{a_i}(p+\xi)}{\Phi_{a_i}(\xi)} - \frac{\Phi_{a_i}(p)\Phi_{a_i}(\xi_j)}{\Phi_{a_i}(p+\xi_j)}\right), \\
        \chi_{ij} - \chi ={}& \left(\prod_{l=1}^3 \left[\frac{\Phi_{a_l}(-p)}{\Phi_{a_l}(p)}\right]^{n_l}\right) \frac{1}{\Phi_{a_j}(p) \Phi_{a_i}(p)}\left[(\Phi_{\xi_{ij}}(p))^2\Phi_{a_i}(-p)\Phi_{a_j}(-p) - (\Phi_{\xi}(p))^2\Phi_{a_i}(p)\Phi_{a_j}(p)\right] \\
        ={}& \left(\prod_{l=1}^3 \left[\frac{\Phi_{a_l}(-p)}{\Phi_{a_l}(p)}\right]^{n_l}\right) \frac{\Phi_{\xi_i}(p) \Phi_{\xi_j}(p)}{\Phi_{a_j}(p) \Phi_{a_i}(p)} (\zeta(\xi) + \zeta(a_i) + \zeta(a_j) - \zeta(\xi_{ij}))\\
        &\times \left(\zeta(p+\xi_i) + \zeta(p+\xi_j) - 2\zeta(p) - \zeta(\xi_{ij}) - \zeta(\xi)\right)\\
        ={}& \left(\prod_{l=1}^3 \left[\frac{\Phi_{a_l}(-p)}{\Phi_{a_l}(p)}\right]^{n_l}\right) \frac{\Phi_{\xi_i}(p) \Phi_{\xi_j}(p)}{\Phi_{a_j}(p) \Phi_{a_i}(p)} \frac{\Phi_{a_i}(a_j)\Phi_{a_i}(\xi)}{\Phi_{a_i}(\xi_j)}\\
        &\times \left(\zeta(p+\xi_j) - \zeta(p) - \zeta(\xi_{ij}) + \zeta(a_i) + \zeta(p+\xi_i) - \zeta(p) - \zeta(\xi) - \zeta(a_i)\right)\\
        ={}& \left(\prod_{l=1}^3 \left[\frac{\Phi_{a_l}(-p)}{\Phi_{a_l}(p)}\right]^{n_l}\right) \frac{\Phi_{\xi_i}(p) \Phi_{\xi_j}(p)}{\Phi_{a_j}(p) \Phi_{a_i}(p)} \frac{\Phi_{a_i}(a_j)\Phi_{a_i}(\xi)}{\Phi_{a_i}(\xi_j)} \left( \frac{\Phi_{a_i}(-p)\Phi_{a_i}(p+\xi_j)}{\Phi_{a_i}(\xi_j)} - \frac{\Phi_{a_i}(p)\Phi_{a_i}(\xi)}{\Phi_{a_i}(p+\xi)} \right).
    \end{align*}}
    Therefore
    \begin{align*}
        \frac{\chi_j - \chi_i}{\chi_{ij} - \chi} ={}& \frac{\Phi_{\xi_{ij}}(p) \Phi_{\xi}(p)}{\Phi_{\xi_{i}}(p) \Phi_{\xi_j}(p)} \frac{\Phi_{a_i}(-a_j)}{\Phi_{a_i}(a_j)}\left(\frac{\Phi_{a_i}(\xi_j)}{\Phi_{a_i}(\xi)}\right)^2 \frac{ \frac{\Phi_{a_i}(-p)\Phi_{a_i}(p+\xi)}{\Phi_{a_i}(\xi)} - \frac{\Phi_{a_i}(p)\Phi_{a_i}(\xi_j)}{\Phi_{a_i}(p+\xi_j)} }{\frac{\Phi_{a_i}(-p)\Phi_{a_i}(p+\xi_j)}{\Phi_{a_i}(\xi_j)} - \frac{\Phi_{a_i}(p)\Phi_{a_i}(\xi)}{\Phi_{a_i}(p+\xi)}} \\
        ={}& \frac{\Phi_{\xi_{ij}}(p) \Phi_{\xi}(p)}{\Phi_{\xi_{i}}(p) \Phi_{\xi_j}(p)} \frac{\Phi_{a_i}(-a_j)}{\Phi_{a_i}(a_j)}\left(\frac{\Phi_{a_i}(\xi_j)}{\Phi_{a_i}(\xi)}\right)^2 \frac{\Phi_{a_i}(\xi_j) \Phi_{a_i}(p+\xi)}{\Phi_{a_i}(\xi)\Phi_{a_i}(p+\xi_j)} \\
        ={}& \frac{\sigma(a_j - a_i)}{\sigma(a_j + a_i)} \frac{\sigma^2(\xi_{ij})\sigma^2(\xi)}{\sigma^2(\xi_i)\sigma^2(\xi_j)}.
    \end{align*}
    This completes the proof of \cref{prop:eigen_func_edbkp}.
\end{proof}

As discussed in \cite{nimmo1997superposition}, let \(a,b\) be solutions of \eqref{edbkp_lax_pair}. Then there exists a discrete potential \(S(a,b)\) such that
\begin{align}\label{pot_S}
    \Delta_i S(a,b) = (\Delta_i a)b - a(\Delta_i b), \quad i = 1,2,3.
\end{align}

\begin{lemma}\label[lemma]{prop:dbkp_potential}
    Given \(\chi(p), \chi(q)\) satisfying \eqref{edbkp_lax_pair}, the following potential
    \begin{align*}
        S(\chi(p),\chi(q)) = c + \left[\wp(p) - \wp(q)\right] \left(\Phi_{\xi}(p+q)\right)^2 \prod_{l=1}^3 \left[\frac{\Phi_{a_l}(-p)}{\Phi_{a_l}(p)}\frac{\Phi_{a_l}(-q)}{\Phi_{a_l}(q)}\right]^{n_l}
    \end{align*}
    satisfies \eqref{pot_S}.
\end{lemma}
\begin{proof}
    By the identity \(\Phi_{\xi}(p)\Phi_{\xi_i}(q) = \Phi_{\xi}(p+q)\Phi_{a_i}(q) - \Phi_{\xi_i}(p+q)\Phi_{a_i}(-p)\), we have
    {\allowdisplaybreaks\begin{align*}
        (\Delta_i \chi(p)) \chi(q) - \chi(p) (\Delta_i \chi(q)) ={}& \chi_i(p)\chi(q) - \chi(p)\chi_i(q)\\
        ={}& \prod_{l=1}^3 \left[\frac{\Phi_{a_l}(-p)}{\Phi_{a_l}(p)}\frac{\Phi_{a_l}(-q)}{\Phi_{a_l}(q)}\right]^{n_l} \frac{1}{\Phi_{a_i}(p)\Phi_{a_i}(q)}\\
        &\times \left[\left(\Phi_{\xi_i}(p)\Phi_{\xi}(q)\right)^2 \Phi_{a_i}(-p)\Phi_{a_i}(q) - \left(\Phi_{\xi}(p)\Phi_{\xi_i}(q)\right)^2 \Phi_{a_i}(p)\Phi_{a_i}(-q)\right]\\
        ={}& \prod_{l=1}^3 \left[\frac{\Phi_{a_l}(-p)}{\Phi_{a_l}(p)}\frac{\Phi_{a_l}(-q)}{\Phi_{a_l}(q)}\right]^{n_l} \frac{1}{\Phi_{a_i}(p)\Phi_{a_i}(q)}\\
        &\times \left[
        \left(\Phi_{\xi}(p+q)\Phi_{a_i}(p) - \Phi_{\xi_i}(p+q)\Phi_{a_i}(-q)\right)^2 \Phi_{a_i}(-p)\Phi_{a_i}(q) \right.\\
        &\quad \left. - \left(\Phi_{\xi}(p+q)\Phi_{a_i}(q) - \Phi_{\xi_i}(p+q)\Phi_{a_i}(-p)\right)^2 \Phi_{a_i}(p)\Phi_{a_i}(-q) \right] \\
        ={}& \prod_{l=1}^3 \left[\frac{\Phi_{a_l}(-p)}{\Phi_{a_l}(p)}\frac{\Phi_{a_l}(-q)}{\Phi_{a_l}(q)}\right]^{n_l} \frac{1}{\Phi_{a_i}(p)\Phi_{a_i}(q)}\\
        &\times \left[\left(\Phi_{\xi_i}(p+q)\right)^2 \Phi_{a_i}(-p) \Phi_{a_i}(-q) -\left(\Phi_{\xi}(p+q)\right)^2 \Phi_{a_i}(p) \Phi_{a_i}(q)\right] \\&\times \left[\Phi_{a_i}(-q)\Phi_{a_i}(q) - \Phi_{a_i}(-p)\Phi_{a_i}(p)\right] \\
        ={}& \prod_{l=1}^3 \left[\frac{\Phi_{a_l}(-p)}{\Phi_{a_l}(p)}\frac{\Phi_{a_l}(-q)}{\Phi_{a_l}(q)}\right]^{n_l} \left[\wp(p) - \wp(q)\right]\\
        &\times \left[\left(\Phi_{\xi_i}(p+q)\right)^2 \frac{\Phi_{a_i}(-p) \Phi_{a_i}(-q)}{\Phi_{a_i}(p)\Phi_{a_i}(q)} -\left(\Phi_{\xi}(p+q)\right)^2 \right] \\
        ={}& S_i(\chi(p),\chi(q)) - S(\chi(p),\chi(q)) = \Delta_i S.
    \end{align*}}
    This completes the proof of \cref{prop:dbkp_potential}.
\end{proof}
Using the \(N\)-fold Moutard transformation for the discrete BKP equation \cite{nimmo1997superposition}, we derive the general pfaffian solution of the elliptic type to \eqref{elliptic_dbkp}.
\begin{theorem}
     Given \(\chi^0, \chi^1, \ldots, \chi^N\) satisfying \eqref{edbkp_lax_pair}, the transformation
    \begin{align}\label{N_moutard}
    \chi^0 \to \frac{S(\chi^0, \chi^1, \ldots, \chi^N)}{S(\chi^1, \ldots, \chi^N)}, \quad \tau_0 \to \tau_0 S(\chi^1, \ldots, \chi^N),
    \end{align}
    preserves the form of the Lax pair \eqref{edbkp_lax_pair_tau}.
    Here, we define
    \begin{align*}
    S(\chi^1, \ldots, \chi^N) = \begin{cases}
      (1,\ldots, N), & N \text{ is even},\\
      (s_0, 1, \ldots, N), & N \text{ is odd},
    \end{cases}
    \end{align*}
    where
    \begin{align*}
    &(s_0, k) = \chi^k, \quad (k,l) = S(\chi^k, \chi^l), \quad \Delta_i S(\chi^k, \chi^l) = \chi_i^k \chi^l - \chi^k \chi_i^l,
    \end{align*}
    for \(k,l=1,\ldots, N\). 
    Based on the seed solution, the eigenfunction and the discrete potential are
    \begin{align*}
      &\chi^k = \Phi_\xi^2 (p_k) \prod_{i=1}^3 \left[\frac{\Phi_{a_i}(-p_k)}{\Phi_{a_i}(p_k)}\right]^{n_i}, \quad \tau_0 = \sigma^2(\xi),\quad \xi = \sum_{i=1}^3 n_i a_i,\\
      &(k,l) = c_{kl} + \left[\wp(p_k) - \wp(p_l)\right] \left(\Phi_{\xi}(p_k+p_l)\right)^2 \prod_{i=1}^3 \left[\frac{\Phi_{a_i}(-p_k)}{\Phi_{a_i}(p_k)}\frac{\Phi_{a_i}(-p_l)}{\Phi_{a_i}(p_l)}\right]^{n_i},
    \end{align*}
    where \(c_{kl}=-c_{lk}\). And the general \(N\)-soliton solution to \eqref{elliptic_dbkp} in pfaffian form is
    \begin{align*}
    \tau = \begin{cases}
            \sigma^2(\xi)(1,\ldots, N) & N \text{ even},\\
            \sigma^2(\xi)(s_0, 1, \ldots, N),    & N \text{ odd}.
    \end{cases}
    \end{align*}
\end{theorem}
We now give the proof of the invariance of the Lax pair \eqref{edbkp_lax_pair_tau} under the transformation \eqref{N_moutard}.
\begin{proof}
  Denote \(G = S(\chi^0, \chi^1, \ldots, \chi^N)\), \(F=S(\chi^1, \ldots, \chi^N)\), \(U = \frac{\sigma(a_j - a_i)}{\sigma(a_j + a_i)} \frac{\sigma^2(\xi)\sigma^2(\xi_{ij})}{\sigma^2(\xi_{i})\sigma^2(\xi_{j})}\).
  After substituting \(\chi=G/F\) and \(\tau=\sigma^2(\xi)F\), Eq.~\eqref{edbkp_lax_pair_tau} is written as
  \begin{align}\label{pf_bkp_moutard}
     U G_{ij} F - G U F_{ij} - G_j F_i + G_i F_j = 0.
  \end{align}
  We show this equation holds under the above setting.  Define the
  pfaffian elements by
  \begin{align*}
    &(s_0,I)=(s_0,J) = 1, \quad (I, k) = \chi_i^k, \quad (J, k) = \chi_j^k, \quad (I,J) = U, \\
    &(s_0, C) = \chi^0, \quad (I,C) = \chi_i^0, \quad (J,C) = \chi_j^0,\quad (k, C) = S(\chi^k, \chi^0).
  \end{align*}
  When \(N\) is an even number, we have
  \begin{align*}
    &F = (1,\ldots, N), \quad G=(s_0, C, 1,\ldots, N), \\
    &F_i = (s_0, I, 1,\ldots, N), \quad G_i = (I, C, 1,\ldots, N),\\
    &U F_{ij} = (I,J,1,\ldots, N), \quad U G_{ij} = (s_0, I, J, C, 1,\ldots,N),
  \end{align*}
  and \eqref{pf_bkp_moutard} is nothing but the result of
  the pfaffian identity
  \[
    (s_0, I, J, C) - (s_0,C)(I,J) - (s_0, I)(J,C) + (s_0, J)(I,C) = 0.
  \]
  When \(N\) is an odd number, we have
  \begin{align*}
    &F = (s_0,1,\ldots, N), \quad G=(C, 1,\ldots, N), \\
    &F_i = (I, 1,\ldots, N), \quad G_i = (s_0, I, C, 1,\ldots, N),\\
    &U F_{ij} = (s_0,I,J,1,\ldots, N), \quad U G_{ij} = (I, J, C, 1,\ldots,N),
  \end{align*}
  and \eqref{pf_bkp_moutard} is nothing but the result of
  the pfaffian expansion
  \[
    (1,s_0,I,J,C,1) = 0.
  \]
  Note that the golden theorem was applied to complete this proof.
\end{proof}

\subsection{General soliton solutions to the discrete BKP hierarchy of the elliptic type}
The general discrete BKP hierarchy of the elliptic type is given by
\begin{align}\label{edbkp_hierarchy}
  &\sum_{i\in A\triangle B}(-1)^{i-1}\,
  S_{A\triangle \{i\}}\,S_{B\triangle \{i\}}\;
  \tau_{A\triangle \{i\}}\,\tau_{B\triangle \{i\}} = 0,\\
  &S_T:=\prod_{\substack{0<j<k\\ j,k\in T}}\frac{\Phi_{a_j}(-a_k)}{\Phi_{a_j}(a_k)} = \prod_{\substack{j<k\\ j,k\in T}}\frac{\sigma(a_k-a_j)}{\sigma(a_j+a_k)}, \quad S_\varnothing=1.\nonumber
\end{align}
Here \(\tau=\tau(n_1,\ldots,n_K)\), and \(\triangle\) denotes the symmetric difference of sets: for sets \(A,B\), we have \(A\Delta B=(A\cup B)\setminus (A\cap B)\). For \(K=I+J\), with \(I\geq1\) and \(J\geq0\), the index sets are
\begin{align}\label{dbkp_index_sets}
  A = \begin{cases}
    \{1,\ldots, I\}, & I \text{ odd},\\
    \{0,1,\ldots, I\}, & I \text{ even},
  \end{cases} \quad
  B = \begin{cases}
    \{I+1,\ldots, K\}, & J \text{ odd},\\
    \{0,I+1,\ldots, K\}, & J \text{ even}.
  \end{cases}
\end{align}
We introduce an auxiliary index \(0\) to handle the odd and even cases uniformly. The tau function is independent of \(n_0\), so shifting \(n_0\) leaves it unchanged: \(\tau_0 = \tau\) and we also set \(a_0=0\).
For a set \(T\subseteq\{0,1,\ldots,K\}\) we denote
\begin{align*}
  \tau_T := \tau|_{n_j\to n_j+1}\ \text{for all } j\in T\setminus\{0\},
  \quad \tau_\varnothing=\tau,
\end{align*}
so that \(\tau_{T\cup\{0\}}=\tau_T\) and \(S_{T\cup\{0\}}=S_T\). It is noted that, as a consequence of \eqref{dbkp_index_sets},
\begin{align*}
    A\triangle B = \begin{cases}
    \{1,\ldots,K\}, & K \text{ even},\\
    \{0,1,\ldots,K\}, & K \text{ odd}.
  \end{cases}
\end{align*}
For example, with \(I=3,J=0\), \eqref{edbkp_hierarchy} gives the discrete BKP equation of the elliptic type \eqref{elliptic_dbkp}.
More examples include
\begin{align}
I=3,\ J=1: \nonumber\\
\begin{split}
    &\Phi_{a_1}(a_2)\Phi_{a_1}(a_3)\Phi_{a_1}(-a_4)
    \Phi_{a_2}(-a_3)\Phi_{a_2}(a_4)\Phi_{a_3}(a_4)\tau_{23}\tau_{14}\\
  &\quad-\Phi_{a_1}(a_2)\Phi_{a_1}(-a_3)\Phi_{a_1}(a_4)
    \Phi_{a_2}(a_3)\Phi_{a_2}(-a_4)\Phi_{a_3}(a_4)\tau_{13}\tau_{24}\\
  &\quad+\Phi_{a_1}(-a_2)\Phi_{a_1}(a_3)\Phi_{a_1}(a_4)
    \Phi_{a_2}(a_3)\Phi_{a_2}(a_4)\Phi_{a_3}(-a_4)\tau_{12}\tau_{34}\\
  &\quad-\Phi_{a_1}(-a_2)\Phi_{a_1}(-a_3)\Phi_{a_1}(-a_4)
    \Phi_{a_2}(-a_3)\Phi_{a_2}(-a_4)\Phi_{a_3}(-a_4)\tau\tau_{1234}=0,
\end{split}\\
I=4,\ J=0:\nonumber\\ 
\begin{split}
  &\Phi_{a_1}(a_2)\Phi_{a_1}(a_3)\Phi_{a_1}(a_4)
    \Phi_{a_2}(-a_3)\Phi_{a_2}(-a_4)\Phi_{a_3}(-a_4)\tau_1\tau_{234}\\
  &\quad-\Phi_{a_1}(a_2)\Phi_{a_1}(-a_3)\Phi_{a_1}(-a_4)
    \Phi_{a_2}(a_3)\Phi_{a_2}(a_4)\Phi_{a_3}(-a_4)\tau_2\tau_{134}\\
  &\quad+\Phi_{a_1}(-a_2)\Phi_{a_1}(a_3)\Phi_{a_1}(-a_4)
    \Phi_{a_2}(a_3)\Phi_{a_2}(-a_4)\Phi_{a_3}(a_4)\tau_3\tau_{124}\\
  &\quad-\Phi_{a_1}(-a_2)\Phi_{a_1}(-a_3)\Phi_{a_1}(a_4)
    \Phi_{a_2}(-a_3)\Phi_{a_2}(a_4)\Phi_{a_3}(a_4)\tau_4\tau_{123}=0,
\end{split}\\
&\ldots\nonumber
\end{align}

\begin{theorem}\label[theorem]{lma:edbkp_hieararchy}
    The pfaffian solution to \eqref{edbkp_hierarchy} is given by
    \begin{align}\label{edbkp_tau}
        \tau = \sigma^2(\xi) S(\chi(p_1), \ldots, \chi(p_N)) = \begin{cases}
            \sigma^2(\xi)(1, 2, \ldots, N), & N \text{ even},\\
            \sigma^2(\xi)(s_0, 1, 2, \ldots, N), & N \text { odd},
        \end{cases}
    \end{align}
    where \(\xi = \sum_{l=1}^K a_l n_l\) and the pfaffian elements are defined as
    \begin{align*}
        &(s_0, k) = \chi^k := \Phi_\xi^2 (p_k) \prod_{i=1}^K \left[\frac{\Phi_{a_i}(-p_k)}{\Phi_{a_i}(p_k)}\right]^{n_i},\\
        &(k,l) = c_{kl} + \left[\wp(p_k) - \wp(p_l)\right] \Phi^2_{\xi}(p_k+p_l) \prod_{i=1}^K \left[\frac{\Phi_{a_i}(-p_k)}{\Phi_{a_i}(p_k)}\frac{\Phi_{a_i}(-p_l)}{\Phi_{a_i}(p_l)}\right]^{n_i},
    \end{align*}
    where \(c_{kl}=-c_{lk}\), and \(1\leq k,l\leq N\).
\end{theorem}

Let us first show the following difference formula.
\begin{lemma}\label[lemma]{prop:edbkp_difference}
  The difference formula for the \(\tau\) function in pfaffian form \eqref{edbkp_tau}
  with one increment in \(K\) different directions is
  \begin{align}\label{edbkp_tau_general}
    \tau(n_1+1,\ldots,n_K+1)\,\frac{\sigma^{2(K-1)}(\xi)}{\prod_{j=1}^K\sigma^2(\xi_j)}
    \prod_{1\leq i<j\leq K}\frac{\Phi_{a_i}(-a_j)}{\Phi_{a_i}(a_j)}
    =\begin{cases}
      (s_1,\ldots,s_K,1,\ldots,N), & K+N \text{ even},\\
      (s_0,s_1,\ldots,s_K,1,\ldots,N), & K+N \text{ odd},
    \end{cases}
  \end{align}
  where
  
  \begin{align*}
    &(s_k,s_l)=U_{kl}, \quad (s_k,l)=\chi^l_k, \quad (s_0,s_l)=1,\\
    &U_{kl}=\frac{\Phi_{a_k}(-a_l)}{\Phi_{a_k}(a_l)}
    \left(\frac{\sigma(\xi)\sigma(\xi_{kl})}{\sigma(\xi_k)\sigma(\xi_l)}\right)^2, \quad 1\leq k<l\leq K.
  \end{align*}
  
\end{lemma}
\begin{proof}
    We prove \eqref{edbkp_tau_general} by induction on \(K\). Use \(\bm\cdot\) to represent the pfaffian indices \(1,\ldots,N\), and let \(X_i=X|_{n_i\to n_i+1}\), where \(X=(\chi^1,\ldots,\chi^N)\).
    For \(K=1\) and even \(N\), by \cref{prop:dbkp_potential}, we have
    \begin{align*}
        \frac{\tau(n_1+1)}{\sigma^2(\xi_1)} &= \operatorname{Pf}_{1\leq k<l\leq N}\left((k,l) + \chi_1^k \chi^l - \chi^k \chi_1^l\right)
        =
        \left. \begin{matrix}
        |1 & X \\
        & X_1 \\
        & (\bm \cdot)\\
        \end{matrix}\right|= (s_0,s_1,1,2, \cdots , N),
    \end{align*}
    and for odd \(N\),
    \begin{align*}
        \frac{\tau(n_1+1)}{\sigma^2(\xi_1)} &= \left. \begin{matrix}
        |1 & 0 & X \\
        & 0 & X_1 \\
        & & X_1 \\
        & & (\bm \cdot)\\
        \end{matrix}\right|
        = \left. \begin{matrix}
        | X_1 \\
         (\bm \cdot)\\
         \end{matrix}\right|
        = (s_1,1,2, \cdots , N).
    \end{align*}
  From Eq.~\eqref{edbkp_lax_pair} and its compatibility for the seed solution \(\tau=\sigma^2(\xi)\), we have
  \begin{align*}
    X_{iK}-X=\frac{1}{U_{iK}}\left(X_K-X_i\right), \quad
    U^K_{ij}=\frac{U_{ij}}{U_{iK}U_{jK}}+\frac{1}{U_{iK}}-\frac{1}{U_{jK}},
  \end{align*}
  where \(U_{ij}^K=U_{ij}|_{n_K\to n_K+1}\). Assume that \eqref{edbkp_tau_general} holds for the \(K-1\) directions \(n_1,\ldots,n_{K-1}\), and evaluate it at \(n_K+1\). For even \(K+N\), the pfaffian row and column operations give
  {\allowdisplaybreaks\begin{align*}
    &\tau(n_1+1,\ldots,n_K+1)\,\frac{\sigma^{2(K-2)}(\xi_K)}{\prod_{j=1}^{K-1}\sigma^2(\xi_{jK})}
    \prod_{1\leq i<j\leq K-1}\frac{\Phi_{a_i}(-a_j)}{\Phi_{a_i}(a_j)}
    = (s_0,s_1,\ldots,s_{K-1},\bm\cdot)|_{n_K\to n_K+1}\\
    ={}&\left.\begin{matrix}
      |1 & 0 & 0 & 0 & \cdots & 0 & X\\
         & 0 & 0 & 0 & \cdots & 0 & X_K\\
         &   & 1 & 1 & \cdots & 1 & X_K\\
         &   &   & U^K_{12} & \cdots & U^K_{1,K-1} & X_{1K}\\
         &   &   &   & \ddots & \vdots & \vdots\\
         &   &   &   &        &        & X_{K-1,K}\\
         &   &   &   &        &        & (\bm\cdot)
    \end{matrix}\right|=\left.\begin{matrix}
      |1 & -1 & -\frac{1}{U_{1K}} & -\frac{1}{U_{2K}} & \cdots & -\frac{1}{U_{K-1,K}} & X\\
         & 0 & 1 & 1 & \cdots & 1 & X_K\\
         &   & 0 & 0 & \cdots & 0 & \bm 0\\
         &   &   & \frac{U_{12}}{U_{1K}U_{2K}} & \cdots & \frac{U_{1,K-1}}{U_{1K}U_{K-1,K}} & -\frac{X_1}{U_{1K}}\\
         &   &   &   & \ddots & \vdots & \vdots\\
         &   &   &   &        &        & -\frac{X_{K-1}}{U_{K-1,K}}\\
         &   &   &   &        &        & (\bm\cdot)
    \end{matrix}\right|\\
    ={}&\prod_{i=1}^{K-1}\frac{1}{-U_{iK}}\times
    \left.\begin{matrix}
      |-U_{1K} & -U_{2K} & \cdots & -U_{K-1,K} & X_K\\
         & U_{12} & \cdots & U_{1,K-1} & X_1\\
         &   & \ddots & \vdots & \vdots\\
         &   &        &        & X_{K-1}\\
         &   &        &        & (\bm\cdot)
    \end{matrix}\right|\\
    ={}&\prod_{i=1}^{K-1}\frac{1}{-U_{iK}}\,(s_K,s_1,\ldots,s_{K-1},\bm\cdot)
    =\prod_{i=1}^{K-1}\frac{1}{U_{iK}}\,(s_1,\ldots,s_{K-1},s_K,\bm\cdot).
  \end{align*}}
  For odd \(K+N\), the same operations give
  {\allowdisplaybreaks\begin{align*}
    &\tau(n_1+1,\ldots,n_K+1)\,\frac{\sigma^{2(K-2)}(\xi_K)}{\prod_{j=1}^{K-1}\sigma^2(\xi_{jK})}
    \prod_{1\leq i<j\leq K-1}\frac{\Phi_{a_i}(-a_j)}{\Phi_{a_i}(a_j)}
    = (s_1,\ldots,s_{K-1},\bm\cdot)|_{n_K\to n_K+1}\\
    ={}&\left.\begin{matrix}
      |1 & 0 & 0 & \cdots & 0 & X\\
         & 0 & 0 & \cdots & 0 & X_K\\
         &   & U^K_{12} & \cdots & U^K_{1,K-1} & X_{1K}\\
         &   &   & \ddots & \vdots & \vdots\\
         &   &   &        &        & X_{K-1,K}\\
         &   &   &        &        & (\bm\cdot)
    \end{matrix}\right|
    =\left.\begin{matrix}
      |1 & -\frac{1}{U_{1K}} & -\frac{1}{U_{2K}} & \cdots & -\frac{1}{U_{K-1,K}} & X\\
         & 1 & 1 & \cdots & 1 & X_K\\
         &   & \frac{U_{12}}{U_{1K}U_{2K}} & \cdots & \frac{U_{1,K-1}}{U_{1K}U_{K-1,K}} & -\frac{X_1}{U_{1K}}\\
         &   &   & \ddots & \vdots & \vdots\\
         &   &   &        &        & -\frac{X_{K-1}}{U_{K-1,K}}\\
         &   &   &        &        & (\bm\cdot)
    \end{matrix}\right|\\
    ={}&\prod_{i=1}^{K-1}\frac{1}{-U_{iK}}\times
    \left.\begin{matrix}
      |1 & 1 & 1 & \cdots & 1 & X\\
         & -U_{1K} & -U_{2K} & \cdots & -U_{K-1,K} & X_K\\
         &   & U_{12} & \cdots & U_{1,K-1} & X_1\\
         &   &   & \ddots & \vdots & \vdots\\
         &   &   &        &        & X_{K-1}\\
         &   &   &        &        & (\bm\cdot)
    \end{matrix}\right|\\
    ={}&\prod_{i=1}^{K-1}\frac{1}{-U_{iK}}\,(s_0,s_K,s_1,\ldots,s_{K-1},\bm\cdot)
    =\prod_{i=1}^{K-1}\frac{1}{U_{iK}}\,(s_0,s_1,\ldots,s_{K-1},s_K,\bm\cdot),
  \end{align*}}
  where the first two rows and columns become those of \(s_0\) and \(s_K\).
  Since
  \begin{align*}
    \prod_{i=1}^{K-1}U_{iK}\times\frac{\sigma^{2(K-2)}(\xi_K)}{\prod_{j=1}^{K-1}\sigma^2(\xi_{jK})}
    \prod_{1\leq i<j\leq K-1}\frac{\Phi_{a_i}(-a_j)}{\Phi_{a_i}(a_j)}
    =\frac{\sigma^{2(K-1)}(\xi)}{\prod_{j=1}^K\sigma^2(\xi_j)}
    \prod_{1\leq i<j\leq K}\frac{\Phi_{a_i}(-a_j)}{\Phi_{a_i}(a_j)},
  \end{align*}
  multiplying both cases by \(\prod_{i=1}^{K-1}U_{iK}\) gives \eqref{edbkp_tau_general} for \(K\) directions.
\end{proof}

Now let us show \cref{lma:edbkp_hieararchy}.
\begin{proof}
    We are now able to derive the hierarchy \eqref{edbkp_hierarchy} using pfaffians. For even \(N\), by \eqref{dbkp_index_sets}, we can denote \(A=\{\alpha_1<\cdots<\alpha_p\}\) and \(B=\{\beta_1<\cdots<\beta_q\}\), such that \(p,q\) are odd, and the expansion of the pfaffians gives
    \begin{align}
      \begin{split}\label{AB_expands}
        (s_{\beta_m},s_{\alpha_1},\ldots,s_{\alpha_p})
        &=\sum_{l=1}^{p}(-1)^{l-1}(s_{\beta_m},s_{\alpha_l})(s_{\alpha_1},\ldots,\hat{s}_{\alpha_l},\ldots,s_{\alpha_p}),\\
        (s_{\alpha_l},s_{\beta_1},\ldots,s_{\beta_q})
        &=\sum_{m=1}^{q}(-1)^{m-1}(s_{\alpha_l},s_{\beta_m})(s_{\beta_1},\ldots,\hat{s}_{\beta_m},\ldots,s_{\beta_q}),
      \end{split}
    \end{align}
    for \(1\leq l\leq p,1\leq m\leq q\), where a hat means that the corresponding index is omitted. Note that
    \begin{align*}
      &\sum_{l=1}^{p}\sum_{m=1}^{q}(-1)^{l+m-2}
      \left[(s_{\alpha_l},s_{\beta_m})+(s_{\beta_m},s_{\alpha_l})\right]\\
      &\quad\times(s_{\alpha_1},\ldots,\hat{s}_{\alpha_l},\ldots,s_{\alpha_p})(s_{\beta_1},\ldots,\hat{s}_{\beta_m},\ldots,s_{\beta_q})=0.
    \end{align*}
    By \eqref{AB_expands}, this is equivalent to the following pfaffian identity by expanding its left-hand side
    \begin{align*}
      &\sum_{l=1}^{p}(-1)^{l-1}(s_{\alpha_1},\ldots,\hat{s}_{\alpha_l},\ldots,s_{\alpha_p})(s_{\alpha_l},s_{\beta_1},\ldots,s_{\beta_q})\\
      &+\sum_{m=1}^{q}(-1)^{m-1}(s_{\beta_1},\ldots,\hat{s}_{\beta_m},\ldots,s_{\beta_q})(s_{\beta_m},s_{\alpha_1},\ldots,s_{\alpha_p})=0.
    \end{align*}
    Apply the golden theorem by appending \(1,\ldots,N\), and with the result \eqref{edbkp_tau_general} one gets Eq.~\eqref{edbkp_hierarchy}.

    Next, for odd \(N\), denote \(A\triangle\{0\}=\{\alpha_1<\cdots<\alpha_p\}\) and \(B\triangle\{0\}=\{\beta_1<\cdots<\beta_q\}\), such that \(p,q\) are even. We can obtain the following pfaffian identity from a similar approach
    \begin{align*}
      &\sum_{l=1}^{p}(-1)^{l-1}
      (s_{\alpha_1},\ldots,\hat{s}_{\alpha_l},\ldots,s_{\alpha_p},1)
      (s_{\alpha_l},s_{\beta_1},\ldots,s_{\beta_q},1)\\
      &+\sum_{m=1}^{q}(-1)^{m-1}
      (s_{\beta_1},\ldots,\hat{s}_{\beta_m},\ldots,s_{\beta_q},1)
      (s_{\beta_m},s_{\alpha_1},\ldots,s_{\alpha_p},1)=0.
    \end{align*}
    We then obtain Eq.~\eqref{edbkp_hierarchy} by applying the golden theorem to append \(2,\ldots,N\).
\end{proof}

\section{Continuous elliptic KP, BKP hierarchies and their tau functions}
The discrete A-, B-, C-, and D-type KP systems provide a unified framework for generating their corresponding continuous integrable hierarchies. Through the Miwa transformation, lattice shifts encode simultaneous translations of infinitely many hierarchy times, so expansions and continuum limits of the discrete equations generate the bilinear differential equations of the associated continuous hierarchies. The discrete AKP (Hirota–Miwa) equation generates the ordinary KP hierarchy, with reductions yielding the KdV, Boussinesq, and other Gel'fand–Dikii hierarchies; the discrete BKP equation generates the B-type KP hierarchy, characterized by Pfaffian structures and reductions such as the Sawada–Kotera hierarchy; and the discrete CKP equation similarly encodes the C-type KP hierarchy and its associated reductions. The D-type KP structure is closely connected with multi-component and coupled KP systems and Pfaffian tau-functions; in particular, discrete coupled KP systems obtained through Pfaffianization provide discrete counterparts whose continuum limits recover Hirota–Ohta-type coupled KP equations. Thus, the discrete A-, B-, C-, and D-type formulations are best viewed not simply as discretizations of individual nonlinear equations, but as compact generating systems for entire continuous KP-type hierarchies, preserving at the discrete level their characteristic determinant, Pfaffian, symmetry, and reduction structures.

Having constructed the discrete AKP and BKP hierarchies of elliptic type and provided their solutions or tau functions, we are now able to derive the continuous counterparts of these hierarchies along with their tau functions. In addition, we will show how to deduce soliton solutions on the elliptic background for some typical nonlinear wave equations such as the KdV, modified KdV and the KP equations.

\subsection{Elliptic KP hierarchy}
Consider the discrete KP equation of the elliptic form \eqref{elliptic_dkp} together with its solutions in \cref{lma:casorati_edkp_hierarchy} and \cref{lma:gram_edkp_hierarchy}. We follow the idea of the Miwa transformation to obtain the KP hierarchy of the elliptic type.
Notice the series expansion with respect to \(a_l\) around \(0\) for \(l=1,2,3\),
\begin{align*}
  &\left[\frac{\sigma(p+a_l)}{\sigma(p)}\right]^{-n_l} = \exp\left(-n_l \left(\zeta(p)a_l - \sum_{\nu=2}^\infty \wp^{(\nu-2)}(p)\frac{a_l^\nu}{\nu!}\right)\right), \\
  &\left[\frac{\sigma(q-a_l)}{\sigma(q)}\right]^{n_l} 
  =\exp\left(n_l \left(-\zeta(q)a_l + \sum_{\nu=2}^\infty (-1)^{\nu+1} \wp^{(\nu-2)}(q)\frac{a_l^\nu}{\nu!}\right)\right),
\end{align*}
let us introduce the Miwa variables \(\bm x = (x_1, x_2, \ldots)\) where
\begin{align*}
  x_1=\sum_{l=1}^3n_la_l,\ \  x_2=\sum_{l=1}^3n_l\frac{a_l^2}{2},\ \cdots,\ \  x_\nu=\sum_{l=1}^3n_l\frac{a_l^\nu}{\nu}.
\end{align*}
Above expansions now give
\begin{align}\label{dn_to_cx}
  \begin{split}
    &\prod_{l=1}^3\left[\sigma(a_l)\Phi_{a_l}(p)\right]^{-n_l} = \exp\left(-\zeta(p)x_1
    +\sum_{\nu=2}^\infty \frac{\wp^{(\nu-2)}(p)}{(\nu-1)!} x_\nu \right)\\
    &\prod_{l=1}^3 \left[\frac{\Phi_{a_l}(-q)}{\Phi_{a_l}(p)}\right]^{n_l}=\exp\left(-(\zeta(p)+\zeta(q))x_1
    +\sum_{\nu=2}^\infty \frac{\wp^{(\nu-2)}(p)+(-1)^{\nu-1}\wp^{(\nu-2)}(q)}{(\nu-1)!} x_\nu\right),
  \end{split}
\end{align}
in which we can rewrite \(\tau = \tau(\bm x)\).
Recall Hirota's bilinear derivatives, defined by
\[
    D_x^n f\cdot g
    :=
    \left.
    \partial_y^n
    \bigl[f(x+y)g(x-y)\bigr]
    \right|_{y=0}.
\]
The corresponding
shift identity is
\begin{align}\label{exp_hirota}
  \exp(\delta D_x)f\cdot g
    =
    \sum_{n=0}^{\infty}
    \frac{\delta^n}{n!}D_x^n f\cdot g
    =
    f(x+\delta)g(x-\delta).
\end{align}
Apply \(n_l \to n_l - \frac 1 2\) in the discrete KP equation \eqref{elliptic_dkp} and use above formula, we can rewrite \eqref{elliptic_dkp} into
\begin{equation}\label{exp_kp_hierarchy}
  \begin{split}
    &\sigma(a_1)\sigma(a_2-a_3)\exp\left(\frac12D_{n_1}-\frac12D_{n_2}-\frac12D_{n_3}\right)\tau\cdot\tau\\
    &+\sigma(a_2)\sigma(a_3-a_1)\exp\left(-\frac12D_{n_1}+\frac12D_{n_2}-\frac12D_{n_3}\right)\tau\cdot\tau\\
    &+\sigma(a_3)\sigma(a_1-a_2)\exp\left(-\frac12D_{n_1}-\frac12D_{n_2}+\frac12D_{n_3}\right)\tau\cdot\tau=0,
  \end{split}
\end{equation}
From the defined Miwa variables we have
\[
\partial_{n_l} = \sum_{\nu=1}^\infty \frac{a_l^\nu}{\nu} \partial_{x_\nu} \quad \text{or} \quad D_{n_l} = \sum_{\nu=1}^\infty \frac{a_l^\nu}{\nu} D_{x_\nu},
\]
for the tau function. Use the elementary Schur polynomial
\begin{align*}
\exp\left(\sum_{n=1}^\infty x_n t^n\right)=\sum_{n=0}^\infty p_n(\bm x)t^n, \quad
p_n(\bm x)=\sum_{\substack{k_1,\ldots,k_n\geq0\\k_1+2k_2+\cdots+n k_n=n}}\frac{x_1^{k_1}x_2^{k_2}\cdots x_n^{k_n}}{k_1!k_2!\cdots k_n!},
\end{align*}
and expanding \eqref{exp_kp_hierarchy} using these polynomials gives
\begin{align}
  \label{elli_KP}
  \begin{split}
    &\left[\sigma(a_1)\sigma(a_2-a_3)\sum_{K,L,M\geq 0} a_1^Ka_2^La_3^M
    p_K\left(\frac12\widetilde{\bm{D}}\right)p_L\left(-\frac12\widetilde{\bm{D}}\right)p_M\left(-\frac12\widetilde{\bm{D}}\right)\right.\\
    &\quad+\sigma(a_2)\sigma(a_3-a_1)\sum_{K,L,M\geq 0} a_1^Ka_2^La_3^M
    p_K\left(-\frac12\widetilde{\bm{D}}\right)p_L\left(\frac12\widetilde{\bm{D}}\right)p_M\left(-\frac12\widetilde{\bm{D}}\right)\\
    &\quad\left.+\sigma(a_3)\sigma(a_1-a_2)\sum_{K,L,M\geq 0} a_1^Ka_2^La_3^M
    p_K\left(-\frac12\widetilde{\bm{D}}\right)p_L\left(-\frac12\widetilde{\bm{D}}\right)p_M\left(\frac12\widetilde{\bm{D}}\right)\right]\tau\cdot\tau=0
  \end{split}
\end{align}
where
\(\widetilde{\bm{D}} = \left(D_{x_1}, \frac{1}{2}D_{x_2}, \frac{1}{3}D_{x_3}, \ldots\right)\). 
Eq.~\eqref{elli_KP} is actually the whole KP hierarchy which consists of infinitely many bilinear equations indexed by triples \((K,L,M)\), see \cite{shigyo2013addition} for a discussion about the usual non-elliptic case.
For example, using the expansion
\begin{align*}
    \sigma(z)=z-\frac{g_2}{240} z^5 - \frac{g_3}{840} z^7 + O(z^9) 
\end{align*}
one can derive some low-order equations by collecting the coefficients of \(a_1, a_2, a_3\) in \eqref{elli_KP}, as listed below.
{\allowdisplaybreaks\begin{align}  &\Big(D_{x_{1}}^{4}-4D_{x_{1}}D_{x_{3}}+3D_{x_{2}}^{2}-g_2\Big) \tau \cdot \tau =0\,, 
\label{Elli-KP1}
\\
& 
\Big(D^3_{x_{1}}D_{x_{2}}
+2D_{x_{2}}D_{x_{3}}-
3D_{x_{1}} D_{x_{4}}\Big) \tau \cdot \tau =0 \,,
\\
&
\Big(D_{x_1}^{6}
+20D_{x_1}^{3}D_{x_3}
+15D_{x_1}^{2}D_{x_2}^{2}-96D_{x_1}D_{x_5}
 \nonumber \\
& \quad +60D_{x_2}D_{x_4}
+3g_2D_{x_1}^{2}-24g_3
\Big)\tau\cdot\tau =0,
\\
&
\Big(D_{x_1}^{6}
+4D_{x_1}^{3}D_{x_3}
-9D_{x_1}^{2}D_{x_2}^{2}
+36D_{x_2}D_{x_4}
 \nonumber \\
& \quad -32D_{x_3}^{2}
+3g_2D_{x_1}^{2}
-24g_3
\Big)\tau\cdot\tau =0,\\
&\Big(
2D_{x_1}^{5}D_{x_2}
+15D_{x_1}^{3}D_{x_4}
+10D_{x_1}^{2}D_{x_2}D_{x_3}
+5D_{x_1}D_{x_2}^{3} \nonumber\\
&\quad
-80D_{x_1}D_{x_6}
+48D_{x_2}D_{x_5}
-g_2D_{x_1}D_{x_2}
\Big)\tau\cdot\tau =0,
\\
&\Big(
D_{x_1}^{5}D_{x_2}
+5D_{x_1}^{3}D_{x_4}
-10D_{x_1}^{2}D_{x_2}D_{x_3}
 \nonumber\\
&\quad
+24D_{x_2}D_{x_5}
-20D_{x_3}D_{x_4}
-2g_2D_{x_1}D_{x_2}
\Big)\tau\cdot\tau =0,
\\
&\Big(
D_{x_1}^{3}D_{x_4}
-2D_{x_1}^{2}D_{x_2}D_{x_3}
+D_{x_1}D_{x_2}^{3}
-g_2D_{x_1}D_{x_2}
\Big)\tau\cdot\tau =0,\\
&\cdots\nonumber
\end{align}}
Setting \(g_2=g_3=0\) gives the corresponding ordinary KP-hierarchy equations in Ref.~\cite{JimboMiwa1983}. 

\begin{theorem}\label{thm:elli_KP_solutions}
    The elliptic KP hierarchy \eqref{elli_KP} admits the tau functions of the following forms.
\begin{enumerate}[(a)]
  \item Casorati determinant
  \begin{align}\label{kp_casorati}
    \tau (\bm x) =  \sigma(\xi) \det\begin{pmatrix}
      \bm\Phi(0) & \bm\Phi(1) &  \cdots & \bm\Phi(N-2) & \bm\Phi(N-1)
    \end{pmatrix}
  \end{align}
  \item Wronskian determinant 
  \begin{align}\label{kp_wronskian}
    \tau (\bm x) =  \sigma(\xi) \det\begin{pmatrix} \bm\Phi(0) & \partial_{x_1}\bm\Phi(0) & \cdots & \partial_{x_1}^{N-1}\bm\Phi(0)\end{pmatrix}
  \end{align}
  \item  Gram-type determinant
  \begin{align}\label{kp_gram}
    \tau (\bm x) =  \sigma(\xi) \det_{1 \leq i,j \le N} \left(m_{ij}\right)
  \end{align}
\end{enumerate}
where \(\xi=x_1+x_0\), and
\begin{align*}
  \bm \Phi(s)&=(\phi_1(s), \phi_2(s), \ldots, \phi_N(s))^{\mathsf T},\\
  \phi_j(s) &=\alpha_j \left[\Phi_{b} (p_j)\right]^{s} \Phi_{\xi-bs}(p_j)
  \exp (\mu_j)
  + \beta_j \left[\Phi_{b} (q_j)\right]^{s} \Phi_{\xi-bs}(q_j) \exp(\eta_j) ,  \\
  \mu_j &= -\zeta(p_j)x_1+\sum_{\nu=2}^\infty\frac{\wp^{(\nu-2)}(p_j)}{(\nu-1)!}x_\nu, \quad
  \eta_j = -\zeta(q_j)x_1+\sum_{\nu=2}^\infty\frac{\wp^{(\nu-2)}(q_j)}{(\nu-1)!}x_\nu,
  \\
  m_{ij} &= c_{ij}+\Phi_\xi(p_i+q_j) \exp(\theta_{ij}), \\
  \theta_{ij} &=-(\zeta (p_i)+\zeta(q_j))x_1 + 
  \sum_{\nu=2}^\infty \frac{\wp^{(\nu-2)} (p_i) + (-1)^{\nu-1} \wp^{(\nu-2)} (q_j)} {(\nu-1)!} x_{\nu}+\mu_{i0}+\eta_{j0}.
\end{align*}
Here \(p_j, q_j\), \(j=1, \cdots, N\) represent wave numbers and
\(b,x_0,\alpha_j,\beta_j,c_{ij},\mu_{i0},\eta_{j0}\) are constants. Note that we have also set \(n=-s\) in \cref{lma:casorati_edkp_hierarchy}.
\end{theorem}
\begin{proof}
The Casorati and Gram solutions are derived from \cref{lma:casorati_edkp_hierarchy} and \cref{lma:gram_edkp_hierarchy} with the use of \eqref{dn_to_cx}. We show the Wronskian solution is the limit \(b \to 0\) of the Casorati solution. Note that for the following formula, expansion around \(b=0\) gives
\begin{align*}
  [\Phi_b(p_j)]^s\exp(\mu_j)&=\frac{1}{\sigma(b)^s}
  \left[\frac{\sigma(p_j+b)}{\sigma(p_j)}\right]^s\exp(\mu_j)\\
  &=\frac{1}{\sigma(b)^s}
  \exp\left(\mu_j+sb\zeta(p_j)-s\sum_{\nu=2}^\infty\frac{b^\nu}{\nu!}\wp^{(\nu-2)}(p_j)\right)\\
  &=\frac{1}{\sigma(b)^s}
  \exp\left(-s\sum_{\nu=1}^\infty\frac{b^\nu}{\nu}\partial_{x_\nu}\right)\exp(\mu_j).
\end{align*}
The same calculation applies to \([\Phi_b(q_j)]^s\exp(\eta_j)\). We obtain
\begin{align*}
 \sigma(b)^s\bm\Phi(s)=\exp\left(-s\sum_{\nu=1}^\infty\frac{b^\nu}{\nu}\partial_{x_\nu}\right)\bm\Phi(0)=\left[1-b\partial_{x_1}+O(b^2)\right]^s\bm\Phi(0),
 \quad s=0,1,\ldots.
\end{align*}
The tau function in Casorati form \eqref{kp_casorati} is now rewritten as
\begin{align*}
  &\sigma(b)^{N(N-1)/2}\tau(\bm x)=\sigma(\xi)
  \det \begin{pmatrix}
    \bm\Phi(0) & \sigma(b)\bm\Phi(1) & \cdots &
    \sigma(b)^{N-1}\bm\Phi(N-1)
  \end{pmatrix}
\end{align*}
At each column \(\sigma(b)^k \bm \Phi(k)\), let us consider
\begin{align*}
  \sigma(b)^k \bm \Phi(k) + \sum_{s=0}^{k-1}(-1)^{k-s}\binom{k}{s}\sigma(b)^s\bm\Phi(s) &= \sum_{s=0}^k(-1)^{k-s}\binom{k}{s}\sigma(b)^s\bm\Phi(s) \\
  &= \left[-1+\left(1-b\partial_{x_1}+O(b^2)\right)\right]^k\bm\Phi(0)\\
  &= \left[-b\partial_{x_1}+O(b^2)\right]^k\bm\Phi(0)\\
  &= (-b)^k\left[\partial_{x_1}^k\bm\Phi(0)+O(b)\right],
\end{align*}
therefore
\begin{align*}
 \sigma(b)^{N(N-1)/2}\tau(\bm x)
 ={}&(-b)^{N(N-1)/2}\sigma(\xi)\left[
  \det \begin{pmatrix}
    \bm\Phi(0) & \partial_{x_1}\bm\Phi(0) & \cdots &
    \partial_{x_1}^{N-1}\bm\Phi(0)
  \end{pmatrix}+O(b)\right],
\end{align*}
take the limit \(b\to 0\)
\begin{align*}
 \lim_{b\to0}\tau(\bm x)
 ={}& (-1)^{N(N-1)/2} \sigma(\xi) \det \begin{pmatrix}
   \bm\Phi(0) & \partial_{x_1}\bm\Phi(0) & \cdots &
   \partial_{x_1}^{N-1}\bm\Phi(0)
 \end{pmatrix}.
\end{align*}
the \((-1)^{N(N-1)/2}\) can be dropped due to the property of the bilinear equations. Finally, we obtain the Wronskian determinant solution \eqref{kp_wronskian}.
\end{proof}


\subsection{Elliptic modified KP hierarchy}
To get the modified KP hierarchy, we perform the Miwa transformation for the discrete variables \(n_1, n_2\) in \eqref{elliptic_dkp}. Denote $n = n_3$, $\tau_n = \tau(n_3)$, and assume \(c = a_3\notin \Lambda\), where \(\Lambda\) is the set of lattice points of Weierstrass functions. We introduce the Miwa variables \(\bm x = (x_1, x_2, \ldots)\) where
\[
    x_\nu=\frac{1}{\nu}\sum_{l=1}^2 n_l a_l^\nu,\quad \nu=1,2,\ldots,
\]
similar to \eqref{dn_to_cx}, we can rewrite \(\tau = \tau_n(\bm x)\).
For nonzero \(c\), apply \((n_1, n_2) \to (n_1 - \frac 12, n_2 - \frac 12)\). By \eqref{exp_hirota}, the discrete KP equation \eqref{elliptic_dkp} becomes
\begin{align*}
    &\sigma(a_1)\sigma(a_2-c)\exp\left(\frac 1 2 D_{n_1} - \frac1 2 D_{n_2}\right)\tau_n \cdot \tau_{n+1}\\
    &\quad+\sigma(a_2)\sigma(c-a_1)\exp\left(-\frac 1 2 D_{n_1} + \frac1 2 D_{n_2}\right) \tau_n \cdot \tau_{n+1}\\
    &\quad+\sigma(c)\sigma(a_1-a_2)\exp\left(\frac 1 2 D_{n_1} + \frac1 2 D_{n_2}\right) \tau_n \cdot \tau_{n+1}=0.
\end{align*}
Expanding the above expression gives the generating equation
\begin{align}\label{elli-mKP}
    \begin{split}
      &\Bigg[\sigma(a_1)\sigma(a_2-c)\sum_{K,L\geq0}a_1^Ka_2^L
    p_K\left(\frac12\widetilde{\bm{D}}\right)p_L\left(-\frac12\widetilde{\bm{D}}\right)\\
    &\quad+\sigma(a_2)\sigma(c-a_1)\sum_{K,L\geq0}a_1^Ka_2^L
    p_K\left(-\frac12\widetilde{\bm{D}}\right)p_L\left(\frac12\widetilde{\bm{D}}\right)\\
    &\quad+\sigma(c)\sigma(a_1-a_2)\sum_{K,L\geq0}a_1^Ka_2^L
    p_K\left(\frac12\widetilde{\bm{D}}\right)p_L\left(\frac12\widetilde{\bm{D}}\right)\Bigg]
    \tau_n\cdot\tau_{n+1}=0,
    \end{split}
\end{align}
where $p_K$ denotes the elementary Schur polynomial and $\widetilde{\bm{D}} = (D_{x_1}, \frac 12 D_{x_2}, \frac 13 D_{x_3}, \ldots)$ as in \eqref{elli_KP}. 
 Above \eqref{elli-mKP} can be viewed as the first modified KP hierarchy. By collecting the coefficients of \(a_1, a_2\) one can obtain its first several members, as listed below.
\begin{align}
    &\bigl(D_{x_1}^2-D_{x_2}+2\zeta(c)D_{x_1}+\zeta^2(c)-\wp(c)\bigr)
    \tau_n\cdot\tau_{n+1}=0,\label{elli-mKP-bl1} \\
    &\Big[D_{x_1}^3+3D_{x_1}D_{x_2}-4D_{x_3}+6\zeta(c)D_{x_2}
    -3\bigl(\zeta^2(c)-\wp(c)\bigr)D_{x_1}\nonumber\\
    &\quad-2\bigl(\zeta^3(c)-3\zeta(c)\wp(c)-\wp'(c)\bigr)\Big]
    \tau_n\cdot\tau_{n+1}=0\label{elli-mKP-bl2},\\
    &\cdots\nonumber
\end{align}

\begin{theorem}\label[theorem]{thm:elliptic-mkp-determinants}
  The modified KP hierarchy of the elliptic type \eqref{elli-mKP} admits the tau functions of the following forms.
  \begin{enumerate}[(a)]
    \item Casorati determinant
    \begin{align}\label{mkp_casorati}
      \tau_n (\bm x) =  \sigma(\xi) \det\begin{pmatrix}
        \bm\Phi(0;n) & \bm\Phi(1;n) &  \cdots & \bm\Phi(N-2;n) & \bm\Phi(N-1;n)
      \end{pmatrix}
    \end{align}
    \item Wronskian determinant 
    \begin{align}\label{mkp_wronskian}
      \tau_n (\bm x) =  \sigma(\xi) \det\begin{pmatrix} \bm\Phi(0;n) & \partial_{x_1}\bm\Phi(0;n) & \cdots & \partial_{x_1}^{N-1}\bm\Phi(0;n)\end{pmatrix}
    \end{align}
    \item  Gram-type determinant
    \begin{align}\label{mkp_gram}
      \tau_n (\bm x) =  \sigma(\xi) \det_{1 \leq i,j \le N} \left(m_{ij}^{(n)}\right)
    \end{align}
  \end{enumerate}
  where \(\xi=x_1+nc+x_0\), and
  \begin{align*}
    \bm \Phi(s;n)&=(\phi_1(s;n), \phi_2(s;n), \ldots, \phi_N(s;n))^{\mathsf T},\\
    \phi_j(s;n) &=\alpha_j \left[\Phi_{b} (p_j)\right]^{s}\left[\Phi_c(p_j)\right]^{-n} \Phi_{\xi-bs}(p_j)
    \exp (\mu_j)
    + \beta_j \left[\Phi_{b} (q_j)\right]^{s}\left[\Phi_c(q_j)\right]^{-n} \Phi_{\xi-bs}(q_j) \exp(\eta_j) ,  \\
    \mu_j &= -\zeta(p_j)x_1+\sum_{\nu=2}^\infty\frac{\wp^{(\nu-2)}(p_j)}{(\nu-1)!}x_\nu, \quad
    \eta_j = -\zeta(q_j)x_1+\sum_{\nu=2}^\infty\frac{\wp^{(\nu-2)}(q_j)}{(\nu-1)!}x_\nu,
    \\
    m_{ij}^{(n)} &= c_{ij}+\Phi_\xi(p_i+q_j)\left[\frac{\Phi_c(-q_j)}{\Phi_c(p_i)}\right]^n \exp(\theta_{ij}), \\
    \theta_{ij} &=-(\zeta (p_i)+\zeta(q_j))x_1 + 
    \sum_{\nu=2}^\infty \frac{\wp^{(\nu-2)} (p_i) + (-1)^{\nu-1} \wp^{(\nu-2)} (q_j)} {(\nu-1)!} x_{\nu}+\mu_{i0}+\eta_{j0}.
  \end{align*}
  Here \(p_j, q_j\), \(j=1, \cdots, N\) represent wave numbers and
  \(b,c,x_0,\alpha_j,\beta_j,c_{ij},\mu_{i0},\eta_{j0}\) are constants.
\end{theorem}
The proof of above theorem is similar to the proof of the solutions to the KP hierarchy \cref{thm:elli_KP_solutions}. Obviously, tau functions of the modified KP hierarchy also solve the KP hierarchy \eqref{elli_KP}.

\subsection{Elliptic BKP hierarchy}
Around $a_l=0, l = 1,2,3$, we have the expansion
\[
 \left[\frac{\Phi_{a_l}(-p_i)}{\Phi_{a_l}(p_i)}\right]^{n_l}
 =\exp\left(-2n_l\zeta(p_i)a_l
       +2n_l\sum_{\mu=2}^{\infty}
        \wp^{(2\mu-3)}(p_i)\frac{a_l^{2\mu-1}}{(2\mu-1)!}\right).
\]
Define the Miwa variables \(\bm x = (x_1, x_3, \ldots)\)
\begin{equation*}
x_1=2\sum^3_{i=1}n_i a_i,\quad x_3=\frac{2}{3}\sum^3_{i=1}n_ia_i^3,\quad \cdots,\quad  x_{2\mu-1}=\frac{2}{2\mu-1}\sum^3_{i=1}n_ia_i^{2\mu-1}
\end{equation*}
such that for \(\tau = \tau(\bm x)\) we have
\[
 \partial_{n_\nu}=2a_\nu\partial_{x_1}
       +\frac{2a_\nu^3}{3}\partial_{x_3}
       +\frac{2a_\nu^5}{5}\partial_{x_5}+\cdots,
 \quad \text{or} \quad
 D_{n_j}=\sum_{\mu\ge1}\frac{2a_j^{2\mu-1}}{2\mu-1}D_{x_{2\mu-1}}.
\]
Then, the discrete BKP equation \eqref{elliptic_dbkp} can be converted into
\begin{align}\label{elli-BKP}
\begin{split}
    &\left[\sigma(a_{2}-a_{3}) \sigma(a_{1}+a_{2}) \sigma(a_{1}+a_{3}) \sum_{K,L,M\geq 0} a_{1}^{K}a_{2}^{L}a_{3}^{M}p_{K}( -%
    \widetilde{\bm D}) p_{L}( \widetilde{\bm D}) p_{M}(
    \widetilde{\bm D})\right. \\
    &\quad + \sigma(a_{3}-a_{1})\sigma(a_{2}+a_{3})\sigma(a_{2}+a_{1}) \sum_{K,L,M\geq 0} a_{1}^{K}a_{2}^{L}a_{3}^{M}p_{K}( %
    \widetilde{\bm D}) p_{L}( -\widetilde{\bm D})
    p_{M}( \widetilde{\bm D})  \\
    &\quad + \sigma(a_{1}-a_{2}) \sigma(a_{3}+a_{1}) \sigma(a_{3}+a_{2}) \sum_{K,L,M\geq 0} a_{1}^{K}a_{2}^{L}a_{3}^{M}p_{K}( %
    \widetilde{\bm D}) p_{L}( \widetilde{\bm D}) p_{M}(-\widetilde{\bm D}) \\
    &\quad\left. + \sigma(a_{1}-a_{2}) \sigma(a_{2}-a_{3}) \sigma(a_{3}-a_{1}) \sum_{K,L,M\geq 0} a_{1}^{K}a_{2}^{L}a_{3}^{M}p_{K}( %
    \widetilde{\bm D}) p_{L}( \widetilde{\bm D}) p_{M}(\widetilde{\bm D})
    \right] \tau \cdot \tau=0
\end{split}
\end{align}
which is the BKP hierarchy.  Here \(p_K\) from the elementary Schur polynomial and $\widetilde{\bm D}=(D_{x_1}, 0, \frac 13 D_{x_3}, 0, \cdots)$. 
Collecting the coefficients of \(a_1, a_2, a_3\), we then have the bilinear equations of the BKP hierarchy
\begin{align}
\Bigg(
&D_{x_1}^{6}
-5D_{x_1}^{3}D_{x_3}
-5D_{x_3}^{2}
+9D_{x_1}D_{x_5}
-\frac{3}{4}g_2D_{x_1}^{2}
-\frac{3}{4}g_3
\Bigg)\tau\cdot\tau =0,\\
\Bigg(
&D_{x_1}^{8}
+7D_{x_1}^{5}D_{x_3}
-21D_{x_1}^{3}D_{x_5}
-35D_{x_1}^{2}D_{x_3}^{2}
+90D_{x_1}D_{x_7} \nonumber\\
&\quad
-42D_{x_3}D_{x_5}
+\frac{7}{4}g_2D_{x_1}^{4}
-7g_2D_{x_1}D_{x_3}
-\frac{39}{4}g_3D_{x_1}^{2}
-\frac{5}{8}g_2^{2}
\Bigg)\tau\cdot\tau =0,\\
&\cdots\nonumber
\end{align}
The following theorem is obtained by the Miwa transformation.
\begin{theorem}\label[theorem]{thm:elliptic-bkp-determinants}
  The BKP hierarchy of the elliptic type \eqref{elli-BKP} admits the tau function as
  \begin{align}\label{bkp_tau}
        \tau = \begin{cases}
            \sigma^2(\xi)(1, 2, \ldots, N), & N \text{ even},\\
            \sigma^2(\xi)(s_0, 1, 2, \ldots, N), & N \text { odd},
        \end{cases}
    \end{align}
    where \(\xi=x_1/2 +x_0\), and the pfaffian elements are defined as
    \begin{align*}
        &(s_0, k) = \Phi^2_\xi(p_k) \exp(\theta_k),\\
        &(k,l) = c_{kl} +[\wp(p_k)-\wp(p_l)]
        \Phi^2_\xi(p_k+p_l)
        \exp(\theta_k+\theta_l), \quad c_{kl}=-c_{lk},\\
        &\theta_k = -\zeta(p_k)x_1+\sum_{\mu=2}^{\infty}
        \frac{\wp^{(2\mu-3)}(p_k)}{(2\mu-2)!}\,x_{2\mu-1}+\theta_{k0},
    \end{align*}
    for \(1\leq k,l\leq N\). Here \(p_k\) represent wave numbers and \(c_{kl},x_0,\theta_{k0}\) are constants.
\end{theorem}

\section{A framework for constructing soliton solutions on elliptic backgrounds via the KP reduction method}
The elliptic KP, modified KP, and BKP hierarchies developed in the previous section
provide a systematic way to construct soliton solutions of integrable nonlinear wave
equations on elliptic background. In this section, we will derive the general $N$-soliton solutions for the KP, KdV, mKdV and the Sawada-Kotera equations. Then we will show and analyze the elliptic seed solutions and $N=1$ soliton solution in detail. 

To analyze the nonsingularity and real-valuedness of the solutions, two types of real lattice will be used throughout this section: the rectangular elliptic lattice \(\Lambda_p = 2\omega_1\mathbb Z + 2\i\omega_2\mathbb Z\) and the rhombic elliptic lattice \(\Lambda_c = 2\omega_1\mathbb Z +  (\omega_1 + \i\omega_2) \mathbb Z\). In particular, the nonsingular elliptic curve corresponding to these lattices satisfies \(g_2, g_3\in \mathbb R\). For further details, see \cref{two_real_lattices}.

\subsection{Line solitons to the KP equation on an elliptic background}
As an immediate result, we can write the determinant solutions to the KP equation
\begin{equation}
 (u_t+6uu_x+u_{xxx})_x+3\epsilon^2u_{yy}=0,
 \quad \epsilon=1,\i, \quad x,y,t\in \mathbb R.
 \label{eq:kp}
\end{equation}
Here \(\epsilon=\i, 1\) correspond to KP-I and KP-II respectively. To derive its solution, set \(x_1=x\), \(x_2=\epsilon y\),
\(x_3=-4t\) in \eqref{Elli-KP1}, we have
\begin{align}\label{Elli-KP1_transfor}
    (D_x^4+D_xD_t+3\epsilon^2D_y^2-g_2)\tau\cdot\tau=0.
\end{align}
Dividing by \(2\tau^2\) gives
\[
(\log\tau)_{xt}+(\log\tau)_{xxxx}
 +6(\log\tau)_{xx}^{2}+3\epsilon^2(\log\tau)_{yy}=\frac{g_2}{2}.
\]
Let \(u(x,y,t)=2(\log\tau)_{xx}\). Eq.~\eqref{eq:kp} is derived from \eqref{Elli-KP1_transfor}. As a result, we have the following theorem.
\begin{theorem}\label[theorem]{thm:kp12-soliton}
The KP equation \eqref{eq:kp} admits the solution \(u=2(\log\tau)_{xx}\), with \(\tau\) given in the following forms.
\begin{enumerate}[(a)]
  \item Casorati determinant
  \begin{align*}
    \tau (x,y,t) = \sigma(\xi)
    \begin{vmatrix}
      \phi_1(0) & \phi_1(1) & \cdots & \phi_1(N-1) \\
      \phi_2(0) & \phi_2(1) & \cdots & \phi_2(N-1) \\
      \vdots & \vdots & \ddots & \vdots \\
      \phi_N(0) & \phi_N(1) & \cdots & \phi_N(N-1)
    \end{vmatrix}
  \end{align*}
  \item Wronskian determinant 
  \begin{align*}
    \tau (x,y,t) = \sigma(\xi)
    \begin{vmatrix}
      \phi_1(0) & \partial_x\phi_1(0) & \cdots & \partial_x^{N-1}\phi_1(0) \\
      \phi_2(0) & \partial_x\phi_2(0) & \cdots & \partial_x^{N-1}\phi_2(0) \\
      \vdots & \vdots & \ddots & \vdots \\
      \phi_N(0) & \partial_x\phi_N(0) & \cdots & \partial_x^{N-1}\phi_N(0)
    \end{vmatrix}
  \end{align*}
  \item  Gram-type determinant
  \begin{align*}
    \tau (x,y,t) = \sigma(\xi) \det_{1 \leq i,j \le N} \left(m_{ij}\right)
  \end{align*}
\end{enumerate}
while \(\xi=x+x_0\) and the corresponding matrix elements are defined as
\begin{align*}
  \phi_j(s) &=\alpha_j \left[\Phi_{b} (p_j)\right]^{s} \Phi_{\xi-bs}(p_j)
  \exp (\mu_j)
  + \beta_j \left[\Phi_{b} (q_j)\right]^{s} \Phi_{\xi-bs}(q_j) \exp(\eta_j) ,  \\
  \mu_j &= -\zeta(p_j)x+\epsilon\wp(p_j)y-2\wp'(p_j)t, \quad
  \eta_j = -\zeta(q_j)x+\epsilon\wp(q_j)y-2\wp'(q_j)t,
  \\
  m_{ij} &= c_{ij}+\Phi_\xi(p_i+q_j) \exp(\theta_{ij}), \\
  \theta_{ij} &= -(\zeta(p_i)+\zeta(q_j))x
  +\epsilon(\wp(p_i)-\wp(q_j))y-2(\wp'(p_i)+\wp'(q_j))t+\mu_{i0}+\eta_{j0}.
\end{align*}
Here \(p_j,q_j\), \(j=1,\ldots,N\), represent wave numbers and
\(b,x_0,\alpha_j,\beta_j,c_{ij},\mu_i^{(0)},\eta_i^{(0)}\) are constants.
\end{theorem}

\begin{remark} The Wronskian solution derived here as the limit of the Casorati solution as \(b \to 0 \)
is the same Wronskian solution given in Theorem~4.1 of \cite{Li2022Elliptic}, while the Gram solution recovers the results in \cite{li2025direct}.
\end{remark}

\begin{corollary}\label[corollary]{kp_soliton_nonsingular}
    To ensure the regularity and real-valuedness in \cref{thm:kp12-soliton}, we give the following sufficient conditions of the \(N\)-soliton solutions of the Gram determinant form under the lattice \(\Lambda=\Lambda_p\).
    \begin{enumerate}
        \item KP-I solution (\(\epsilon=\i\)). Take \(p_i-p_j\notin\Lambda_p\) for \(i\ne j\),
        \begin{align}\label{KP-I_nonsingular_cond}
            x_0 = \i \omega_2, \quad q_i = \overline{p_i}, \quad 0< \Re p_i < \omega_1, \quad \eta_{i0}=\overline{\mu_{i0}}-2\zeta(\i\omega_2)\overline{p_i},
        \end{align}
        and \(\mat_{1\leq i,j\leq N}(c_{ij})\) to be a constant Hermitian positive semidefinite matrix.
        \item KP-II solution (\(\epsilon=1\)). Take 
        \begin{align}
        \begin{split}\label{KP-II_nonsingular_cond}
            &x_0 = \i \omega_2, \quad 0 < p_1 < \cdots < p_N, \quad 0 < q_1 < \cdots < q_N, \quad p_N+q_N< 2\omega_1, \\
            &c_{ij}=\delta_{ij}c_i, \quad c_i > 0, \quad \mu_{i0}+\eta_{i0}+\zeta(\i\omega_2)(p_i+q_i)\in\mathbb R.
        \end{split}
        \end{align}
    \end{enumerate}
    Above conditions ensure that for real \(x,y,t\), the solution \(u(x,y,t)=2(\log\tau)_{xx}\) is real and nonsingular.
\end{corollary}
\begin{proof}
    First, we show the solution to KP-I is real and nonsingular under
    condition \eqref{KP-I_nonsingular_cond}.
    For each \(h=1,\ldots,N\), by \eqref{sigma_positive} we have
    \begin{align*}
        &A =
        \frac{\sigma\bigl(\xi+\sum_{i=1}^h(p_i+\overline{p_i})\bigr)}
        {\sigma(\xi)}
        \exp\left(-\zeta(\i\omega_2)
        \sum_{i=1}^h(p_i+\overline{p_i})\right)>0,\\
        &\sum_{i=1}^h
        \left[\theta_{ii}+\zeta(\i\omega_2)(p_i+\overline{p_i})\right]
        =
        2\Re\sum_{i=1}^h
        \left[
        -\zeta(p_i)x+\i\wp(p_i)y-2\wp'(p_i)t
        +\mu_{i0}+\zeta(\i\omega_2)p_i
        \right]\in\mathbb R.
    \end{align*}
    Applying the elliptic Cauchy determinant identity
    \eqref{elliptic_cauchy_det} to the leading principal minors gives
    \begin{align*}
        \det_{1\leq i,j\leq h}
        \left(\Phi_{\xi}(p_i+\overline{p_j})e^{\theta_{ij}}\right)&=
        \exp\left(\sum_{i=1}^h\theta_{ii}\right)
        \det_{1\leq i,j\leq h}
        \left(\Phi_{\xi}(p_i+\overline{p_j})\right)\\
        &=
        A\exp\left(
        \sum_{i=1}^h
        \left[\theta_{ii}+\zeta(\i\omega_2)(p_i+\overline{p_i})\right]
        \right)
        \frac{
        \prod\limits_{1\leq i<j\leq h}|\sigma(p_j-p_i)|^2
        }{
        \prod\limits_{i=1}^h\sigma(2\Re p_i)
        \prod\limits_{1\leq i<j\leq h}
        |\sigma(p_i+\overline{p_j})|^2
        }>0.
    \end{align*}
    Next, by
    \begin{align*}
        \overline{\Phi_{\xi}(p_j+\overline{p_i})e^{\theta_{ji}}}&=\exp\left(
        -2\zeta(\i\omega_2)(p_i+\overline{p_j})
        +\overline{\mu_{j0}}+\overline{\eta_{i0}}
        -\mu_{i0}-\eta_{j0}
        \right)
        \Phi_{\xi}(p_i+\overline{p_j})e^{\theta_{ij}}\\
        &=\Phi_{\xi}(p_i+\overline{p_j})e^{\theta_{ij}},
    \end{align*}
    we have
    \[
        \mat(\Phi_{\xi}(p_i+\overline{p_j})e^{\theta_{ij}})^\dagger
        =
        \mat(\Phi_{\xi}(p_i+\overline{p_j})e^{\theta_{ij}}).
    \]
    Thus all leading principal minors of this Hermitian matrix are
    positive, so it is positive definite by Sylvester's criterion.
    Therefore, for every nonzero column vector
    \(\bm x\in\mathbb C^N\),
    \[
        \bm x^\dagger\mat(m_{ij})\bm x
        =
        \bm x^\dagger\mat(c_{ij})\bm x
        +
        \bm x^\dagger
        \mat(\Phi_{\xi}(p_i+\overline{p_j})e^{\theta_{ij}})
        \bm x>0.
    \]
    Thus \(\tau/\sigma(\xi)=\det_{1\leq i,j\leq N}(m_{ij})>0\),
    and hence
    \(u=-2\wp(x+\i\omega_2)+2\partial_x^2\log\det(m_{ij})\)
    is real-valued and nonsingular.

    Next, for KP-II, since \(c_{ij}=\delta_{ij}c_i\), expansion of
    \(\tau/\sigma(\xi)\) along the diagonal entries gives
    \begin{align*}
        \det_{1\leq i,j\leq N}(m_{ij})
        =
        \sum_{I\subseteq\{1,\ldots,N\}}
        \left(\prod_{j\notin I}c_j\right)
        \det_{i,j\in I}
        \left(\Phi_{\xi}(p_i+q_j)e^{\theta_{ij}}\right)>0,
    \end{align*}
    as \(c_i>0\) and
    \(\det_{i,j\in I}
    (\Phi_{\xi}(p_i+q_j)e^{\theta_{ij}})>0\)
    for every nonempty \(I\), by
    \eqref{elliptic_cauchy_det}, \eqref{sigma_positive},
    and the ordering and phase conditions in
    \eqref{KP-II_nonsingular_cond}.
\end{proof}

\subsection{Lump solution to the KP-I equation on an elliptic background}
Lump solutions to the KP-I equation were first constructed in \cite{Manakov1977TwoDimensional,SatsumaAblowitz1979Lumps}. The higher order lump solution was firstly derived by Pelinovsky and Stepanyants in
\cite{PelinovskyStepanyants1993}. One of the authors numerically simulated the fundamental lump and discovered the interesting collision of two lumps without being aware of the work by Pelinovsky et al. \cite{FengMitsui1998,FengKawaharaMitsui1999}. 
Higher-order lumps and their patterns were investigated in \cite{YangYang2022Lumps,YangYang2025Concentric} along with many papers for the lump chain and their higher order lump solutions \cite{Chang2018Asymptotic,ChakravartyZowada2022Dynamics,ChakravartyZowada2022Classification,ChakravartyZowada2023Patterns,SinghStepanyants2016,HuEtAl2018MultiLumps,ZhangEtAl2022Degenerate,ZhangGuoStepanyants2023,ZhangEtAl2023MultiLump,LesterEtAl2021LumpChains,ZhangEtAl2022ResonantLumpChains,ZhangEtAl2023MultiLump,ZhangGuoStepanyants2023}. Connections between higher-order KP-I lump patterns and
rogue-wave patterns are discussed in
\cite[Chapter~5]{YangYang2024RogueWaves}.
These connections provide a broader context for the study
of localized structures through explicit solution formulas.
Here we use spectral differentiation of an elliptic Gram
kernel to construct KP-I solutions on elliptic backgrounds,
including lump-type examples.

\begin{theorem}
\label{thm:kpi-lump-general}
    The lump solutions on an elliptic background to the KP-I equation \eqref{eq:kp} are given by
    \begin{align}
    u=2(N-1)\wp(\xi)-2N\wp(\xi + 2a)+2\partial_x^2\log \tilde \tau, \quad \tilde \tau &=\det_{1\leq k,l\leq N}(m_{n_k n_l}),
 \label{eq:kpi-lump-general-field}
\end{align}
    where $\xi=x+x_0$, and \(n_k,n_l\) are distinct integers such that $0\leq n_1<\cdots<n_N$, with
    \begin{align*}
        m(p,q)&=\Phi_{\xi}(p+q)\exp\Big(-[\zeta(p)+\zeta(q)]x
               +\i[\wp(p)-\wp(q)]y-2[\wp'(p)+\wp'(q)]t\Big),\\
         m_{ij}&=\left. \frac{(\partial_p+c)^i}{i!}\frac{(\partial_q+c)^j}{j!} \frac{m(p,q)}
          {m(a,a)}\right|_{p=q=a}.
    \end{align*}
    Here \(a, c\) are constants.
\end{theorem}
\begin{proof}
 As in the proof of Lemma~3.2 in \cite{ohta2012general},
the differential relations underlying the Gram determinant construction
are preserved by the spectral differential operators \(\frac{(\partial_p+c)^i}{i!}\) and \(\frac{(\partial_q+c)^j}{j!}\). Hence
\begin{align*}
\tau=\sigma(\xi)m(a,a)^N\widetilde{\tau}=\sigma(\xi)\det_{1\leq k,l\leq N}
\left(
\left.
\frac{(\partial_p+c)^{n_k}}{n_k!}
\frac{(\partial_q+c)^{n_l}}{n_l!}
m(p,q)
\right|_{p=q=a}
\right)
\end{align*}
is a solution to \eqref{Elli-KP1_transfor} when \(\epsilon=\i\). We can easily show
\begin{align*}
\partial_x^2\log m(a,a)=\wp(\xi)-\wp(\xi+2a),
\end{align*}
thus
\begin{align*}
    u = 2\partial_x^2\log \tau =2(N-1)\wp(\xi)-2N\wp(\xi+2a)
  +2\partial_x^2\log\widetilde{\tau}
\end{align*}
is a solution to the KP-I equation \eqref{eq:kp}.
\end{proof}

\begin{corollary}\label[corollary]{cor:kp1_lump_nonsingular}
    The following parameter choice 
    \begin{align*}
        \Lambda = \Lambda_p,\quad  0 < a < \omega_1, \quad
        x_0=b+\i\omega_2,\quad c=d-\zeta(\i \omega_2),
        \quad b,d\in\mathbb R
    \end{align*}
    gives a real-valued and nonsingular solution in \cref{thm:kpi-lump-general}. 
\end{corollary}
\begin{proof}
    We first show the case \(b=0\), i.e., \(x_0=\i\omega_2\). 
    Note that 
    \begin{align*}
        m_{ij}= \left. \frac{(\partial_p+c)^i}{i!}\frac{(\partial_q+c)^j}{j!} \frac{m(p,q)}
          {m(a,a)}\right|_{p=q=a} = \left.
        \frac{\partial_p^i\partial_q^j}{i! j!}
        \left[
        e^{c(p+q-2a)}
        \frac{m(p,q)}{m(a,a)}
        \right]
        \right|_{p=q=a}.
    \end{align*}
    The tau function in \cref{thm:kpi-lump-general} can be obtained by taking limits for spectral parameters in \cref{thm:kp12-soliton}. Pick distinct \(p_0,\ldots,p_{n_N}\in(0,\omega_1)\), and set \(c_{ij}=0, q_k=p_k, \mu_{k0}=\eta_{k0}=d(p_k-a)-p_k\zeta(\mathrm{i}\omega_2)\). By \cref{kp_soliton_nonsingular} we have
    \begin{align*}
        \overline{e^{c(p+q-2a)}
        \frac{m(p,q)}{m(a,a)}} = e^{c(p+q-2a)}
        \frac{m(q,p)}{m(a,a)},
    \end{align*}
    therefore \(\overline{m_{ji}}=m_{ij}\).
    Let us consider the leading principal minors of the matrix \(\mat_{0\le i,j\le n_N}(m_{ij})\): for \(h=0,1,\ldots,n_N\), applying the elliptic Cauchy determinant identity \eqref{elliptic_cauchy_det} gives
    \begin{align*}
        \det_{0\le i,j\le h}(m_{ij})
={}&
\lim_{p_0,\ldots,p_{h}\to a}
\frac{
\det\limits_{0\le k,l\le h}
\left(
e^{c(p_k+p_l-2a)}
\frac{m(p_k,p_l)}{m(a,a)}
\right)
}{
\displaystyle\prod_{0\le k<l\le h}(p_k-p_l)^2
} \\
={}& \lim_{p_0,\ldots,p_{h}\to a}
\frac{
\exp\left(
2\sum_{k=0}^{h}
\bigl[c(p_k-a)-x\zeta(p_k)-2t\wp'(p_k)\bigr]
\right)
}{
m(a,a)^{h+1}\displaystyle\prod_{0\le k<l\le h}(p_k-p_l)^2
}
\det_{0\le k,l\le h}
\left(
\Phi_{\xi}(p_k+q_l)
\right) \\
={}& \lim_{p_0,\ldots,p_{h}\to a}
\frac{
\exp\left(
2\sum_{k=0}^{h}
\bigl[c(p_k-a)-x\zeta(p_k)-2t\wp'(p_k)\bigr]
\right)
}{
m(a,a)^{h+1}\displaystyle\prod_{0\le k<l\le h}(p_k-p_l)^2
} \frac{\sigma\!\left(\xi+2\sum_{k=0}^{h}p_k\right)}
{\sigma(\xi)}
\frac{\displaystyle\prod_{k<l}\sigma(p_k-p_l)^2}
{\displaystyle\prod_{k,l}\sigma(p_k+p_l)},
\end{align*}
and note that
\(
\lim\limits_{p_0,\ldots,p_{h}\to a}
\prod_{k<l}
\left(
\frac{\sigma(p_k-p_l)}{p_k-p_l}
\right)^2=1,
\)
by \eqref{sigma_positive} we have
\begin{align*}
    \det_{0\le i,j\le h}(m_{ij})
={}&
\frac{
e^{-2(h+1)\zeta(a)x-4(h+1)\wp'(a)t}
}{
m(a,a)^{h+1}
}
\frac{
\sigma\bigl(\xi+2(h+1)a\bigr)
}{
\sigma(\xi)\sigma(2a)^{(h+1)^2}
}\\
={}&
\frac{
\sigma\bigl(\xi+2(h+1)a\bigr)\sigma(\xi)^{h}
}{
\sigma(\xi+2a)^{h+1}
\sigma(2a)^{h(h+1)}
} =
\frac{
\displaystyle
e^{-2(h+1)a\zeta(\mathrm i\omega_2)}
\frac{\sigma\bigl(\xi+2(h+1)a\bigr)}{\sigma(\xi)}
}{
\displaystyle
\left[
e^{-2a\zeta(\mathrm i\omega_2)}
\frac{\sigma(\xi+2a)}{\sigma(\xi)}
\right]^{h+1}
\sigma(2a)^{h(h+1)}
}
>0.
\end{align*}
Hence \(\mat_{0\le i,j\le n_N}(m_{ij})\) is Hermitian positive definite by Sylvester’s criterion, and subsequently we have \(\tilde \tau > 0\). For \(b\neq 0\) case, one can perform the same argument but starting with \(x_0 = b+\i \omega_2\).
\end{proof}

Next, let us consider some examples from \cref{thm:kpi-lump-general} with parameters satisfying \cref{cor:kp1_lump_nonsingular} and \(b=-2a,d=\zeta(2a)\).

\begin{figure}[!htbp]
  \centering
  \subfloat[Lump surface.]{%
    \includegraphics[width=0.48\textwidth]{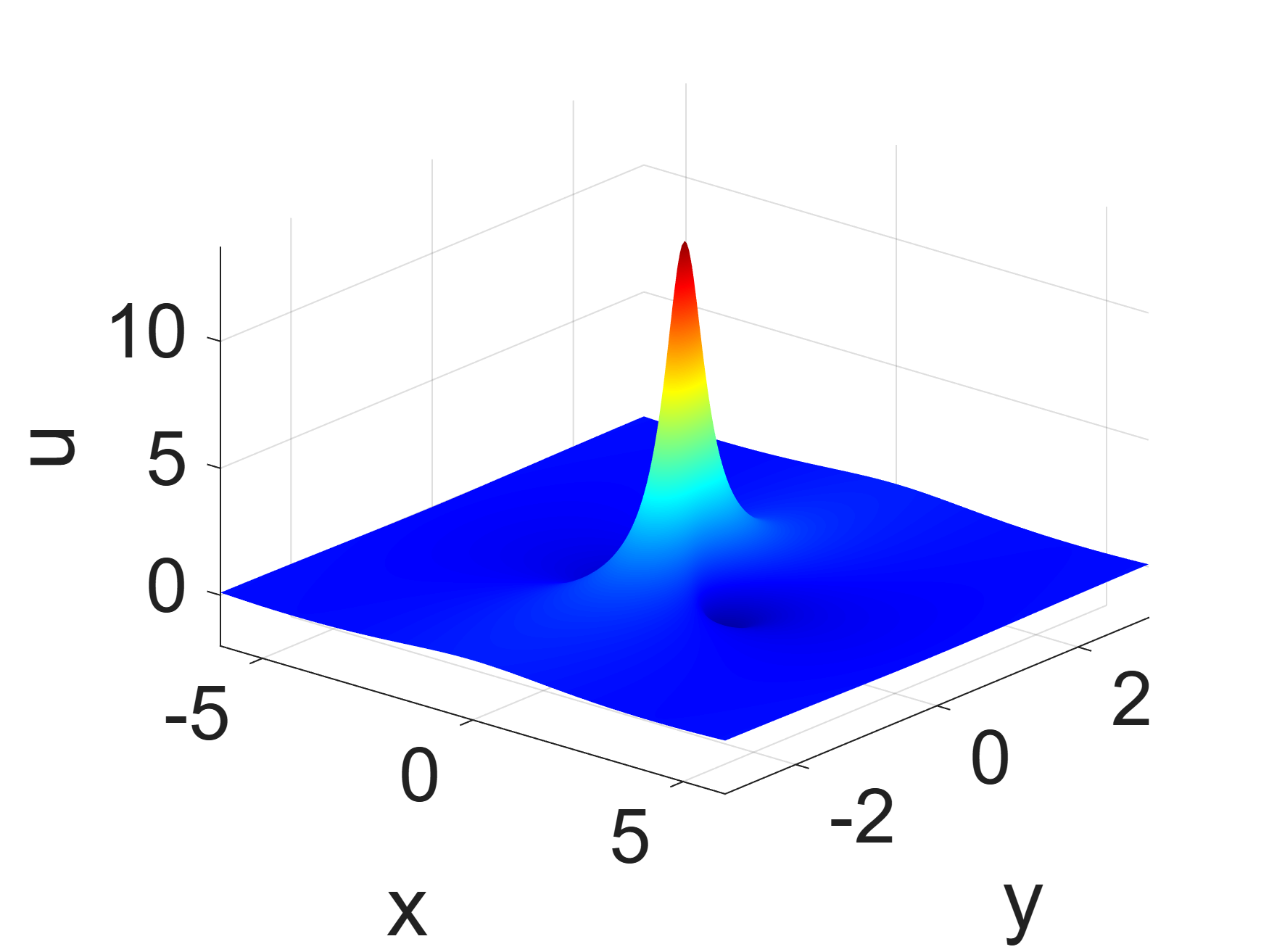}}
  \hfill
  \subfloat[Slice at $y=t=0$.]{%
    \includegraphics[width=0.48\textwidth]{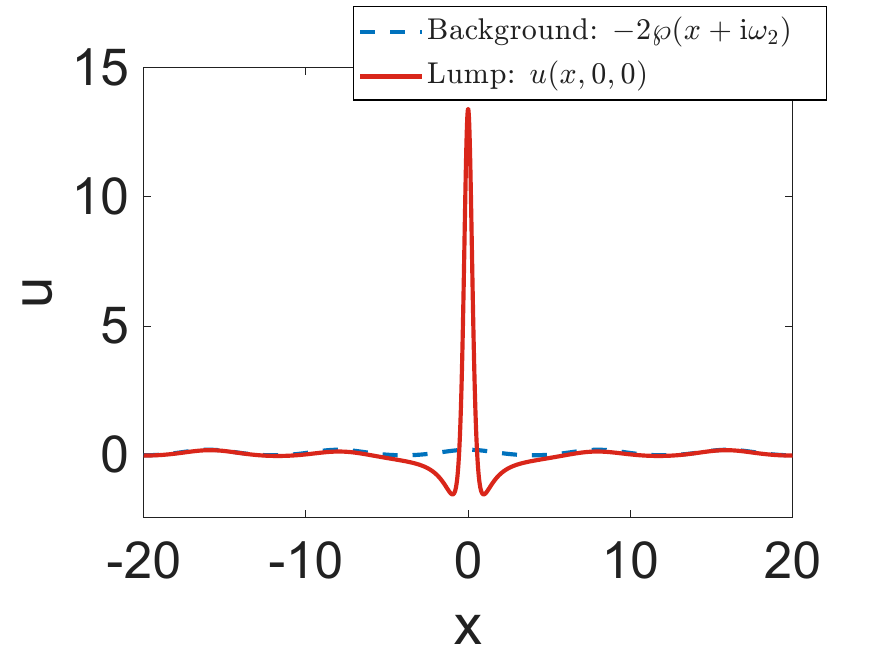}}
  \caption{The fundamental KP-I lump $u(x,y,0)$ with $N=1$, $n_1=1$,
  $\Lambda=\Lambda_p$, $\omega_1=\omega_2=4$, $a=1$,
  $x_0=-2+4\i$, and $c=\zeta(2)-\zeta(4\i)$.}
  \label{fig:kpi_lump_n1}
\end{figure}

\begin{figure}[!htbp]
  \centering
  \subfloat[Before interaction, $t=-0.75$.]{%
    \includegraphics[width=0.32\textwidth]{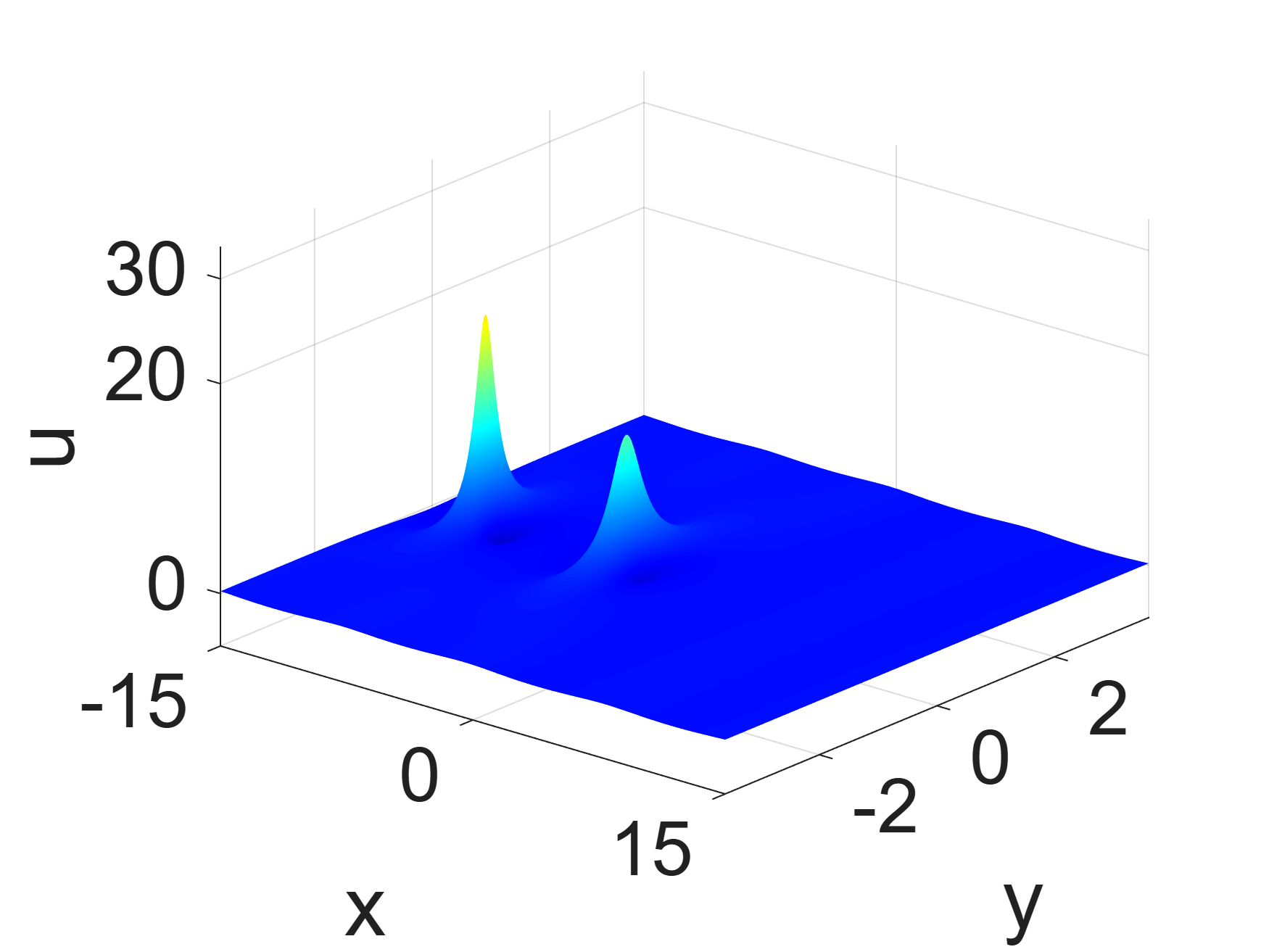}}
  \hfill
  \subfloat[During collision, $t=0$.]{%
    \includegraphics[width=0.32\textwidth]{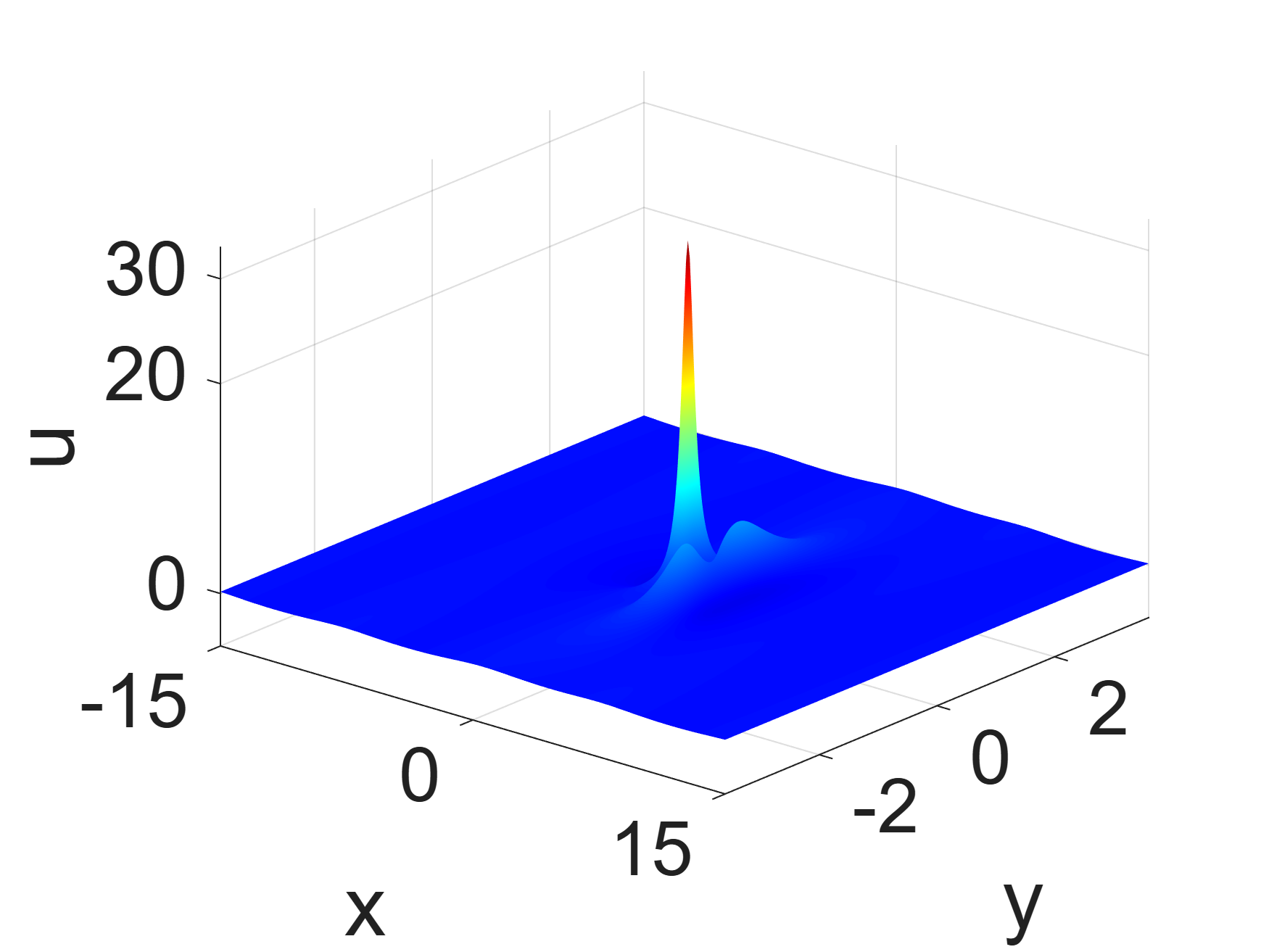}}
  \hfill
  \subfloat[After interaction, $t=0.75$.]{%
    \includegraphics[width=0.32\textwidth]{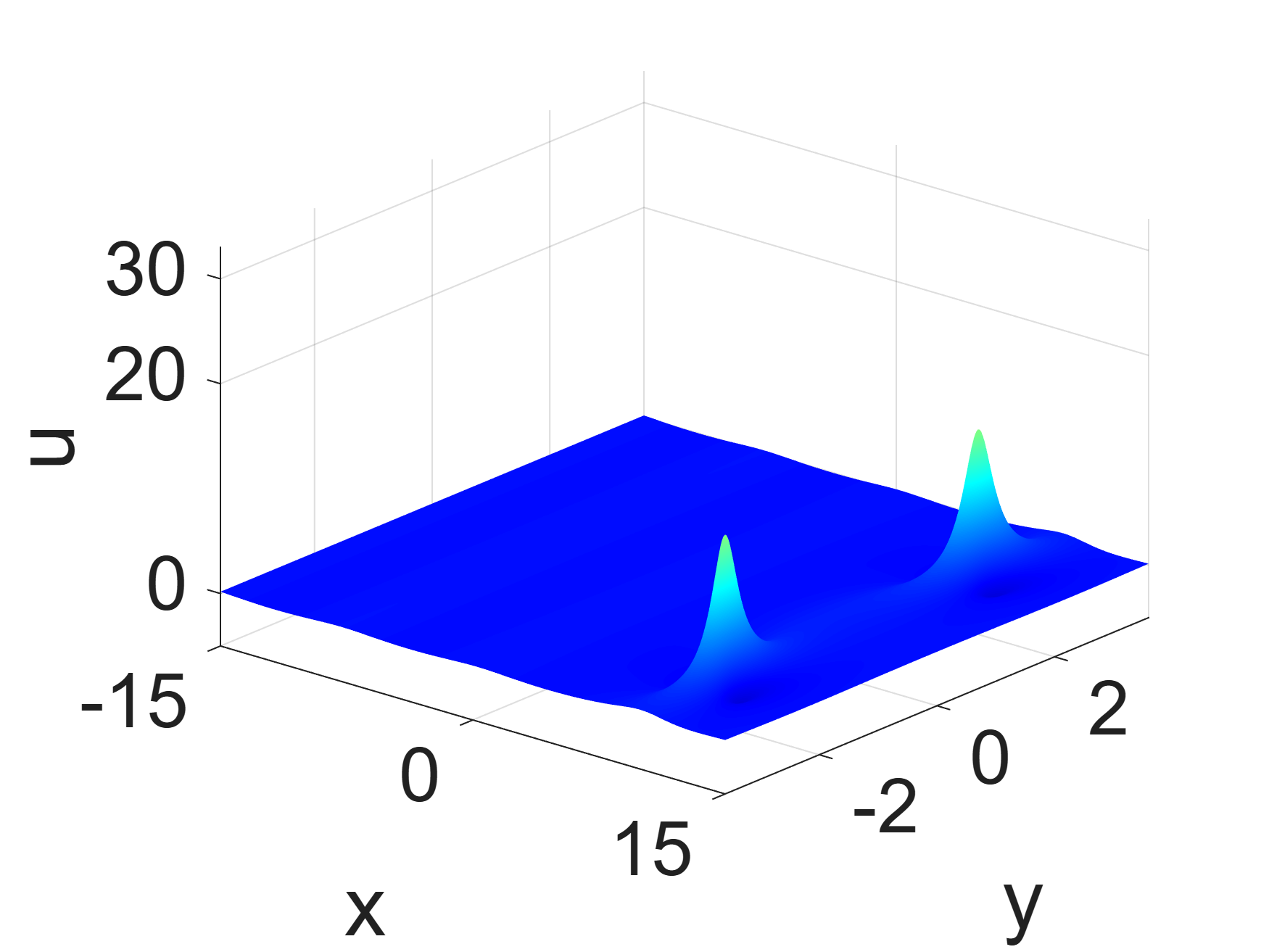}}
  \caption{The second-order $(2,2)$ KP-I lump $u(x,y,t)$ with $N=1$,
  $n_1=2$, $\Lambda=\Lambda_p$, $\omega_1=\omega_2=4$, $a=1$,
  $x_0=-2+4\i$, and $c=\zeta(2)-\zeta(4\i)$.}
  \label{fig:kpi_lump_order2}
\end{figure}

\paragraph{The fundamental lump: $(1,1)$.}
Take $N=1$ and $n_1=1$. 
The fundamental
lump is
\begin{align*}
\widetilde\tau
={}&
\left[
\wp(a)x-2\wp''(a)t
+\zeta(x+\mathrm i\omega_2)-\zeta(\mathrm i\omega_2)
\right]^2+[\wp'(a)]^2y^2
+\wp(2a)-\wp(x+\mathrm i\omega_2),\nonumber\\
 u={}&-2\wp(x+\i \omega_2)+2\partial_x^2\log \tilde \tau,
\end{align*}
The fundamental lump is illustrated in \cref{fig:kpi_lump_n1}.
Under the limit \(\omega_1, \omega_2 \to \infty\) or \(g_2,g_3\to 0\), we have \(
\sigma(z)\to z,
\zeta(z)\to\frac1z,
\wp(z)\to\frac1{z^2},
\)
and subsequently
\[
u=
4\frac{\displaystyle
\frac{4y^2}{a^2}+\frac{a^2}{4}
-\left(x-\frac{12t}{a^2}\right)^2
}{\displaystyle
\left[
\frac{4y^2}{a^2}
+\frac{a^2}{4}+\left(x-\frac{12t}{a^2}\right)^2
\right]^2
}.
\]
\paragraph{A higher-order lump: $(2,2)$.}
Now take $N=1$ and $n_1=2$.
The explicit formula is
\begin{align*}
\widetilde \tau
={}&
\frac14
\Bigg\{
\left[
\wp(a)x-2\wp''(a)t
+\zeta(x+\mathrm i\omega_2)-\zeta(\mathrm i\omega_2)
\right]^2
-[\wp'(a)]^2y^2
+\wp'(a)x-2\wp'''(a)t
+\wp(2a)-\wp(x+\mathrm i\omega_2)
\Bigg\}^{2}
\\
&+\frac{y^2}{4}
\Bigg\{
2\wp'(a)
\left[
\wp(a)x-2\wp''(a)t
+\zeta(x+\mathrm i\omega_2)-\zeta(\mathrm i\omega_2)
\right]
+\wp''(a)
\Bigg\}^{2}
\\
&+\left[\wp(2a)-\wp(x+\mathrm i\omega_2)\right]
\Bigg\{
\left[
\wp(a)x-2\wp''(a)t
+\zeta(x+\mathrm i\omega_2)-\zeta(\mathrm i\omega_2)
\right]^2+[\wp'(a)]^2y^2
+2\wp(2a)+\wp(x+\mathrm i\omega_2)
\Bigg\}
\\
&+\left[\wp'(2a)-\wp'(x+\mathrm i\omega_2)\right]
\left[
\wp(a)x-2\wp''(a)t
+\zeta(x+\mathrm i\omega_2)-\zeta(\mathrm i\omega_2)
\right],\\
 u={}&-2\wp(x+\i \omega_2)+2\partial_x^2\log \tilde \tau.
\end{align*}
The second-order lump before, during, and after interaction is shown in
\cref{fig:kpi_lump_order2}.
\begin{remark}
We comment here that this interesting solution \cref{fig:kpi_lump_order2} under zero background was firstly derived by Pelinovsky and Stepanyants in
\cite{PelinovskyStepanyants1993}. One of the authors numerically simulated the fundamental lump (\(1,1\)) and  (\(2,2\))-lump \cite{FengMitsui1998,FengKawaharaMitsui1999}. 
\end{remark}
\begin{remark}
 Elliptic analogues of more higher order lump  and lump chain solutions can be further studied by following the result in \cref{thm:kpi-lump-general}. 
\end{remark}

\subsection{Solitons to the KdV equations on an elliptic background}
The KdV equation
\begin{equation}\label{eq:kdv}
 u_t+6uu_x+u_{xxx}=0, \quad  x,t\in \mathbb R,
\end{equation}
is the period-two reduction of the KP equation.  At the level of the
tau function, the reduction means that the \(x_2\)-dependence is absent
up to an exponential gauge, in other words, tau functions of the KP hierarchy need to satisfy 
\begin{align}\label{kdv_2_reduction}
  \partial_{x_2} \tau(\bm x) = C\tau(\bm x), \quad C\text{ is constant}.
\end{align}
For Casorati and Wronskian solutions \eqref{kp_casorati}-\eqref{kp_wronskian}, condition \eqref{kdv_2_reduction} is satisfied by taking 
\begin{equation*}
 q_j=-p_j,\quad j=1,\ldots,N,
\end{equation*}
while for the Gram solutions \eqref{kp_gram}, condition \eqref{kdv_2_reduction} is achieved by taking
\begin{equation*}
 c_{ij} = \delta_{ij} c_i, \quad q_j=p_j,\quad j=1,\ldots,N,
\end{equation*}
where \(\delta_{ij}\) is the Kronecker delta. In summary, the $N$-soliton solution to the KdV equation on the elliptic background can be summarized by the following theorem.
\begin{theorem}\label[theorem]{cor:KdV-elliptic-determinants} The KdV
equation ~\eqref{eq:kdv} admits a general $N$-soliton solution \(u=2(\log \tau)_{xx}\) where \(\tau\) has one of the following three forms
\begin{enumerate}[(a)]
  \item Casorati determinant
  \begin{align*}
    \tau(x,t) = \sigma(\xi)
    \begin{vmatrix}
      \hat\phi_1(0) & \hat\phi_1(1) & \cdots & \hat\phi_1(N-1) \\
      \hat\phi_2(0) & \hat\phi_2(1) & \cdots & \hat\phi_2(N-1) \\
      \vdots & \vdots & \ddots & \vdots \\
      \hat\phi_N(0) & \hat\phi_N(1) & \cdots & \hat\phi_N(N-1)
    \end{vmatrix}
  \end{align*}
  \item Wronskian determinant
  \begin{align*}
    \tau(x,t) = \sigma(\xi)
    \begin{vmatrix}
      \hat\phi_1(0) & \partial_x\hat\phi_1(0) & \cdots & \partial_x^{N-1}\hat\phi_1(0) \\
      \hat\phi_2(0) & \partial_x\hat\phi_2(0) & \cdots & \partial_x^{N-1}\hat\phi_2(0) \\
      \vdots & \vdots & \ddots & \vdots \\
      \hat\phi_N(0) & \partial_x\hat\phi_N(0) & \cdots & \partial_x^{N-1}\hat\phi_N(0)
    \end{vmatrix}
  \end{align*}
  \item Gram-type determinant
  \begin{align*}
    \tau(x,t) = \sigma(\xi)
    \det_{1\leq i,j\leq N}\left(\hat m_{ij}\right)
  \end{align*}
\end{enumerate}
while \(\xi=x+x_0\) and the corresponding matrix elements are defined as
\begin{align*}
  \hat\phi_j(s)
  &=\alpha_j\left[\Phi_b(p_j)\right]^s
    \Phi_{\xi-bs}(p_j)\exp(\mu_j)
    +\beta_j\left[\Phi_b(-p_j)\right]^s
    \Phi_{\xi-bs}(-p_j)\exp(-\mu_j),\\
  \mu_j
  &=-\zeta(p_j)x-2\wp'(p_j)t,\\
  \hat m_{ij}
  &=\delta_{ij}c_i+\Phi_\xi(p_i+p_j)\exp(\theta_i),\\
  \theta_i
  &=-2\zeta(p_i)x-4\wp'(p_i)t+\theta_{i0}.
\end{align*}
Here \(p_j\), \(j=1,\cdots,N\) represent wave numbers and
\(b,\alpha_j,\beta_j,c_i,\theta_{i0}\) are constants.
\end{theorem}
\begin{corollary}
For Gram determinant solution, take $\Lambda = \Lambda_p, x_0=\i\omega_2$ and distinct spectral points
\begin{equation*}
 p_i\in(0,\omega_1)\ \cup\
              \bigl(\i\omega_2+(0,\omega_1)\bigr),\quad
 c_i>0,\quad \theta_{i0}+2\zeta(\i \omega_2)p_i\in\mathbb R.
\end{equation*}
Then $\tau/\sigma(\xi)>0$ everywhere on $\mathbb R^2$.
Thus $u=2(\log\tau)_{xx}$ is a real-valued, nonsingular KdV solution.
\end{corollary}

\paragraph{Elliptic background solution (\(N=0\)).} Let us consider the seed solution for \(N=0\) in \cref{cor:KdV-elliptic-determinants}, which gives the stationary elliptic solution
\begin{equation*}
  u(x)=-2\wp(x+\i \omega_2).
\end{equation*}
Using the relation between the Weierstrass \(\wp\) and Jacobi \(\cn\) (see Appendix), 
and applying the Galilean transformation \(u(x,t)\mapsto u(x-12e_3t,t)+2e_3\), we obtain the elliptic wave solution in the form
\begin{equation}\label{eq:kdv_traveling_solution}
  u(x,t)=2(e_3-e_2)\cn^2\big(\nu(x-12e_3t),\kappa\big),  \quad \nu = \sqrt{e_1 - e_2}, \quad \kappa = \sqrt{\frac{e_3-e_2}{e_1-e_2}}.
\end{equation}

\paragraph{One soliton on elliptic background (\(N=1\)).} The Gram determinant in \cref{cor:KdV-elliptic-determinants} gives
\begin{align*}
  u(x,t)&=-2\wp(\xi)
+2\partial_x^2\log\left[c_1+\Phi_\xi(2p_1)e^{-2\zeta(p_1)x-4\wp'(p_1)t+\theta_{10}}\right], \quad \quad \xi=x+x_0.
\end{align*}
On the rectangular lattice \(\Lambda_p\), choose
\[
  x_0=\i\omega_2,\quad c_1>0,\quad
 \theta_{10}=-2p_1\zeta(\i\omega_2)
\]
And the two nonsingular branches are
\begin{enumerate}[(a)]
  \item Bright breather: \(0<p_1<\omega_1\). 
  \item Dark breather: \(p_1=a+\i\omega_2\), \(0<a<\omega_1\). 
\end{enumerate}

\begin{figure}[!htbp]
  \centering
  \subfloat[Bright breather.]{%
    \includegraphics[width=0.48\textwidth]{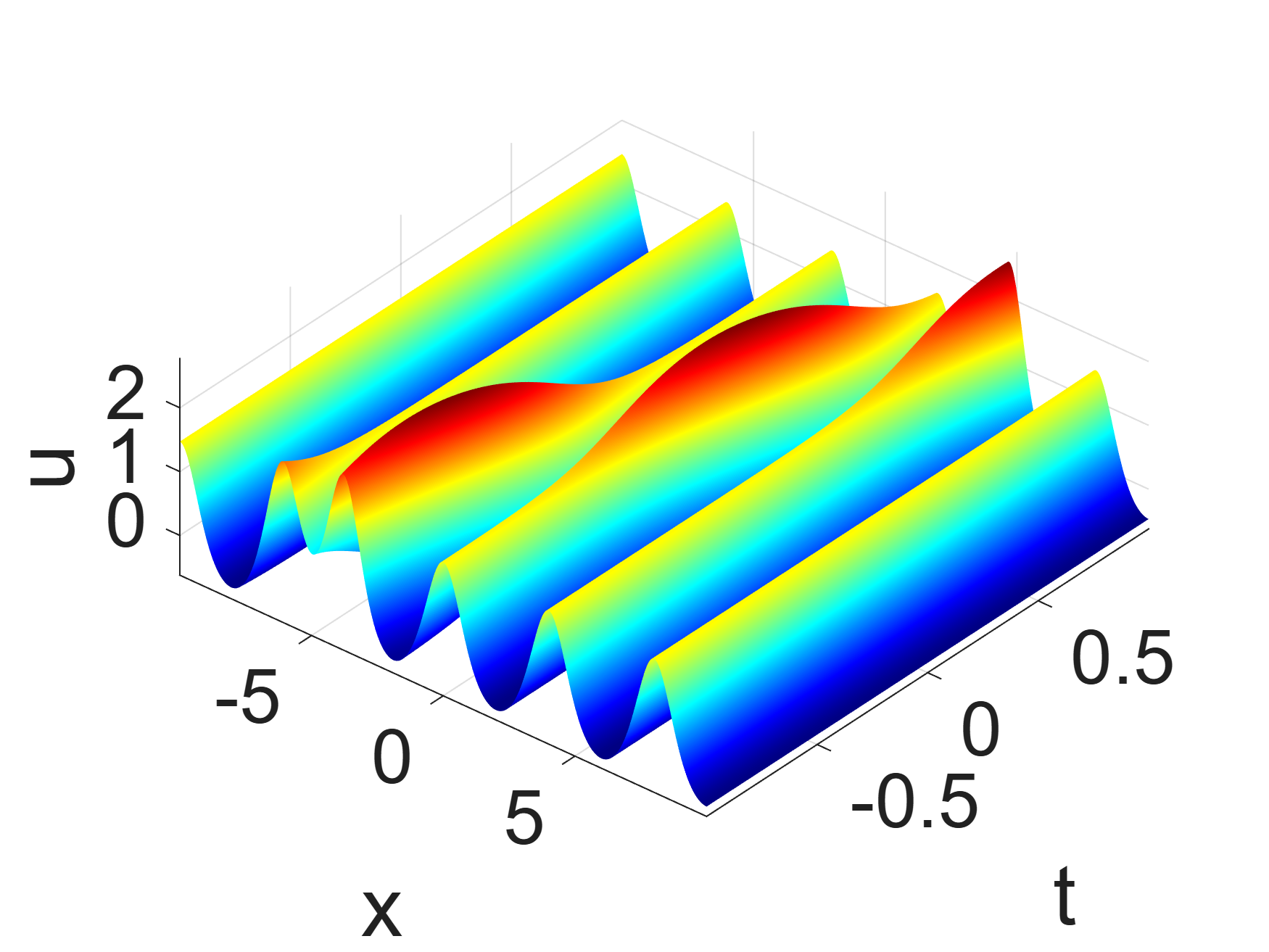}}
  \hfill
  \subfloat[Dark breather.]{%
    \includegraphics[width=0.48\textwidth]{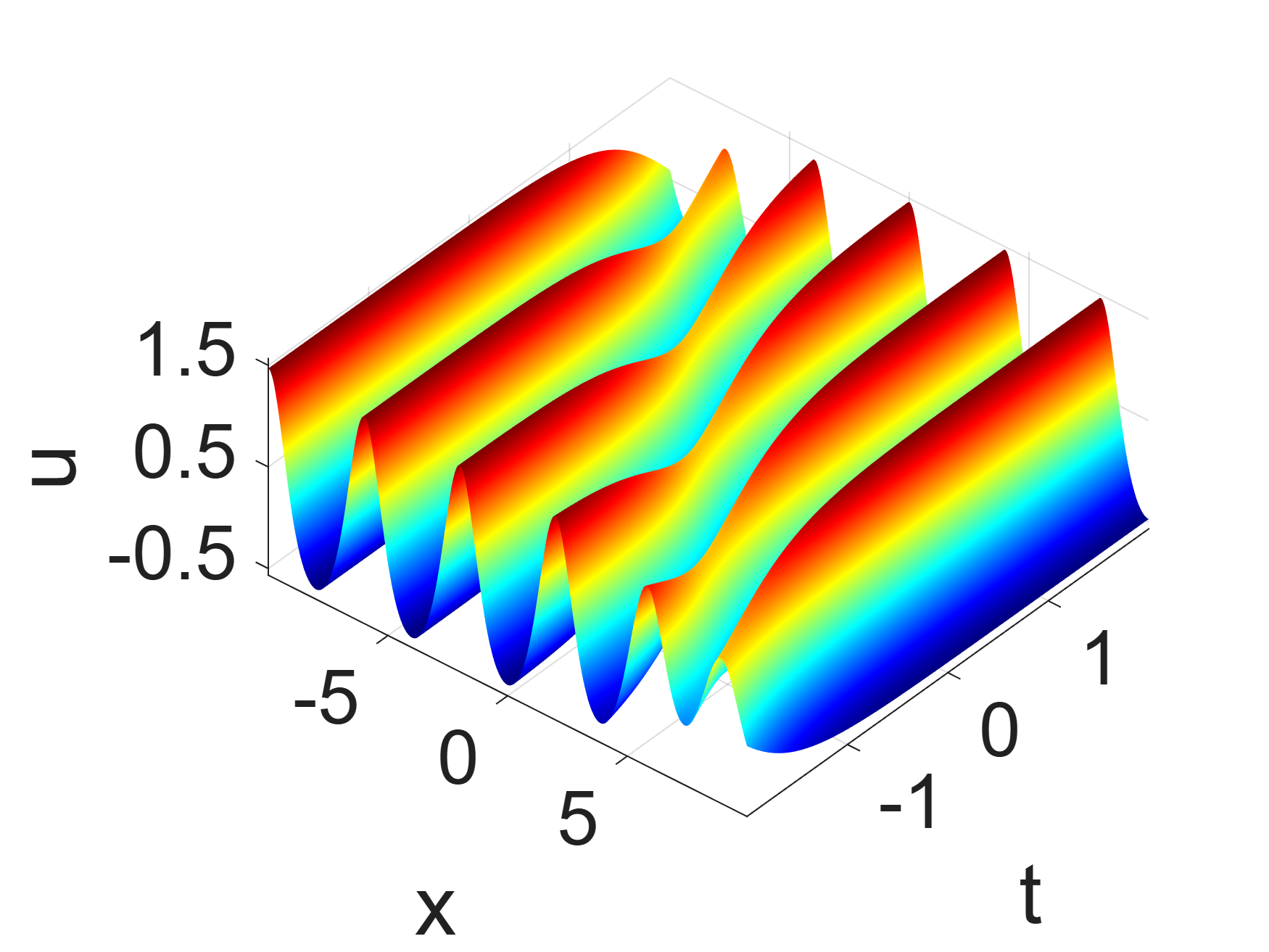}}
  \caption{The $N=1$ KdV solutions $u(x,t)$ with $\Lambda=\Lambda_p$,
  $\omega_1=2$, $\omega_2=3/2$, $x_0=3\i/2$, $c_1=1$, and
  $\theta_{10}=-2p_1\zeta(3\i/2)$.
  (a) $p_1=1$; (b) $p_1=1+3\i/2$.}
  \label{fig:kdv_n1}
\end{figure}

The two $N=1$ branches are illustrated in \cref{fig:kdv_n1}.

\begin{figure}[!htbp]
  \centering
  \subfloat[Bright--bright interaction.]{%
    \includegraphics[width=0.48\textwidth]{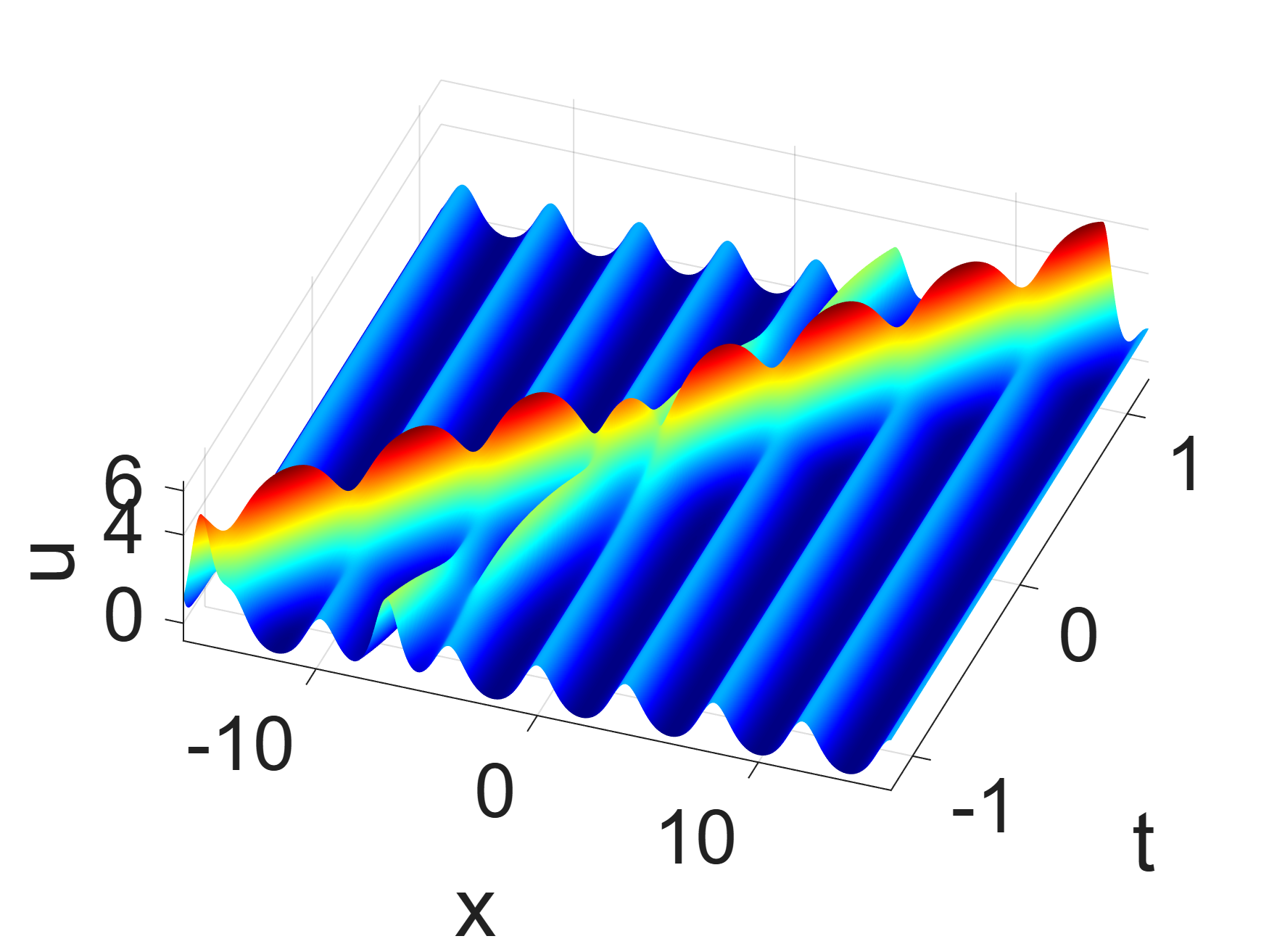}}
  \hfill
  \subfloat[Bright--dark interaction.]{%
    \includegraphics[width=0.48\textwidth]{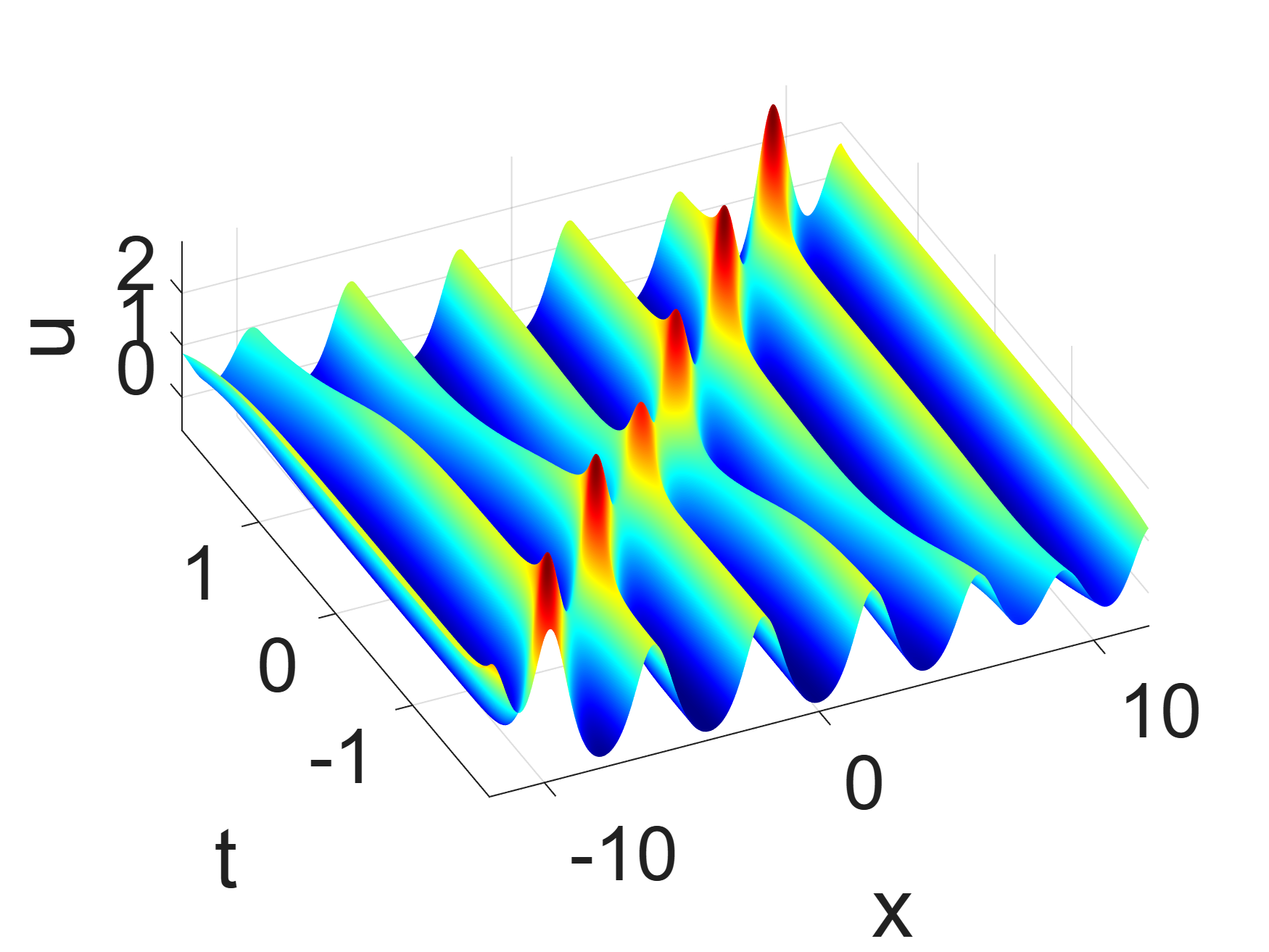}}
  \caption{The $N=2$ KdV solutions $u(x,t)$ with $\Lambda=\Lambda_p$,
  $\omega_1=2$, $\omega_2=3/2$, $x_0=3\i/2$, $c_1=c_2=1$, and
  $\theta_{j0}=-2p_j\zeta(3\i/2)$ for $j=1,2$.
  (a) $(p_1,p_2)=(0.6,0.9)$; (b) $(p_1,p_2)=(0.95,1+3\i/2)$.}
  \label{fig:kdv_n2}
\end{figure}

Multi-breather solutions (\(N=2,3,\ldots\)) can be obtained from
\cref{cor:KdV-elliptic-determinants}. Bright--bright and bright--dark
interactions are illustrated in \cref{fig:kdv_n2}.

\begin{remark}\label[remark]{rem:KdV-Hoefer-conversion}
  It is noted that under transformation
  \begin{align*}
    x \to \frac1 \nu X, \quad t \to \frac1{\nu^3}T, \quad u(x,t) \to \nu^2 U(X,T),
  \end{align*}
  \eqref{eq:kdv_traveling_solution} becomes
  \begin{align*}
    U(X,T)=2\kappa^2\cn^2(X-c_0T,\kappa), \quad c_0=\frac{12e_3}{\nu^2}.
  \end{align*}
  This is a solution to the KdV equation \(U_T+6UU_X+U_{XXX}=0\). Here \(U(X,T)\) is exactly Eq.~(4) in \cite{Hoefer_2023}.
  For above \(N=1\) case, our bright and dark breather solutions can be transformed into the same form of Eqs.~(6) and (7) in \cite{Hoefer_2023} via the relationship between Weierstrass \(\sigma\) and Jacobi \(\vartheta_4\)
  \begin{align*}
    \frac{\sigma(x+\i\omega_2\pm p_1)}{\sigma(x+\i\omega_2)}e^{\mp p_1\zeta(\i\omega_2)}
    &=\exp\left(\frac{\zeta(\omega_1)}{2\omega_1}(p_1^2\pm2xp_1)\right)
    \frac{\Theta(\nu x\pm\nu p_1)}{\Theta(\nu x)},\\
    \frac{\sigma(a+\i\omega_2\pm x)}{\sigma(a+\i\omega_2)}
    e^{\mp\zeta(a+\i\omega_2)x}
    &=
    \exp\left[
    \frac{\zeta(\omega_1)}{2\omega_1}(x^2\pm2ax)
    \mp\bigl(\zeta(a+\i\omega_2)-\zeta(\i\omega_2)\bigr)x
    \right]
    \frac{\Theta(\nu x\pm\nu a)}{\Theta(\nu a)},
  \end{align*}
  where \(\Lambda=\Lambda_p\) and
  \begin{align*}
    &\Theta(z):=\vartheta_4\!\left(\frac{\pi z}{2K},q\right), \quad \vartheta_4(z,q)=1+2\sum_{n=1}^{\infty}(-1)^n q^{n^2}\cos(2nz), \\
    &q=e^{-\pi\omega_2/\omega_1},\quad K= \nu\omega_1, \quad \nu=\sqrt{e_1-e_2}.
  \end{align*}
\end{remark}

\subsection{Solitons to the mKdV equation on elliptic backgrounds}

As a direct application to the modified KP hierarchy of the elliptic type, we consider the solution to the mKdV equation
\begin{equation}\label{elli-mKdV}
   u_t+ u_{xxx}-6\epsilon^2 u^2 u_x=0, \quad \epsilon = 1,\i, \quad  x,t\in \mathbb R,
\end{equation}
here \(\epsilon = 1\) for the defocusing case and \(\epsilon = \i\) for the focusing case.
Starting from \eqref{elli-mKP-bl1} and \eqref{elli-mKP-bl2}, for the $2$-reduction, i.e., condition 
\[
\partial_{x_2} \tau_n(\bm x) = C\tau_n(\bm x), \quad C\text{ is constant},
\]
to be satisfied, take $q_j=-p_j$ for \eqref{mkp_casorati} and \eqref{mkp_wronskian}, and $q_j=p_j,c_{ij}=\delta_{ij}c_i$ for \eqref{mkp_gram}. Setting $x_1=x$, $x_3=-4t$, and \(n=0\), \eqref{elli-mKP-bl1}--\eqref{elli-mKP-bl2} reduce to
\begin{align}
  &\bigl(D_x^2+2\zeta(c)D_x+\zeta^2(c)-\wp(c)\bigr)
  \tau_0\cdot\tau_{1}=0,\label{elli-mKdV-bl1}\\
  &\Big[D_x^3+D_t-3\bigl(\zeta^2(c)-\wp(c)\bigr)D_x-2\bigl(\zeta^3(c)-3\zeta(c)\wp(c)-\wp'(c)\bigr)\Big]
  \tau_0\cdot\tau_{1}=0.\label{elli-mKdV-bl2}
\end{align}
Eqs. \eqref{elli-mKdV-bl1}--\eqref{elli-mKdV-bl2} give the mKdV equation \eqref{elli-mKdV} by the transformation
\begin{equation}\label{mkdv_u}
    u(x,t)=\epsilon\left[\zeta(c)+\partial_x \log\frac{\tau_0(x+6\wp(c)t,t)}{\tau_{1}(x+6\wp(c)t,t)}\right].
\end{equation}

Therefore, the general $N$-soliton solution to the mKdV equation is given by the following tau function.
\begin{theorem}\label[theorem]{cor:mKdV-elliptic-determinants}
The tau functions of the solution \eqref{mkdv_u} to the mKdV equation are given in the following forms.
\begin{enumerate}[(a)]
  \item Casorati determinant
  \begin{align*}
    \tau_n(x,t) = \sigma(\xi)
    \begin{vmatrix}
      \hat\phi_1(0;n) & \hat\phi_1(1;n) & \cdots & \hat\phi_1(N-1;n) \\
      \hat\phi_2(0;n) & \hat\phi_2(1;n) & \cdots & \hat\phi_2(N-1;n) \\
      \vdots & \vdots & \ddots & \vdots \\
      \hat\phi_N(0;n) & \hat\phi_N(1;n) & \cdots & \hat\phi_N(N-1;n)
    \end{vmatrix}
  \end{align*}
  \item Wronskian determinant
  \begin{align*}
    \tau_n(x,t) = \sigma(\xi)
    \begin{vmatrix}
      \hat\phi_1(0;n) & \partial_x\hat\phi_1(0;n) & \cdots & \partial_x^{N-1}\hat\phi_1(0;n) \\
      \hat\phi_2(0;n) & \partial_x\hat\phi_2(0;n) & \cdots & \partial_x^{N-1}\hat\phi_2(0;n) \\
      \vdots & \vdots & \ddots & \vdots \\
      \hat\phi_N(0;n) & \partial_x\hat\phi_N(0;n) & \cdots & \partial_x^{N-1}\hat\phi_N(0;n)
    \end{vmatrix}
  \end{align*}
  \item Gram-type determinant
  \begin{align*}
    \tau_n(x,t) = \sigma(\xi)
    \det_{1\leq i,j\leq N}\left(\hat m_{ij}^{(n)}\right)
  \end{align*}
\end{enumerate}
while \(\xi=x+nc+x_0\) and the corresponding matrix elements are defined as
\begin{align*}
  \hat\phi_j(s;n)
  &=\alpha_j\left[\Phi_b(p_j)\right]^s
    \left[\Phi_c(p_j)\right]^{-n}
    \Phi_{\xi-bs}(p_j)\exp(\mu_j)+\beta_j\left[\Phi_b(-p_j)\right]^s
    \left[\Phi_c(-p_j)\right]^{-n}
    \Phi_{\xi-bs}(-p_j)\exp(-\mu_j),\\
  \mu_j
  &=-\zeta(p_j)x-2\wp'(p_j)t,\\
  \hat m_{ij}^{(n)}
  &=\delta_{ij}c_i+\Phi_\xi(p_i+p_j)
    \left[\frac{\Phi_c(-p_j)}{\Phi_c(p_i)}\right]^n
    \exp(\theta_{ij}),\\
  \theta_{ij}
  &=-(\zeta(p_i)+\zeta(p_j))x
    -2(\wp'(p_i)+\wp'(p_j))t+\mu_{i0}+\eta_{j0}.
\end{align*}
Here \(p_j\), \(j=1,\cdots,N\) represent wave numbers and
\(b,c,x_0,\alpha_j,\beta_j,c_i,\mu_{i0},\eta_{j0}\) are constants.
\end{theorem}

\begin{corollary}\label[corollary]{cor:mkdv_nonsingular}
For the Gram determinant form, if \(\mu_{i0}+\eta_{i0}\in\mathbb R\), then the following choices give a real-valued, nonsingular mKdV solution.
\begin{enumerate}
    \item Defocusing mKdV ($\epsilon=1$):  $\Lambda=\Lambda_p$, $0<c\leq\omega_1$, $x_0=\i\omega_2$, and
    distinct \(p_i\) satisfying
    \begin{equation*}
     p_i\in(c,\omega_1)\ \cup\
              \bigl(\i\omega_2+(0,\omega_1)\bigr),
     \quad c_i e^{-2\zeta(\i \omega_2)p_i}>0.
    \end{equation*}
    \item Focusing mKdV ($\epsilon=\i$): $c=\i\gamma$, $0<\gamma<2\omega_2$, and 
    distinct \(p_i\) satisfying
    \begin{equation*}
     p_i\in(0,\omega_1)\ \cup\
                    \bigl(\i\omega_2+(0,\omega_1)\bigr),
    \end{equation*}
    in this case, both \(\Lambda_p\) and \(\Lambda_c\) occur
    \begin{align*}
     &\Lambda = \Lambda_p,\quad  x_0=-\frac c2:&&
            c_i\Phi_c(p_i)\in\i\mathbb R\setminus\{0\},\\
    &\Lambda = \Lambda_c, \quad p_i+p_j\notin\Lambda_c, \quad x_0=-\frac c2:&&
            c_i\Phi_c(p_i)\in\i\mathbb R\setminus\{0\},\\
    &\Lambda = \Lambda_p, \quad x_0=\i\omega_2-\frac c2:&&
            c_i\Phi_c(p_i)e^{-2\zeta(\i \omega_2)p_i}
        \in\i\mathbb R\setminus\{0\}.
    \end{align*}
\end{enumerate}
\end{corollary}


Let us consider the elliptic background (\(N=0\)) and $N=1$ soliton solutions.
\paragraph{Elliptic background solution (\(N=0\)).} 
In this case, \(\tau_n(X,t) = \sigma(X+nc+ x_0)\). Thus the traveling seed solution is
\begin{equation}\label{mkdv_n0_seed}
  u(x,t)
  = \epsilon\left[\zeta(c)+\zeta(X+x_0)-\zeta(X+c+x_0)\right].
\end{equation}
We can identify four nonsingular solutions by \cref{cor:mkdv_nonsingular}
\begin{enumerate}
\item Defocusing case (\(\epsilon=1\)). (i) \(\Lambda = \Lambda_p\), \(0<c\leq\omega_1\) and \(x_0=\i\omega_2\). 
  \item Focusing case (\(\epsilon=\i\)). 
  (ii) \(\Lambda = \Lambda_p\), \(c=\i\gamma\), \(x_0 = -\frac c2\); (iii) \(\Lambda = \Lambda_p\), \(c=\i\gamma\), \(x_0=\i\omega_2-\frac c2\); (iv) \(\Lambda = \Lambda_c\), \(c=\i\gamma\), \(x_0=-\frac c2\).
\end{enumerate}
In particular, with a special choice \(c=\omega_1\) for defocusing case and \(c=\i\omega_2\) for focusing case, we can rewrite the solution \eqref{mkdv_n0_seed} as \(u(x,t)=a U(ax,a^3t)\) where
\begin{align*}
    &\text{(i)}&  U(x,t)&=k\operatorname{sn}\left(x+(1+k^2)t,k\right), \quad  k=\frac{\sqrt{e_1-e_2}-\sqrt{e_1-e_3}}{\sqrt{e_1-e_2}+\sqrt{e_1-e_3}}, \quad a=\sqrt{e_1-e_2}+\sqrt{e_1-e_3},\\
    &\text{(ii)}&  U(x,t)&=-\operatorname{dn}\left(x-(2-k^2)t,k\right),\quad k=\frac{2\bigl[(e_1-e_2)(e_3-e_2)\bigr]^{1/4}}{\sqrt{e_1-e_2}+\sqrt{e_3-e_2}}, \quad a = \sqrt{e_1-e_2}+\sqrt{e_3-e_2},\\
    &\text{(iii)}&  U(x,t)&=\operatorname{dn}\left(x-(2-k^2)t,k\right),\quad k=\frac{2\bigl[(e_1-e_2)(e_3-e_2)\bigr]^{1/4}}{\sqrt{e_1-e_2}+\sqrt{e_3-e_2}}, \quad a = \sqrt{e_1-e_2}+\sqrt{e_3-e_2},\\
    &\text{(iv)}& U(x,t)&=-k\operatorname{cn}\!\left(x-(2k^2-1)t,k\right), \quad k=\sqrt{\frac12-\frac{3e_1}{4\sqrt{(e_1-e_2)(e_1-e_3)}}}, \quad a=2\bigl[(e_1-e_2)(e_1-e_3)\bigr]^{1/4}.
\end{align*}

\paragraph{One soliton on elliptic background (\(N=1\)).}
Setting \(\mu_{10}=\eta_{10}=0\) in the Gram determinant solution of \cref{cor:mKdV-elliptic-determinants}, we obtain
\begin{align}\label{eq:mkdv_n1_tau}
  \tau_n(x,t)&=c_1\sigma(\xi)
    +\left[\frac{\Phi_c(-p_1)}{\Phi_c(p_1)}\right]^n
      \frac{\sigma(\xi+2p_1)}{\sigma(2p_1)}e^{-2(\zeta(p_1)x+2\wp'(p_1)t)},
    \quad \xi=x+nc+x_0,
\end{align}
The solution is
\[
  u(x,t)=\epsilon\left[\zeta(c)+\partial_X
    \log\frac{\tau_0(X,t)}{\tau_1(X,t)}\right],
  \quad X=x+6\wp(c)t.
\]

\begin{enumerate}[(a)]
    \item Defocusing case (\(\epsilon=1\)). Consider the rectangular lattice \(\Lambda_p\). Assume
    \(c_1\ne0\), \(2p_1\notin\Lambda_p\), and
    \(c+ p_1\notin\Lambda_p\). We have the following three different solutions.
    \begin{itemize}
         \item Bright breather: \(0<c<p_1<\omega_1\),
  \(x_0=\i\omega_2\), and \(c_1\sigma(2p_1)e^{-2p_1\zeta(\i\omega_2)}>0\).

  \item Dark breather: \(0<c<\omega_1\),
  \(p_1=a+\i\omega_2\), \(0<a<\omega_1\), \(x_0=\i\omega_2\), and \(c_1e^{-2p_1\zeta(i\omega_2)}>0\).

  \item Kink breather: \(0<c<\omega_1, p_1 = c, x_0 = \i\omega_2\) and \(c_1=\frac{e^{2c\zeta(\i\omega_2)}}{\sigma(2c)}\), the kink solution takes an alternative form
     \begin{align*}
      u(x,t)={}&\zeta(c)+\zeta(X+\i\omega_2)
        -\zeta(X+\i\omega_2+c)\\
      &+\partial_X\log\left[
        1+\frac{\sigma(X+\i\omega_2+2c)}
                 {\sigma(X+\i\omega_2)}
        e^{-2c\zeta(\i\omega_2)-2\zeta(c)X-4\wp'(c)t}
        \right].
    \end{align*}

The bright, dark, and kink breathers are illustrated in \cref{fig:mkdv_n1_defocusing}.
\begin{figure}[!htbp]
  \centering
  \subfloat[Bright breather.\label{fig:mkdv_n1_defocusing_real}]{%
    \includegraphics[width=0.32\textwidth]{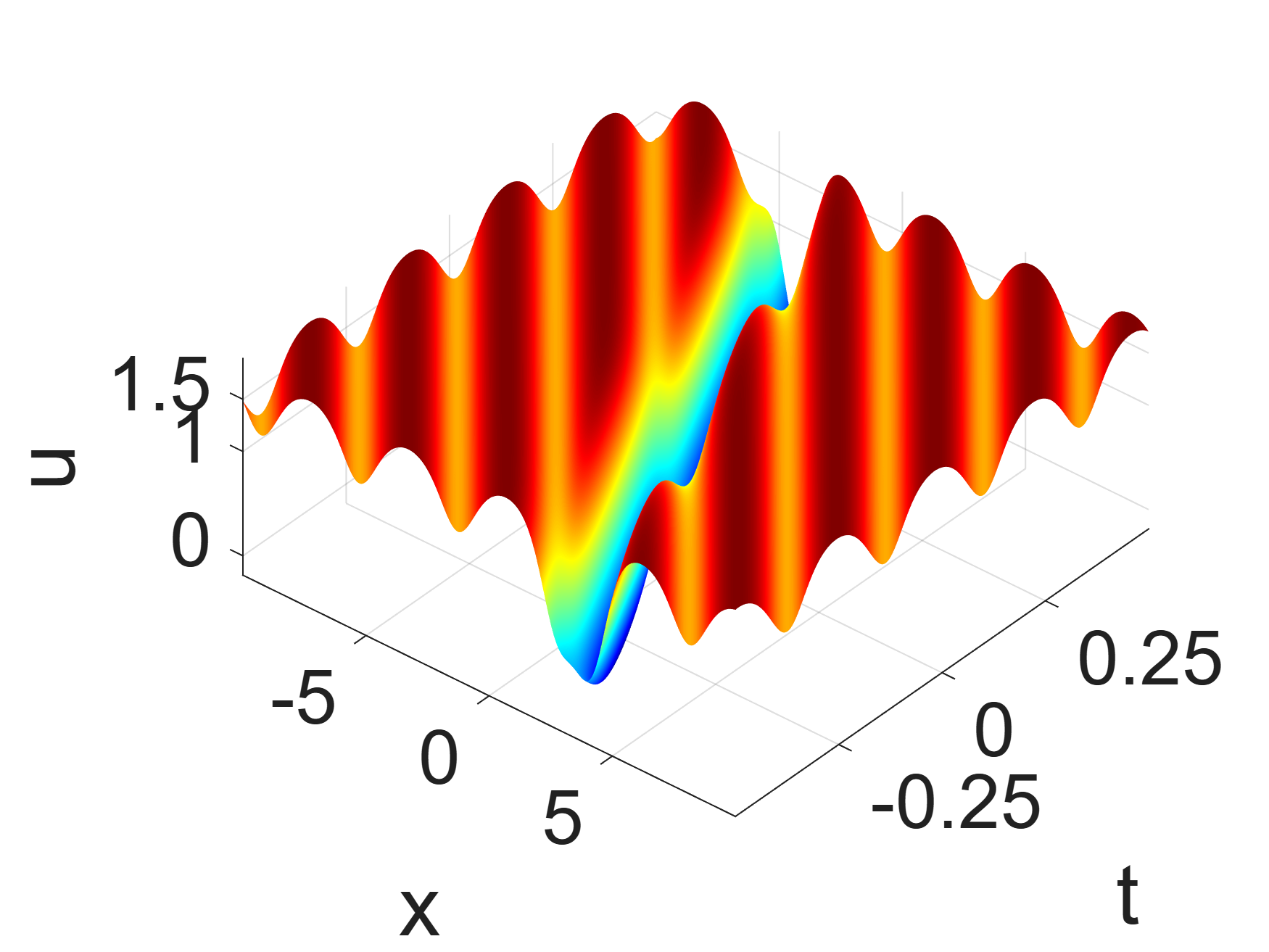}}
  \hfill
  \subfloat[Dark breather.\label{fig:mkdv_n1_defocusing_shifted}]{%
    \includegraphics[width=0.32\textwidth]{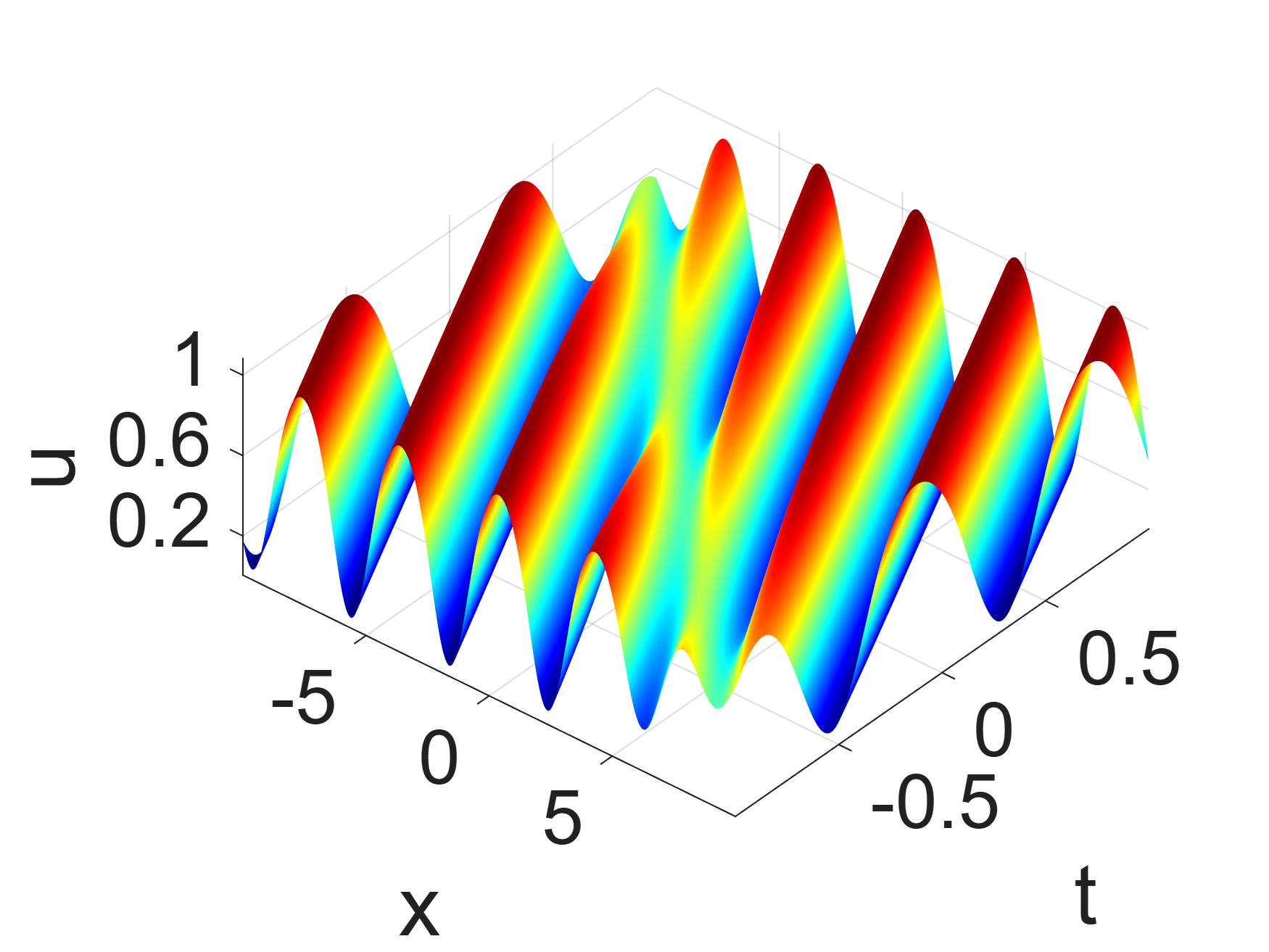}}
  \hfill
  \subfloat[Kink breather.\label{fig:mkdv_n1_defocusing_kink}]{%
    \includegraphics[width=0.32\textwidth]{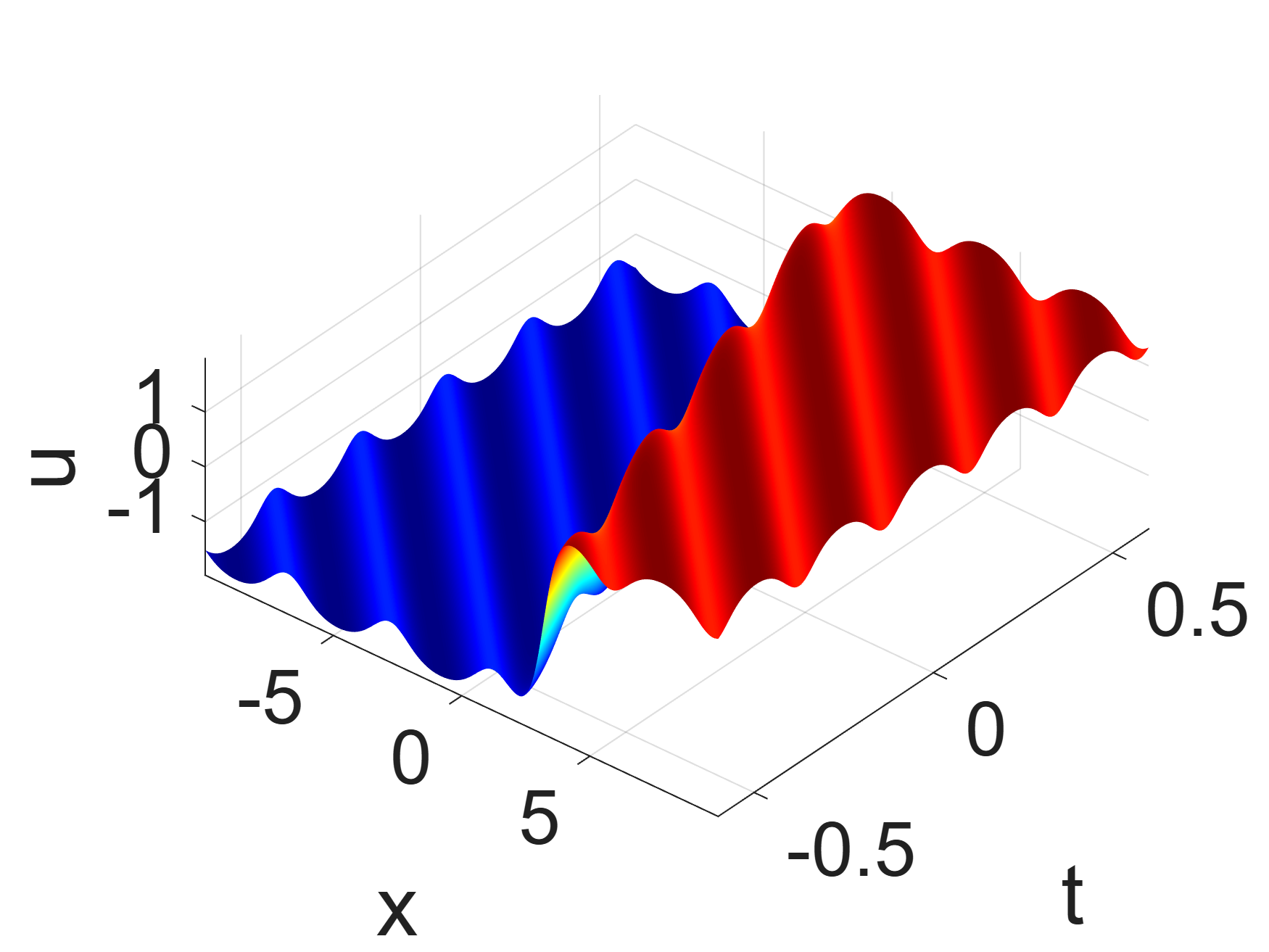}}
  \caption{The defocusing $N=1$ mKdV solutions $u(x,t)$ with $\epsilon=1$,
  $\Lambda=\Lambda_p$, $\omega_1=2$, $\omega_2=3/2$, $\mu_{10}=\eta_{10}=0$, and $x_0=3\i/2$.
  (a) $c=0.6$, $p_1=0.7$, and $c_1=e^{2p_1\zeta(3\i/2)}$.
  (b) $c=1.2$, $p_1=0.9+3\i/2$, and $c_1=e^{2p_1\zeta(3\i/2)}$.
  (c) $c=p_1=0.6$ and $c_1=e^{2c\zeta(3\i/2)}/\sigma(2c)$.}
  \label{fig:mkdv_n1_defocusing}
\end{figure}
    \end{itemize}

    \item Focusing case (\(\epsilon=\i\)). Similar to the seed solution, nonsingular solutions can be obtained on both rectangular
and rhombic lattices. Take \(c=\i\gamma\), \(0<\gamma<2\omega_2\).

\begin{itemize}
    \item Rectangular lattice: \(\Lambda=\Lambda_p\), \(0< p_1 < \omega_1\), \(x_0=-\dfrac c2\), \(c_1\Phi_c(p_1)\in\i\mathbb R\setminus\{0\}\).

    \item Half-period limit: \(\Lambda=\Lambda_p\), \(p_1\to\omega_1\), \(x_0=-\dfrac c2\), and \(c_1=-\frac{\i\sqrt{\wp(\omega_1)-\wp(c)}}{2(\omega_1-p_1)\Phi_c(p_1)}\), the resulting nonsingular solution can be written as
    \begin{align*}
      u(x,t)=\mathrm i\Biggl[
      &\zeta(c)+\zeta\!\left(X-\frac c2\right)
      -\zeta\!\left(X+\frac c2\right)\\
      &+\partial_X\log
      \frac{
      \displaystyle
      \zeta\!\left(X-\frac c2\right)
      +\wp(\omega_1)X-2\wp''(\omega_1)t
      +\frac12\bigl[\zeta(c+\omega_1)-\zeta(\omega_1)\bigr]
      }{
      \displaystyle
      \zeta\!\left(X+\frac c2\right)
      +\wp(\omega_1)X-2\wp''(\omega_1)t
      -\frac12\bigl[\zeta(c+\omega_1)-\zeta(\omega_1)\bigr]
      }
      \Biggr],
    \end{align*}
    \item Rhombic lattice: \(\Lambda=\Lambda_c\), \(0< p_1 < \omega_1\), \(x_0=-\frac c2,\
    c_1\Phi_c(p_1)\in\i\mathbb R\setminus\{0\}.
    \)
    When $\zeta(p_1)-p_1\zeta(\omega_1)/\omega_1=0$, the solution is
    spatially periodic. This zero-rate case is shown in
    \cref{fig:mkdv_n1_focusing_zero}.
\end{itemize}
The three cases are illustrated in \cref{fig:mkdv_n1_focusing}.
\begin{figure}[!htbp]
  \centering
  \subfloat[Rectangular lattice.\label{fig:mkdv_n1_focusing_real}]{%
    \includegraphics[width=0.32\textwidth]{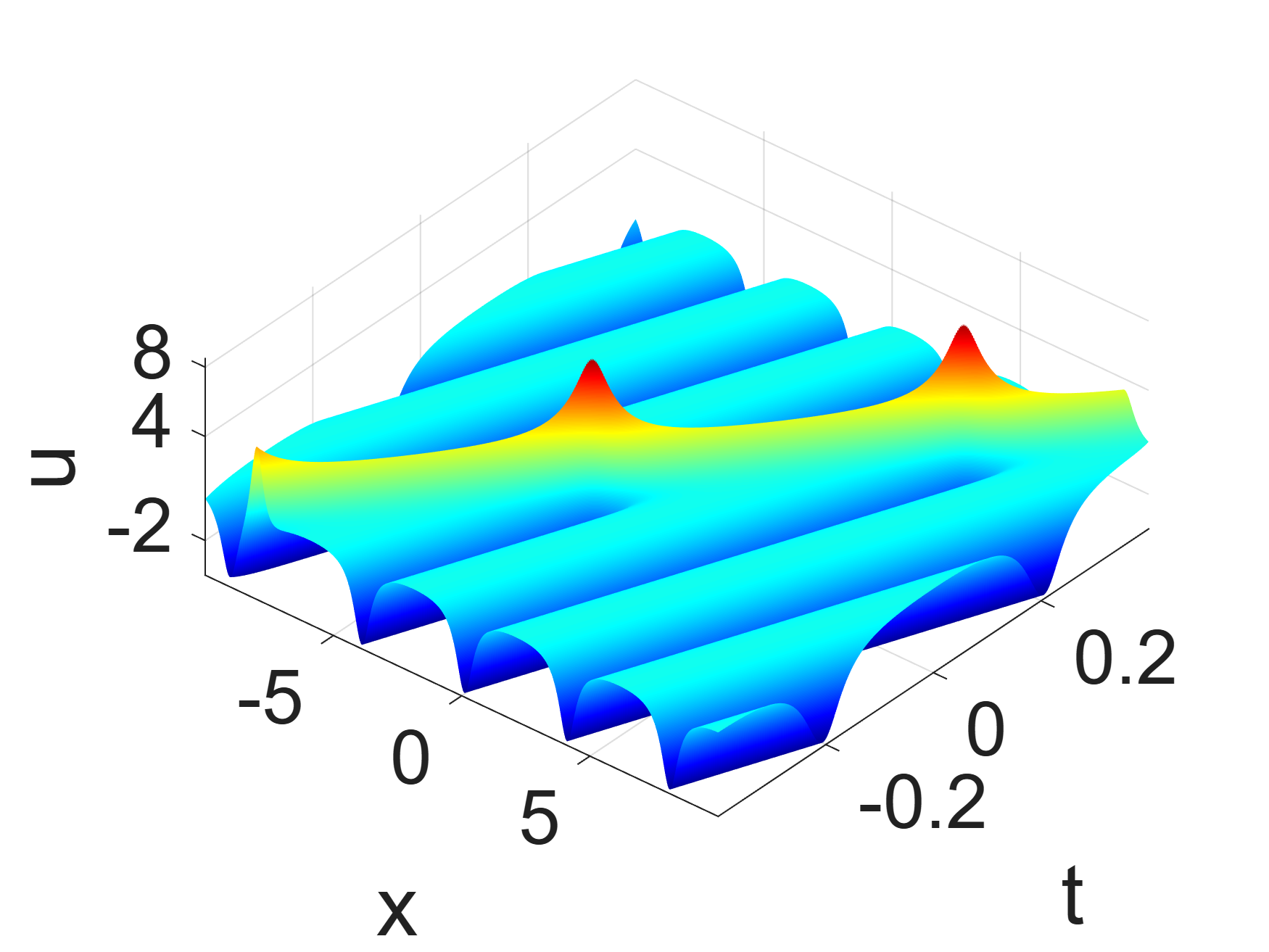}}
  \hfill
  \subfloat[Half-period limit.\label{fig:mkdv_n1_focusing_limit}]{%
    \includegraphics[width=0.32\textwidth]{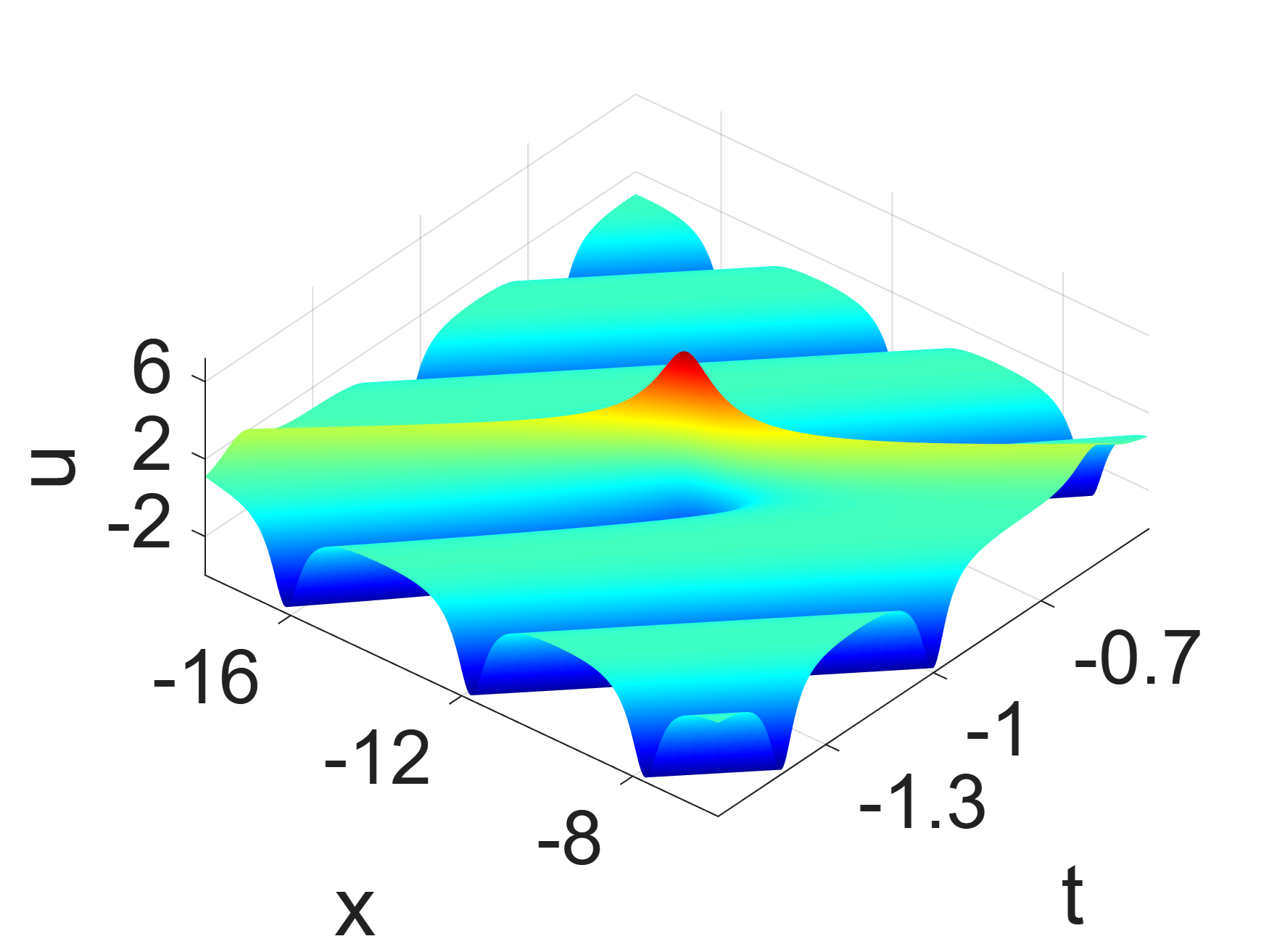}}
  \hfill
  \subfloat[Rhombic zero-rate case.\label{fig:mkdv_n1_focusing_zero}]{%
    \includegraphics[width=0.32\textwidth]{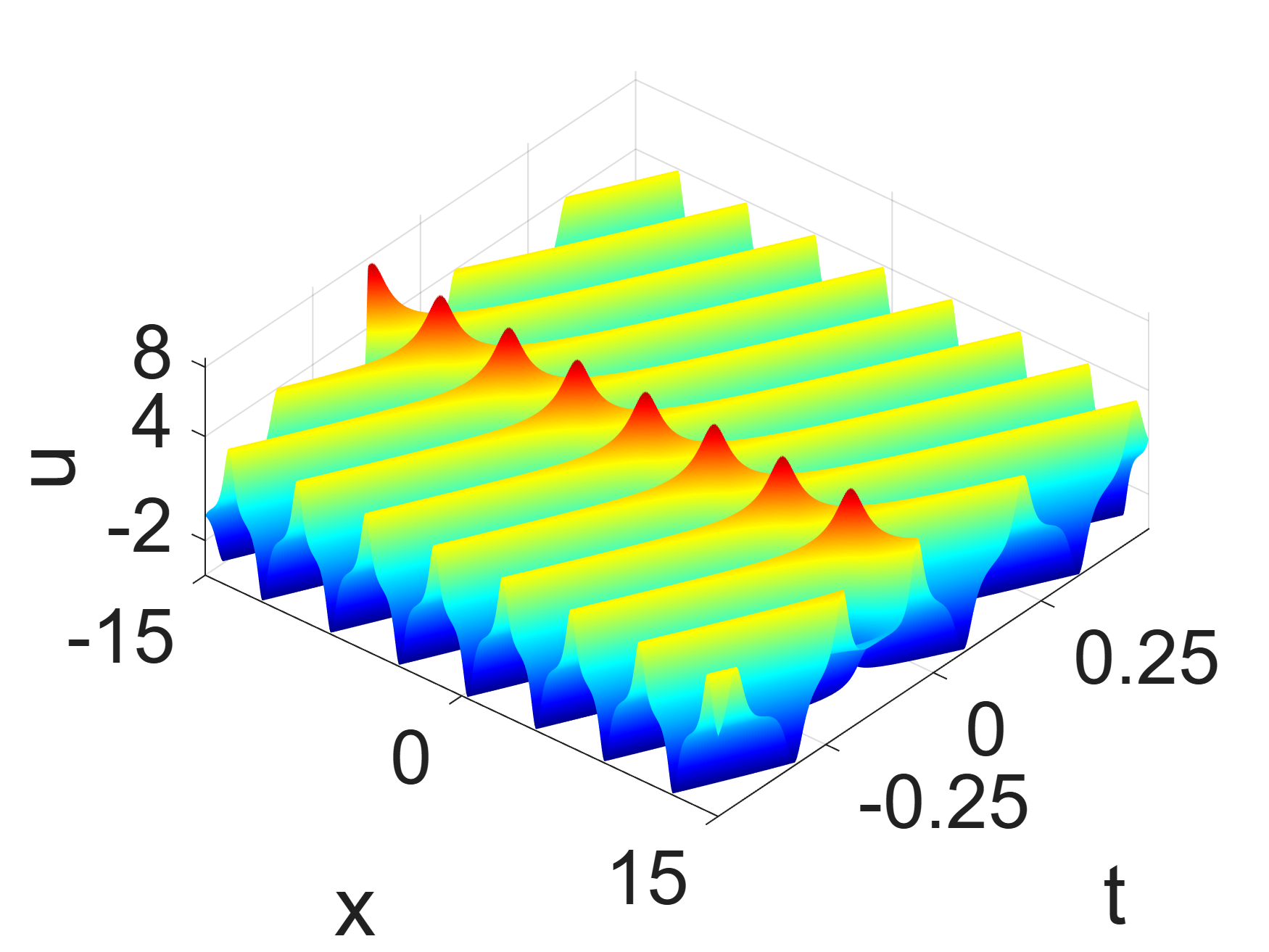}}
  \caption{The focusing $N=1$ mKdV solutions $u(x,t)$ with
  $\epsilon=\i$, $\omega_1=2$, $c=0.8\i$, $\mu_{10}=\eta_{10}=0$, and $x_0=-c/2$.
  (a) $\Lambda=\Lambda_p$, $\omega_2=3/2$, $p_1=0.6$, and
  $c_1=-\i\sqrt{\wp(\omega_1)-\wp(c)}/[2(\omega_1-p_1)\Phi_c(p_1)]$.
  (b) The limit $p_1\to\omega_1=2$ of (a).
  (c) $\Lambda=\Lambda_c$, $\omega_2=0.8$,
  $p_1\simeq1.081863359241729$ satisfying $\zeta(p_1)-p_1\zeta(\omega_1)/\omega_1=0$,
  and $c_1=-\i\rho/\Phi_c(p_1)$ with $\rho\simeq1.19588536655$.}
  \label{fig:mkdv_n1_focusing}
\end{figure}

\end{enumerate}

\subsection{Solitons to the Sawada--Kotera equation on an elliptic background}
\label{sec:sk-report}
The continuous BKP hierarchy admits a reduction to the Sawada--Kotera
equation \cite{Hirota1989BKP,Hirota2004Direct}. The corresponding reduction of the elliptic BKP hierarchy gives solutions on an elliptic
background of the Sawada--Kotera (SK) equation
\begin{equation}
9u_t+u_{xxxxx}+15uu_{xxx}+15u_xu_{xx}+45u^2u_x=0,
 \quad x,t\in\mathbb R.
 \label{eq:sk-report-equation}
\end{equation}
The following $N$-soliton solution on an elliptic background can be derived.
\begin{theorem}
\label{thm:sk-report-general}
Choose $2N$ spectral parameters $p_1,\ldots,p_{2N}$ satisfying
\begin{equation}
 \wp'(p_k)+\wp'(p_{2N+1-k})=0,\quad
 \wp(p_k)\ne\wp(p_{2N+1-k}),
 \label{eq:sk-report-general-pairs}
\end{equation}
where \(k=1,\ldots,2N\), and set
\begin{equation}\label{eq:sk-report-normalization}
 \theta_k=-\zeta(p_k)x+\frac{\wp'''(p_k)}{24}t+\theta_{k0},\quad  \xi=\frac{x}{2}+x_0, 
 \end{equation}
where $x_0$, $\theta_{k0}$ and $c_k$ are constants, with
$c_k=-c_{2N+1-k}$.
Define the Pfaffian entries by
\[
 (k,l)=\delta_{l,2N+1-k}c_l
 +[\wp(p_k)-\wp(p_l)]
 \Phi_\xi(p_k+p_l)^2e^{\theta_k+\theta_l},
 \quad 1\leq k<l\leq2N,
\]
and set \(\tau(x,t)=\sigma^2(\xi)(1,2,\ldots,2N)\).
Then \eqref{eq:sk-report-equation} admits
the solution
\[
u(x,t)=2\partial_x^2\log\tau(X,t), \quad X=x-\frac{g_2}{12}t.
\]
\end{theorem}

\begin{proof}
By Theorem~\ref{thm:elliptic-bkp-determinants}, the elliptic BKP hierarchy has the Pfaffian tau function
\begin{align*}
 \tau&=\sigma^2(\xi)(1,2,\ldots,2N),\quad (k,l)=c_{kl}+[\wp(p_k)-\wp(p_l)]
\Phi_\xi(p_k+p_l)^2e^{\theta_k+\theta_l},\\
 \theta_k&=
-\zeta(p_k)x_1
+\frac12\wp'(p_k)x_3
+\frac{\wp'''(p_k)}{24}x_5+\theta_{k0}, \quad \xi = \frac{x_1}{2} + x_0.
\end{align*}
In particular, $\tau$ satisfies the first bilinear equation of
the hierarchy,
\[
 \left(D_{x_1}^6-5D_{x_1}^3D_{x_3}-5D_{x_3}^2
 +9D_{x_1}D_{x_5}
 -\frac34g_2D_{x_1}^2-\frac34g_3\right)\tau\cdot\tau=0.
\]
Under the condition \eqref{eq:sk-report-general-pairs} and taking $c_{kl}=\delta_{l,2N+1-k}c_{l}$, $c_{l}=-c_{2N+1-l}$, 3-reduction is realized and 
the terms involving $D_{x_3}$ disappear, which leads to
\[
 \left(D_{x_1}^6+9D_{x_1}D_{x_5}
 -\frac34g_2D_{x_1}^2-\frac34g_3\right)\tau\cdot\tau=0.
\]
By setting $x_1=X=x-g_2t/12$, $x_3=0$, $x_5=t$, it is easy to show that
\[
 u(x,t)=2\partial_x^2\log\tau(X,t)
\]
solves \eqref{eq:sk-report-equation}.
\end{proof}
\begin{corollary}\label[corollary]{cor:sk_nonsingular}
For the Pfaffian solution in \cref{thm:sk-report-general}, take
$\Lambda=\Lambda_p$, $x_0=\i\omega_2$, and distinct
spectral points $p_1,\ldots,p_N$ satisfying
\[
0<\Re p_k<\omega_1,\quad
0<\Im p_k<\omega_2,\quad
\Re\wp'(p_k)=0
\]
where \(k=1,\ldots,N\), together with the conditions
\begin{align*}
&c_{2N+1-k}>0,\quad c_k=-c_{2N+1-k},
\quad p_{2N+1-k}=\overline{p}_k, \\
&\theta_{k0}+\theta_{2N+1-k,0}
 +2\zeta(\i\omega_2)(p_k+\overline{p}_k)
 -\frac{\i\pi}{2}\in\mathbb R.
\end{align*}
Then the corresponding solution \(u(x,t)\) is real-valued and
nonsingular.
\end{corollary}

Let us consider the elliptic background (\(N=0\)) and \(N=1\) soliton solutions based on \cref{cor:sk_nonsingular}.
\paragraph{Elliptic background solution (\(N=0\)).}
Taking $N=0$ in \cref{thm:sk-report-general} gives
\begin{equation*}
\tau(X,t)=\sigma^2\!\left(\frac{X}{2}+\i\omega_2\right),\quad
 u(x,t)=-\wp\!\left(\frac{X}{2}+\i\omega_2\right).
\end{equation*}
By the relation between the Weierstrass $\wp$ and Jacobi $\cn$
functions, with $e_1>e_3>e_2$ the roots of $4w^3-g_2w-g_3=0$ as in \cref{two_real_lattices},
\begin{equation*}
 u(x,t)=-e_3+(e_3-e_2)\cn^2\!\left(\frac{\nu X}{2},\kappa\right),
 \quad \nu^2=e_1-e_2,\quad \kappa^2=\frac{e_3-e_2}{e_1-e_2}.
\end{equation*}

\paragraph{One soliton on elliptic background (\(N=1\)).}
For $N=1$ in \cref{thm:sk-report-general}, with $q_1=p_2$
and $c=c_2=-c_1$, the tau function is
\begin{equation*}
 \tau(X,t)=c\sigma^2\left(\frac X2+\i\omega_2\right)
  +\frac{\wp(p_1)-\wp(q_1)}{\sigma^2(p_1+q_1)}
        \sigma^2\left(\frac X2+\i\omega_2+p_1+q_1\right)e^{\theta_1+\theta_2}.
\end{equation*}
Here the phases are
\begin{equation*}
 \theta_k=-\zeta(p_k)X+\frac{\wp'''(p_k)}{24}t+\theta_{k0},
 \quad k=1,2,
\end{equation*}
and the spectral pair satisfies
\begin{equation*}
 \wp'(p_1)+\wp'(q_1)=0,\quad \wp(p_1)\ne\wp(q_1).
\end{equation*}

\section{Conclusion and further topics}
We have developed elliptic analogues of the discrete AKP and BKP
hierarchies and constructed their determinant and Pfaffian tau
functions through Darboux, binary Darboux, and Moutard transformations.
Using determinant and Pfaffian identities, we have directly proved
that these tau functions satisfy the higher-order discrete hierarchy
equations. Through the Miwa transformation, we obtain the continuous
elliptic KP, modified KP, and BKP hierarchies together with their tau
functions. Applications include soliton solutions of the KP, KdV,
focusing and defocusing mKdV, and Sawada--Kotera equations on elliptic
backgrounds. Spectral differentiation further yields KP-I lump
solutions, including fundamental and higher-order examples.
Sufficient reality and regularity conditions are also given for
several determinant solution families.

The present paper paves the way for constructing various soliton
solutions on elliptic backgrounds within a framework of
bilinear hierarchies and their reductions. Existing approaches include partial degeneration of higher-genus spectral
curves~\cite{BertolaJenkinsTovbis2023}, Darboux transformation based on elliptic seed
solutions~\cite{FengLingTakahashi2020}, or direct linearization using
elliptic Cauchy kernels~\cite{yoo2013elliptic}. 
However, the discrete and continuous AKP and BKP hierarchies of elliptic type presented here  
allow us to obtain solutions of different nonlinear wave
equations systematically through suitable 
reductions.

Several directions call for further investigation. First, we need to extend the present method to integrable PDEs arising from
reductions of the KP--Toda hierarchy involving negative flows such as the NLS and sine--Gordon equations. The
main obstacle within our framework is to construct an elliptic
analogue of the KP--Toda hierarchy that incorporates compatible
positive and negative flows, together with explicit tau functions
satisfying the associated bilinear identities. Such a construction
would substantially broaden the range of equations accessible
through elliptic hierarchy reductions.

Second, an important direction is to construct soliton solutions on
elliptic backgrounds for coupled systems, including the Manakov
system and the coupled Sasa--Satsuma equation. For the Manakov system,
the $3\times3$ Lax problem and the associated trigonal spectral curves
introduce additional spectral structure~\cite{ElginEnolskiiIts2007},
while explicit elliptic waves~\cite{Wright2013Elliptic} and their
density and polarization dynamics~\cite{Kamchatnov2014Periodic}
have already been studied. More recently, Liu et
al.~\cite{LiuEtAl2026ManakovElliptic} constructed vector localized
states on elliptic backgrounds by Darboux transformations, including
 bright--dark solitons and breathers. 

Third, the method could be extended to other $(2+1)$-dimensional
integrable systems, particularly the Davey--Stewartson (DS) equations.
Relevant developments include the elliptic travelling waves
classified by Biondini and Kireyev~\cite{BiondiniKireyev2022DS}, DS-I
rogue lumps constructed through multicomponent KP reductions by Rao,
He, and Cheng~\cite{RaoHeCheng2022DSI}, resonant DS-II solutions studied
by Rao et al.~\cite{RaoEtAl2024DSII}, and spatially bounded DS-I rogue
waves obtained by Yang and Yang~\cite{YangYang2026DSI}. In a recent
preprint, Biondini, Chernyavsky, Lin, and
Ringland~\cite{BiondiniEtAl2026DSWhitham} derive Whitham modulation
equations for the DS systems and analyze the linear stability of
their periodic travelling waves. These studies motivate developing
an elliptic tau-function framework for DS solitons and localized
waves, with particular attention to localization in both spatial
directions, global regularity, and the influence of background
modulation and stability on the resulting wave dynamics.

Fourth, it would be intriguing to construct the discrete CKP and DKP hierarchies of elliptic type, the corresponding continuous hierarchies by the Miwa transformation and the reductions to the (1+1)-dimensional integrable systems such as the 
Kaup–Kupershmidt equation \cite{Kaup1980}. 
An elliptic discrete CKP equation has already been constructed in \cite{Sun2026Elliptic}.

Finally, the discrete hierarchies developed here provide a starting
point for investigating soliton solutions on elliptic backgrounds in
integrable lattice systems, including the scalar and coupled
 Ablowitz--Ladik lattices. 
Their continuum
limits would also clarify the relationship between discrete and
continuous soliton dynamics on elliptic backgrounds.

\section*{Acknowledgments}
B.-F.~Feng expresses his sincere gratitude to Yasuhiro Ohta
and Kenichi Maruno for many years of fruitful collaboration
on integrable systems, particularly discrete integrable
systems. He also thanks Mark Ablowitz, Gino Biondini, Mark Hoefer, Frank Nijhoff, Dmitry Pelinovsky, Jianke Yang and 
Da-jun Zhang
for valuable discussions
about soliton solutions on elliptic backgrounds.
B.-F. ~Feng gratefully acknowledges the support from School of Mathematical and Statistical Sciences (SMSS) Research Collaboration Enhancement Program.

The authors acknowledge supplementary assistance from ChatGPT. The main ideas and proofs were developed and completed by the authors, while ChatGPT assisted with the detailed calculations involved in deriving the seed and \(N=1\) soliton solutions, the development of code for generating the figures, and the polishing of the English in the manuscript.

\appendix
\section{Elliptic function preliminaries}

\subsection{Definitions of Weierstrass functions}
The Weierstrass functions \(\wp(z): \mathbb C \to \mathbb C\cup \{\infty\}\), \(\sigma(z): \mathbb C \to \mathbb C\), and \(\zeta(z): \mathbb C \to \mathbb C\cup \{\infty\}\) are defined by
\begin{align*}
  \wp(z;{\Lambda}) &:= \wp(z) = \begin{cases}\displaystyle
    \frac{1}{z^2} + \sum_{\omega \in {\Lambda}\setminus \{0\}} \left[\frac{1}{(z-\omega)^2} - \frac{1}{\omega^2}\right] & \text{for } z\notin {\Lambda},\\
    \infty & \text{for } z\in {\Lambda},
  \end{cases}\\
  \sigma(z;\Lambda) &:= \sigma(z) = z \prod_{\omega \in {\Lambda}\setminus \{0\}} \left(1-\frac{z}{\omega}\right) \exp\left[\frac{z}{\omega}+\frac{1}{2} \left(\frac{z}{\omega}\right)^2\right], \\
  \zeta{(z;\Lambda )}&:= \zeta(z)={\frac {\sigma '(z;\Lambda )}{\sigma (z;\Lambda )}}={\frac {1}{z}}+\sum _{w\in \Lambda\setminus\{0\} }\left({\frac {1}{z-w}}+{\frac {1}{w}}+{\frac {z}{w^{2}}}\right), \quad \zeta'(z)=-\wp(z),
\end{align*}
where the lattice is given by
\begin{gather*}
  \Lambda=2\Omega_1\mathbb Z+2\Omega_2\mathbb Z
  =\{2m\Omega_1+2n\Omega_2:m,n\in\mathbb Z\},\quad 
  \Omega_1, \Omega_2 \in \mathbb C
  \text{ are linearly independent over }\mathbb R.
\end{gather*}
Here, the Weierstrass \(\wp\) function is doubly periodic with respect to the lattice, whereas the sigma and the zeta functions are quasi-periodic:
\begin{align*}
\wp(z+2m\Omega_1+2n\Omega_2)
&=\wp(z),\\
\sigma(z+2m\Omega_1+2n\Omega_2)
&=(-1)^{m+n+mn}\sigma(z)\exp\left[
2\left(m\zeta(\Omega_1)+n\zeta(\Omega_2)\right)\left(z+m\Omega_1+n\Omega_2\right)
\right],\\
\zeta(z+2m\Omega_1+2n\Omega_2)
&=\zeta(z)+2m\zeta(\Omega_1)+2n\zeta(\Omega_2).
\end{align*}
Moreover, we have $\sigma(-z) = -\sigma(z)$, $\zeta(-z)=-\zeta(z)$, and $\wp(-z)=\wp(z)$. With the derivative of \(\wp\), there is an algebraic differential equation
\begin{align}
  \wp'(z)^2 = 4\wp(z)^3 - g_2 \wp(z) - g_3 = 4(\wp(z) - e_1)(\wp(z) - e_2)(\wp(z) - e_3),\label{elliptic_curve}
\end{align}
where \(g_2 = 60\sum_{\omega\in\Lambda\setminus\{0\} }\omega^{-4}, g_3 = 140\sum_{\omega\in\Lambda\setminus\{0\}} \omega^{-6}\), and the three half-period values are defined by
\begin{align*}
    e_1=\wp(\Omega_1;\Lambda),\qquad
    e_2=\wp(\Omega_2;\Lambda),\qquad
    e_3=\wp(\Omega_1+\Omega_2;\Lambda).
\end{align*}
For real lattices, the indices $1,2,3$ are fixed by the conventions in \cref{two_real_lattices}, and these labels are used throughout the paper. All Weierstrass functions, invariants, and half-period values in a given formula refer to the same lattice.

\subsection{Formulas for Weierstrass functions and elliptic Cauchy determinant}
There are some useful relationships between those Weierstrass functions
\begin{align*}
  \begin{split}
    &\sigma(x + a)\sigma(x - a)\sigma(y + b)\sigma(y - b) - \sigma(x + b)\sigma(x - b)\sigma(y + a)\sigma(y - a) \\
    &= \sigma(x + y)\sigma(x - y)\sigma(a + b)\sigma(a - b),
    \end{split}\\
    &\zeta(\alpha) + \zeta(\beta) + \zeta(\gamma) - \zeta(\alpha + \beta + \gamma) = \frac{\sigma(\alpha + \beta)\sigma(\beta + \gamma)\sigma(\gamma + \alpha)}{\sigma(\alpha)\sigma(\beta)\sigma(\gamma)\sigma(\alpha + \beta + \gamma)},\\
    &\zeta(\alpha + \beta) - \zeta(\alpha) - \zeta(\beta) = \frac{1}{2} \frac{\wp'(\alpha) - \wp'(\beta)}{\wp(\alpha) - \wp(\beta)},\\
    &\left[\zeta(\alpha + \beta) - \zeta(\alpha) - \zeta(\beta)\right]^2 = \wp(\alpha) + \wp(\beta) + \wp(\alpha + \beta),\\
    &\frac{\sigma(x + y)\sigma(x - y)}{\sigma^2(x)\sigma^2(y)} = \wp(y) - \wp(x),\\
    &\bigl(\wp(x) - \wp(a)\bigr)\bigl(\wp(y) - \wp(b)\bigr) - \bigl(\wp(x) - \wp(b)\bigr)\bigl(\wp(y) - \wp(a)\bigr)= \bigl(\wp(x) - \wp(y)\bigr)\bigl(\wp(a) - \wp(b)\bigr),\\
    &\wp(a+b)
    =\frac14\left[\frac{\wp'(a)-\wp'(b)}{\wp(a)-\wp(b)}\right]^2
    -\wp(a)-\wp(b),\\
    &\wp(2z)
    =\frac14\left[\frac{\wp''(z)}{\wp'(z)}\right]^2-2\wp(z)
    =\frac{(\wp(z)^2+g_2/4)^2+2g_3\wp(z)}
    {4\wp(z)^3-g_2\wp(z)-g_3}.
\end{align*}
With the definition of \(\Phi_a(b) = \frac{\sigma(a+b)}{\sigma(a)\sigma(b)}\), we have the following equalities
\begin{align*}
    &\Phi_\kappa(x)\Phi_\kappa(y) = \Phi_\kappa(x + y)\bigl[\zeta(\kappa) + \zeta(x) + \zeta(y) - \zeta(\kappa + x + y)\bigr],\\
    &\Phi_\kappa(x)\Phi_\lambda(y) = \Phi_\kappa(x - y)\Phi_{\kappa + \lambda}(y) + \Phi_{\kappa + \lambda}(x)\Phi_\lambda(y - x),\\
    &\Phi_\kappa(x)\Phi_{-\kappa}(x) = \wp(x) - \wp(\kappa).
\end{align*}
Another important result is the Frobenius formula \cite{frobenius1882ueber}, which is also known as the elliptic Cauchy determinant identity, for example see \cite{NijhoffDelice2018,zabrodin2024elliptic}
\begin{align}\label{elliptic_cauchy_det}
\det_{1\leq k,l\leq N}
\left(
\Phi_{\xi}(p_k+q_l)
\right)=
\frac{\sigma\!\left(\xi+\sum_{k=1}^{N}(p_k+q_k)\right)}
{\sigma(\xi)}
\frac{\displaystyle\prod_{1\leq k<l\leq N}\sigma(p_k-p_l)\sigma(q_k-q_l)}
{\displaystyle\prod_{k,l=1}^{N}\sigma(p_k+q_l)}.
\end{align}

\subsection{Weierstrass functions under real lattices}\label{two_real_lattices}
From complex analysis we know the following statements are equivalent.
\begin{enumerate}
    \item The lattice \(\Lambda\) is real, i.e., \(\omega \in \Lambda\) iff \(\overline \omega \in \Lambda\). 
    \item \(g_2, g_3 \in \mathbb R\).
    \item \(\wp(\bar z) = \overline{\wp(z)}\).
\end{enumerate}
The algebraic curve \eqref{elliptic_curve} is nonsingular if its discriminant \(\Delta = g_2^3 - 27g_3^2\) is nonzero. We assume \eqref{elliptic_curve} is nonsingular. When \(g_2, g_3 \in \mathbb R\), the discriminant naturally gives two types of real lattice: $\Delta > 0$ for the rectangular lattice \(\Lambda_p\), and $\Delta < 0$ for the rhombic lattice \(\Lambda_c\). These lattices can be expressed in the following forms
\begin{align*}
    \Lambda_p &= 2\omega_1\mathbb Z + 2\i\omega_2\mathbb Z, \\
    \Lambda_c &= 2\omega_1\mathbb Z + (\omega_1 +\i \omega_2)\mathbb Z,
\end{align*}
where $\omega_1,\omega_2>0$. The property \(\wp(\bar z) = \overline{\wp(z)}\) allows the following conclusion for real-valuedness of Weierstrass functions.
\begin{enumerate}[(a)]
\item The rectangular lattice $\Lambda_p$.
The half-period values satisfy
\[
e_1>e_3>e_2,\quad e_1>0>e_2,\quad
e_3g_3<0 \ (g_3\ne0).
\]
The reality valuedness of Weierstrass functions under the rectangular lattice is given by
\begin{align*}
\wp(z)&\in\mathbb R,
&&z\in
\Big[(\omega_1\mathbb Z+\i\mathbb R)
\cup(\mathbb R+\i\omega_2\mathbb Z)\Big]\setminus\Lambda_p,\\
\zeta(x)&\in\mathbb R, \zeta(\i y)\in\i\mathbb R,
&&x\in\mathbb R\setminus2\omega_1\mathbb Z,y\in\mathbb R\setminus2\omega_2\mathbb Z,\\
\sigma(x)&\in\mathbb R, \sigma(\i y)\in\i\mathbb R,
&&x,y\in\mathbb R.
\end{align*}
For $x,y\in\mathbb R$,
\[
\zeta(x+\i\omega_2)-\zeta(y+\i\omega_2)
=-\int_y^x\wp(t+\i\omega_2)\,dt\in\mathbb R.
\]
Hence, for $u,v,y\in\mathbb R$,
\begin{align}\label{sigma_positive}
&\frac{\sigma(v+\i\omega_2)}{\sigma(u+\i\omega_2)}
 e^{-(v-u)\zeta(y+\i\omega_2)}=
\exp\left\{
\int_u^v
\bigl[\zeta(x+\i\omega_2)-\zeta(y+\i\omega_2)\bigr]\,dx
\right\}>0.
\end{align}

\item The rhombic lattice $\Lambda_c$.
The half-period values satisfy
\[
e_1\in\mathbb R,\quad
e_2=\overline{e_3}\notin\mathbb R,\quad
g_3=4|e_2|^2e_1,\quad
e_1g_3>0\quad(g_3\ne0).
\]
The reality valuedness of Weierstrass functions under the rhombic lattice is given by
\begin{align*}
\wp(z)&\in\mathbb R,
&&z\in
\Big[(\omega_1\mathbb Z+\i\mathbb R)
\cup(\mathbb R+\i\omega_2\mathbb Z)\Big]\setminus\Lambda_c,\\
\zeta(x)&\in\mathbb R, \zeta(\i y)\in\i\mathbb R,
&&x\in\mathbb R\setminus2\omega_1\mathbb Z,y\in\mathbb R\setminus2\omega_2\mathbb Z,\\
\sigma(x)&\in\mathbb R, \sigma(\i y)\in\i\mathbb R,
&&x,y\in\mathbb R.
\end{align*}
Moreover \(\wp(z+\i\omega_2)=\wp(z+\omega_1)\),
and
\[
\zeta(x+\i\omega_2)-\zeta(\i\omega_2)\in\mathbb R,
\quad x\notin(2\mathbb Z+1)\omega_1.
\]
For $m\in\mathbb Z$ and
$u,v,y\in((2m-1)\omega_1,(2m+1)\omega_1)$,
\begin{align*}
&\frac{\sigma(v+\i\omega_2)}{\sigma(u+\i\omega_2)}
 e^{-(v-u)\zeta(y+\i\omega_2)}=
\exp\left\{
\int_u^v
\bigl[\zeta(x+\i\omega_2)-\zeta(y+\i\omega_2)\bigr]\,dx
\right\}>0.
\end{align*}
Finally, for $x,y\in\mathbb R$ with $0<|y|<\omega_2$,
\begin{gather*}
\sigma(x+\i y)\sigma(x-\i y)=|\sigma(x+\i y)|^2>0, \quad 
\zeta(x+\i y)-\zeta(x-\i y)\in\i\mathbb R.
\end{gather*}
\end{enumerate}

\subsection{Jacobi elliptic and theta functions}
In this section, we use the half-period labels fixed in \cref{two_real_lattices}, and the lattice $\Lambda=\Lambda_p$ for Weierstrass functions unless specified otherwise.
For $0<\kappa<1$, define locally at $z=0$,
\begin{gather*}
z=\int_0^{\sn(z,\kappa)}
 \frac{ds}{\sqrt{(1-s^2)(1-\kappa^2s^2)}},\\
\cn^2(z,\kappa)=1-\sn^2(z,\kappa),\quad
\dn^2(z,\kappa)=1-\kappa^2\sn^2(z,\kappa),\\
\sn(0,\kappa)=0,\quad \cn(0,\kappa)=\dn(0,\kappa)=1,\quad
K=\int_0^{\pi/2}\frac{d\theta}{\sqrt{1-\kappa^2\sin^2\theta}}.
\end{gather*}
The square root equals $1$ at $s=0$; continuation is meromorphic.
\begin{align*}
\frac{d}{dz}\sn(z,\kappa)&=\cn(z,\kappa)\dn(z,\kappa),
&\sn(z+2K,\kappa)&=-\sn(z,\kappa),\\
\frac{d}{dz}\cn(z,\kappa)&=-\sn(z,\kappa)\dn(z,\kappa),
&\cn(z+2K,\kappa)&=-\cn(z,\kappa),\\
\frac{d}{dz}\dn(z,\kappa)&=-\kappa^2\sn(z,\kappa)\cn(z,\kappa),
&\dn(z+2K,\kappa)&=\dn(z,\kappa).
\end{align*}

For $\Lambda=\Lambda_p$, set
\begin{equation}\label{append:jacobi_rect_parameters}
\nu=\sqrt{e_1-e_2},\quad
\kappa^2=\frac{e_3-e_2}{e_1-e_2}.
\end{equation}
Then
\begin{align*}
\wp(z)-e_2&=\frac{\nu^2}{\sn^2(\nu z,\kappa)},\quad
\wp(z)-e_1=\nu^2\frac{\cn^2(\nu z,\kappa)}{\sn^2(\nu z,\kappa)}, \quad \wp(z)-e_3=\nu^2\frac{\dn^2(\nu z,\kappa)}{\sn^2(\nu z,\kappa)},\\
\wp'(z)&=-2\nu^3
 \frac{\cn(\nu z,\kappa)\dn(\nu z,\kappa)}{\sn^3(\nu z,\kappa)}.
\end{align*}
The half-period shift gives
\begin{align}
\wp(z+\i\omega_2)
 &=e_2+(e_3-e_2)\sn^2(\nu z,\kappa)\nonumber\\
 &=e_3-(e_3-e_2)\cn^2(\nu z,\kappa)
 =e_1-\nu^2\dn^2(\nu z,\kappa),
 \label{append:jacobi_shift}\\
\wp'(z+\i\omega_2)
 &=2\nu(e_3-e_2)
 \sn(\nu z,\kappa)\cn(\nu z,\kappa)\dn(\nu z,\kappa).
 \nonumber
\end{align}

For $\Lambda=\Lambda_c$, set
\[
\nu=\bigl[(e_1-e_2)(e_1-e_3)\bigr]^{1/4}>0,
\quad \kappa^2=\frac12-\frac{3e_1}{4\nu^2}\in(0,1).
\]
Then
\begin{equation}\label{append:jacobi_rhomb}
\wp(z)=e_1+\nu^2
 \frac{1+\cn(2\nu z,\kappa)}{1-\cn(2\nu z,\kappa)}.
\end{equation}
In both cases,
\[
K=\nu\omega_1,\quad
\nu\omega_2=\int_0^{\pi/2}
 \frac{d\theta}{\sqrt{1-(1-\kappa^2)\sin^2\theta}}.
\]

For the remaining theta formulas, take $\Lambda=\Lambda_p$, with $\nu,\kappa$ from
\eqref{append:jacobi_rect_parameters} and $K=\nu\omega_1$. Set
\begin{align*}
q&=e^{-\pi\omega_2/\omega_1}\in(0,1),\quad
\Theta(z)=\vartheta_4\!\left(\frac{\pi z}{2K},q\right),\\
\vartheta_4(z,q)
 &=1+2\sum_{n=1}^{\infty}(-1)^nq^{n^2}\cos(2nz)\\
 &=\prod_{n=1}^{\infty}(1-q^{2n})
 \bigl(1-2q^{2n-1}\cos(2z)+q^{4n-2}\bigr).
\end{align*}
Thus
\[
\Theta(-z)=\Theta(z),\quad
\Theta(z+2K)=\Theta(z),\quad
\Theta(x)>0\quad(x\in\mathbb R).
\]
The sigma--theta relation is
\begin{equation}\label{append:sigma_theta}
e^{-\zeta(\i\omega_2)z}
\frac{\sigma(z+\i\omega_2)}{\sigma(\i\omega_2)}
=\exp\!\left(\frac{\zeta(\omega_1)}{2\omega_1}z^2\right)
 \frac{\Theta(\nu z)}{\Theta(0)}.
\end{equation}
Consequently,
\begin{align*}
\zeta(z+\i\omega_2)-\zeta(\i\omega_2)
 &=\frac{\zeta(\omega_1)}{\omega_1}z
 +\nu\frac{\Theta'(\nu z)}{\Theta(\nu z)},\\
\wp(z+\i\omega_2)
 &=-\frac{\zeta(\omega_1)}{\omega_1}
 -\nu^2\left[
 \frac{\Theta''(\nu z)}{\Theta(\nu z)}
 -\left(\frac{\Theta'(\nu z)}{\Theta(\nu z)}\right)^2\right],\\
e^{-h\zeta(\i\omega_2)}
\frac{\sigma(x+h+\i\omega_2)}{\sigma(x+\i\omega_2)}
 &=\exp\!\left[\frac{\zeta(\omega_1)}{2\omega_1}(h^2+2xh)\right]
 \frac{\Theta(\nu x+\nu h)}{\Theta(\nu x)}.
\end{align*}
Primes on $\Theta$ denote derivatives with respect to its argument.


\bibliography{citev2}
\end{document}